\documentclass[abstract=on,floatfix,superscriptaddress,twocolumn,aps,pre]{revtex4-2}

\usepackage{graphicx}
\usepackage{bm}
\usepackage{physics}
\usepackage[mathlines]{lineno}
\usepackage{setspace}
\usepackage[utf8]{inputenc}
\usepackage[english]{babel}
\usepackage[T1]{fontenc}
\usepackage[dvipsnames]{xcolor}
\usepackage{amssymb}
\usepackage{amsmath}
\usepackage{mathtools}
\usepackage{comment}

\usepackage{multirow}
\usepackage{array}
\newcolumntype{P}[1]{>{\centering\arraybackslash}p{#1}}
\newcolumntype{M}[1]{>{\centering\arraybackslash}m{#1}}

\usepackage{amsmath}

\DeclareMathOperator*{\argmin}{arg\,min}
\DeclareUnicodeCharacter{00A0}{~}

\usepackage{lineno}

\begin{document}


\title{Methods for energy dispersive x-ray spectroscopy with photon-counting and deconvolution techniques}

\author{Alessandro Forte}
\email{alessandro.forte@physics.ox.ac.uk}
\affiliation{Department of Physics, Clarendon Laboratory, University of Oxford, Parks Road, Oxford OX1 3PU, UK}

\author{Thomas Gawne}
\affiliation{Center for Advanced Systems Understanding (CASUS), Görlitz, D-02826, Germany}
\affiliation{Helmholtz-Zentrum Dresden-Rossendorf (HZDR), Dresden, D-01328, Germany}

\author{Oliver S. Humphries}
\affiliation{European XFEL, Holzkoppel 4, 22869 Schenefeld, Germany}

\author{Thomas Campbell}
\affiliation{Department of Physics, Clarendon Laboratory, University of Oxford, Parks Road, Oxford OX1 3PU, UK}

\author{Yuanfeng Shi}
\affiliation{Department of Physics, Clarendon Laboratory, University of Oxford, Parks Road, Oxford OX1 3PU, UK}

\author{Sam M. Vinko}
\email{sam.vinko@physics.ox.ac.uk}
\affiliation{Department of Physics, Clarendon Laboratory, University of Oxford, Parks Road, Oxford OX1 3PU, UK}
\affiliation{Central Laser Facility, STFC Rutherford Appleton Laboratory, Didcot OX11 0QX, UK}


\date{\today}

\begin{abstract}
Spectroscopic techniques are essential for studying material properties, but the small cross-sections of some methods may result in low signal-to-noise ratios (SNRs) in the collected spectra. In this article we present methods, based on combining Bragg spectroscopy with photon counting and deconvolution algorithms, which increase the SNRs, making the spectra better suited to further analysis. We aim to provide a comprehensive guide for constructing spectra from camera images. The efficacy of these methods is validated on synthetic and experimental data, the latter coming from the field of high-energy density (HED) science, where x-ray spectroscopy is essential for the understanding of materials under extreme thermodynamic conditions.
\end{abstract}


\maketitle


\section{\label{Introduction} Introduction}

The birth of spectroscopy as a scientific discipline can be traced back to Joseph von Fraunhofer's work with diffraction gratings, which were the first instruments to systematically divide the spectral component of the electromagnetic radiation, dispersing them into distinct regions of space~\cite{born2013principles}. Following this, the invention of the photographic plate in the mid-19th century enabled the first permanent recording of spectral lines, representing the earliest example of a light detector. The following discovery of the photoelectric effect, along with the development of quantum mechanics to support the underlying theory, led to the development of modern detectors, such as photomultiplier tubes (PMTs)~\cite{foord1969use} and later, semiconductor-based detectors like charge-coupled devices (CCDs) and finally hybrid detectors~\cite{peterson2001works}, which are the most commonly used nowadays and will be the detectors considered in this work.

X-ray spectroscopy is an essential tool across multiple areas of physics as it enables the experimental examination of numerous materials properties, including chemical composition, electronic structure, and dynamic processes at the atomic and molecular levels~\cite{svanberg2012atomic}. The deep penetration of x-rays allows for the exploration of bulk material properties, while their short wavelengths enable the excitation of inner-shell electrons and detailed probing of material structure, expanding the range of spectroscopic techniques available. Furthermore, the typical short pulse durations of x-ray sources allow for time-resolved studies of dynamic processes, usually driven by other coupled instruments. Common x-ray sources include synchrotrons~\cite{willmott2019introduction}, laser-plasma generated sources~\cite{jaroszynski2006radiation}, and more recently, x-ray free electron lasers (XFEL)~\cite{schmuser2014free}, which are characterised by ultrashort pulse durations (10s fs), tunable photon energies, narrow bandwidths (<0.3\% of photon energy), and which have achieved ultra-high x-ray intensities up to $10^{22}$ Wcm$^{-2}$~\cite{yamada2024extreme}.

Spectroscopy is typically conducted through a dispersive crystal coupled with a detector in various geometries. The crystal disperses the spectral components of the radiation emitted by the target, which are then collected by the detector. In the following sections we will separately analyze two of the most common configurations: the flat crystal and the von H\'amos geometries. Additionally, we will focus on how to treat camera images characterized by low photon counts, except for small regions around transition lines -- a common occurrence in x-ray spectroscopy. These low photon numbers typically arise from the low cross-sections of the spectroscopic processes employed, short exposure times, and limited solid angle coverage. If the photon density is low everywhere on the camera and resolution requirements are not stringent, single-photon counting spectroscopy (SPCS)~\cite{basden2003photon}, a method suited to pixel detectors, becomes preferable to dispersive spectroscopy. In SPCS, photon energies are determined based on the signal generated by each photon hit, rather than their position on the detector. Although SPCS is widely used with modern detectors, it falls outside the scope of this discussion and will not be covered in this article.

As an illustrative case, we will apply our methods to images from high-energy density (HED) science, an area of growing importance due to its relevance to astrophysical phenomena~\cite{eggert2010melting} and practical applications such as inertial confinement fusion~\cite{Zylstra2022-kp,LawsonCriterion-2022,Gain-2024} and material development~\cite{Kraus2016-hc,Miao2020-ew}. In fact, the experimental realizations of such systems are extremely short-lived, implying brief exposure times and the necessity of \textit{in-situ} diagnostics to evaluate key physical properties of interest such as structure factors~\cite{Toleikis_2010,Fletcher2015}, ionization potential depression~\cite{PhysRevLett.109.065002,Gawne_PRE_2023}, x-ray opacities~\cite{PhysRevE.100.043207,PhysRevLett.119.085001,PhysRevLett.124.225002}, and collisional ionization rates~\cite{NatCommun-colls,PhysRevLett.120.055002}.
XFELs, thanks to their unique characteristics,  are typically used to probe these extreme states of matter, which are produced by irradiating the material either with powerful optical lasers, which compress the sample, or with the XFEL itself in the so-called ``self-scattering'' experiments~\cite{willmott2019introduction,vinko2012creation}.

The objective of this work is first to provide a step-by-step guide on how to construct a spectrum from the initial detector images~\cite{instruments}, a resource which, to the best of our knowledge, is not found in existing literature as a single comprehensive document, but rather scattered across various sources. In doing this, we will also present methods for enhancing the signal-to-noise ratio (SNR) and the resolution of a camera image, capable of simultaneously treating regions with a high and low density of photons.
Despite these features being common and completely general, we will showcase these techniques on experimental data coming from HED science.

\section{\label{Methods} Methods}

\begin{figure*}
    \centering
    \includegraphics[width=\textwidth,keepaspectratio]{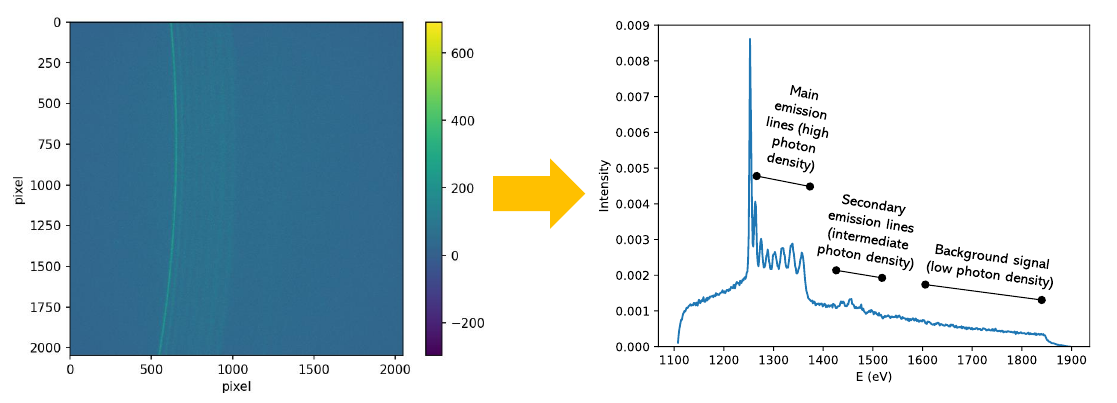}
    \caption{Example of the construction of a spectrum (right) from the corresponding camera image (left), which is a 2048x2048 pixel array. The image presents areas with different photon densities, labelled in the associated spectrum. In this example, the signal is the emission spectrum of warm-dense MgF$_2$ from Ref.~\cite{Gawne_PRE_2023}. The bright arcs in the image correspond to the $K\alpha$ and satellite emission lines, emitted from different ionization states of the Mg ions.
    }
    \label{fig:general process}
\end{figure*}

In this section, we outline the techniques used to construct the spectrum from the initial camera image. For completeness, we will begin with a brief overview of how a semiconductor pixel detector works, which will be beneficial for subsequent discussions. Next, we will detail the process of finding the energy map for two of the most common types of spectrometers. Finally, we will explain how to go from the Analog-to-Digital Units (ADU) map to the distribution of photon hits by means of single photon counting algorithms and image deconvolution techniques. In the following, we will disregard the physical processes which produce the photons arriving at the spectrometer, and focus only on the problem of constructing a spectrum from the initial matrix $A_{ij}$, whose elements represent the total signal on pixel $(i,j)$ (see Fig.~\ref{fig:general process}). Here, the pixels form a two-dimensional array labelled by $i=1,\dots, N_y$ and $j=1,\dots, N_x$, with $N_y$ and $N_x$ representing the number of pixels in the vertical and horizontal directions, respectively.
 
The construction of the spectrum can be divided into three sequential steps:
\begin{enumerate}
    \item \textbf{Computing the energy map of the camera.} This process involves determining the energy associated with each pixel, $E_{ij}$. As we will see, these energies represent averaged values for each pixel, accounting for effects such as source broadening and finite pixel size. These effects mix the energies of photons hitting a single pixel, ultimately reducing the maximum achievable spectral resolution. Section \ref{Energy calibration} details the techniques for computing this energy map.
    \item \textbf{Estimating the photon hits on the detector.} For a given camera image, $A_{ij}$, this steps calculates the most likely distribution of photon hits on the detector, $N^{\gamma}_{ij}$, providing a good estimate for the number of photons hitting each pixel $(i,j)$. The methods for performing this operation will be discussed in section \ref{Estimate of the photon hits distribution}.
    \item \textbf{Constructing the spectrum.} This final step involves placing $N^{\gamma}_{ij}$ into the appropriate energy bin according to the energy map $E_{ij}$, with corrections applied for the varying solid angles associated with each energy level.
\end{enumerate}
The details of how the image is converted to a spectrum depends on the specific geometry of the spectrometer. While all the spectrometers relevant to this work are based on Bragg's law and this is ultimately used to construct the spectrum, in practice specific geometries (such as the von H\'amos geometry, which is given some attention later) may allow for simpler dispersion equations.

\subsection{\label{CCDs} Pixel Detectors}

\begin{figure}
    \centering
    \includegraphics[width=\columnwidth,keepaspectratio]{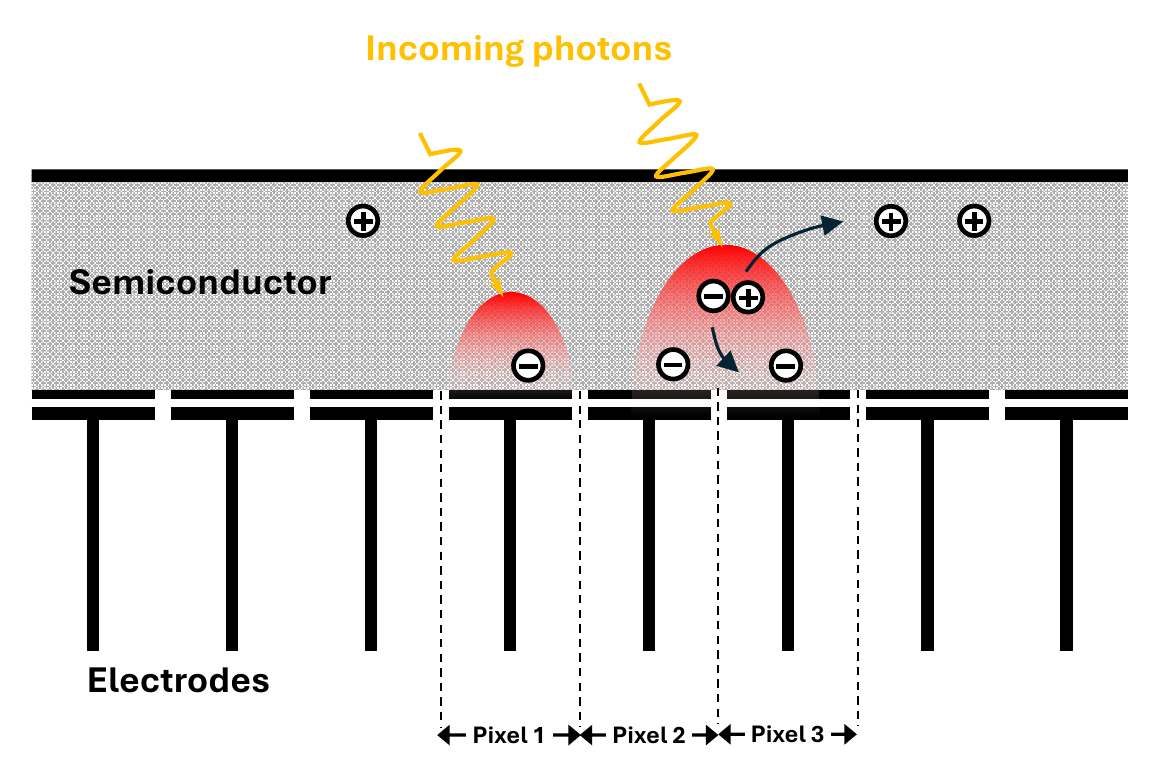}
    \caption{Simplified scheme of a pixel detector section illustrating how charge spreading can distribute a photon across multiple pixels.
    }
    \label{ccd}
\end{figure}

Pixel detectors are integral components in the detection and measurement of light in various scientific applications, including spectroscopy, astronomy, and medical imaging. These detectors work by converting incoming photons into electronic signals that can be quantified and analysed. The general functioning of a pixel detector involves several stages: photon absorption, charge generation and collection, charge transfer, and signal readout. 

\begin{enumerate}
    \item \textbf{Photon absorption.} The process begins when photons enter the detector and strike the photosensitive surface, usually made of a doped or compound semiconductor (e.g. doped silicon, GaAs). At x-ray energies, photons interact with the semiconductor atoms via photoionization, generating electron-hole pairs. The number of these pairs is approximately proportional to the energy of the photon hitting the surface, with random fluctuations quantified by the Fano factor \cite{lewenkopf2008numerical}. The efficiency in converting the photon energy into electrons energy is instead measured by the quantum efficiency (QE), which is essentially the probability that a photon will be photoabsorbed, creating a photocurrent.
    \item \textbf{Charge Generation and Collection.} The generated electrons are collected in potential wells created by an array of electrodes on the detector surface. Each pixel in the detector corresponds to one potential well. The number of electrons collected in each well is roughly proportional to the energy contained in the light hitting that pixel, making the detector highly effective for capturing detailed images. However, thermal noise, which is intrinsic to the detector and depends on its temperature, can also generate electron-hole pairs, known as ``dark current'', which disturbs the measurement~\cite{newberry1991signal}. Additionally, the electron cloud generated by a photon absorption can diffuse over multiple pixels, causing image blurring. This diffusion is quantified by the Point Spread Function (PSF)~\cite{du2004effects}, which is a two-dimensional decaying function (typically a Gaussian, but sometimes distorted) with a characteristic width referred to as charge spreading radius ($R_s$)~\cite{abboud2013sub, hagino2019measurement}. The spreading radius can be estimated both theoretically and experimentally ~\cite{gatti1987dynamics, hagino2019measurement}, and depends on various parameters, including detector thickness and incoming photon energy. While reducing the sensor thickness can mitigate this spreading effect, it also decreases the probability of photon absorption, thereby impacting the quantum efficiency. Furthermore, even though $R_s$ is rigorously dependent on the photon energy — affecting the depth at which the photon is absorbed (see Fig. \ref{ccd}) — we assume it to be constant in this article, as we can restrict our analysis to a portion of the detector illuminated by a sufficiently narrow photon energy range, thanks to Bragg spectroscopy.
    \item \textbf{Charge Transfer.} Once the exposure is complete, the collected charge needs to be transferred to the readout electronics. For modern hybrid detectors, this is done \textit{in situ} with the readout electronics connected to each individual pixel. In the case of CCDs, this is achieved through a process known as charge transfer, where the charges are moved sequentially through the CCD's structure to the output register. This transfer is carefully controlled to preserve the spatial information of the image. The loss of electrons occurring in this operation, which is not present in hybrid detectors, is measured by the so-called Charge Transfer Efficiency (CTE)
    \item \textbf{Signal Readout.} In the final stage, the collected charges are read out by converting them into a voltage signal. This is done using a charge amplifier that translates the charge into a corresponding voltage. The voltage signal is then sent to an Analog-to-Digital Converter (ADC), which converts it into digital counts known as Analog-to-Digital units (ADU). These ADUs represent the total energy of the photons incident on each pixel, combined with noise, and form the basis of the digital image.
\end{enumerate}

For our analysis, we are interested in knowing the following detector characteristics: $R_s$, the level of noise of each pixel, and the ADU counts associated to a photon of energy $E$, $ADU_{sp}(E)$, which will, in general, depend on the pixel. The noise on each pixel is measured simply by running the detector multiple times while the source is blocked (known as ``dark runs''~\cite{mullikin1994methods}), and then looking at the statistics of the empty frames for each pixel across the different shots. The dark runs are taken regularly during an experiment to measure the pixel noise as the detector conditions naturally change over time, and their mean, usually computed excluding tail values to account for anomalies (e.g. stray photons), is then subtracted from the raw image to remove any bias on the ADU values due to thermal noise. Typically, this subtraction is executed by the facility's software, as part of automatic post-processing. Additionally, it is now common for facilities to provide the detector images, with the background subtracted, in energy units. The ADU to energy unit conversion of detectors, i.e. $ADU_{sp}(E)$, is already regularly measured as part of performance testing, and can automatically be applied to the raw ADU image in a post-processing step. However, in this article, we will always assume to work with background-subtracted images, but in ADU units, and we will show how to compute $ADU_{sp}(E)$ directly from the experimental images. While $R_s$ is typically given in the camera documentation, accounting for the charge spread between pixels remains a task for the user as it is computationally quite intensive, and not always necessary depending on the detector geometry or experiment. Lastly, image post-processing often includes masking or correcting faulty pixels, with special attention to edge pixels, where leakage currents can reduce charge collection efficiency~\cite{shor2004edge}.

In the regions with low photon densities, which cover most of the detector in our cases, we are able to identify a photon hit by searching clusters of neighbouring pixels with ADU counts consistently above the noise. This technique, known as single photon counting, allows us to count the photons on the detector, thereby reducing the uncertainties in estimating $N^{\gamma}_{ij}$ by exploiting the spatial correlation between ADU values of neighbouring pixels. The small regions with medium to high photon densities require instead more elaborated techniques to improve signal quality, which will be discussed in the following. Although it is possible to estimate the energy of the photon directly from the ADU count, as done in SPCS~\cite{basden2003photon, morton1968photon}, working in dispersive (or Bragg) spectroscopy enables us to determine the energy of the photons by their positions, facilitating the use of the photon counting techniques.





\subsection{\label{Energy calibration} Energy calibration}

For both spectrometers of interest here, the energy map depends on the geometrical parameters of our setups (e.g. source-detector distance): 

\begin{eqnarray}
E_{ij} = E(x'_{ij},y'_{ij}; \boldsymbol{\Lambda}),
\label{energy_map}
\end{eqnarray}
where $x'_{ij}$ and $y'_{ij}$ are the coordinates of the centre of pixel $(ij)$ on the planar detector and $\Lambda$ collects all the geometrical parameters. These parameters will be used as fitting variables to find the energy map, ensuring that it is consistent with known experimental lines. Let us now separately examine the two types of geometries.

\subsubsection{\label{flat crystal spectrometer}Flat crystal spectrometer}

\begin{figure*}
    \centering
    \includegraphics[width=\textwidth,keepaspectratio]{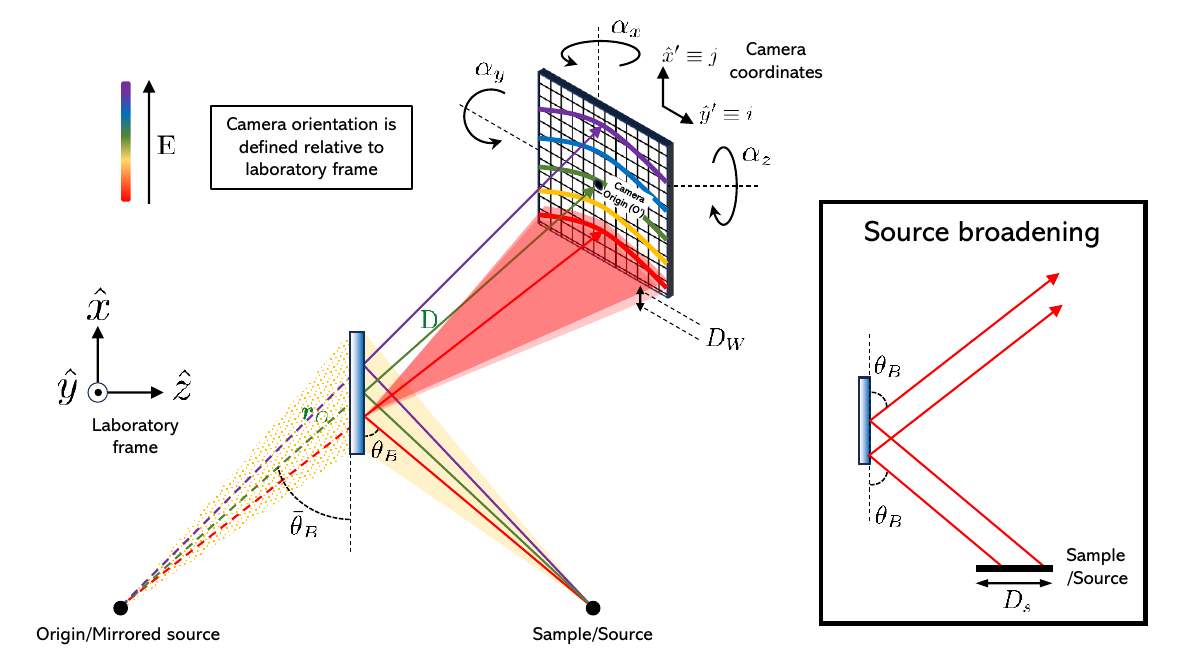}
    \caption{Flat crystal (FC) spectrometer geometry with an illustration of source broadening (inset). The laboratory frame is defined by $\hat{z}$ being the normal of the crystal, $\hat{x}$ such that $\boldsymbol{r_0}$ is contained in the $\hat{x}-\hat{z}$ plane, and $\hat{y}$ chosen for the frame to be an orthonormal right-handed triad. The camera frame is defined with $\hat{x}$ and $\hat{y}$ aligned with the pixel array, $\hat{z}$ orthogonal to the camera plane, and the origin $O'$ at the centre of the detector.
    }
    \label{Flat crystal spec}
\end{figure*}

We begin by assuming that the rocking curve of the reflecting crystal~\cite{zachariasen1967general} is a Dirac delta centred at the Bragg angle, $\theta_B$, and that the sample acts as a point source, thereby neglecting source broadening.
Under these assumptions, using Bragg's law~\cite{zachariasen1967general} and referring to Fig. \ref{Flat crystal spec}, one can derive the following relation between $E_{ij}$, the coordinates of the pixel and the geometrical parameters~\cite{humphries2020isochoric}:

\begin{equation}
\begin{split}
E_{ij} & = E(\theta_{B}(x'_{ij},y'_{ij}; \boldsymbol{\Lambda})) \\
& =  \frac{hc}{2d \sin(\theta_{B}(x'_{ij},y'_{ij}; \boldsymbol{\Lambda}))} \\
& = \frac{hc}{2d}\frac{\|\boldsymbol{r_0}+\boldsymbol{Rr'_{ij}}\|}{\boldsymbol{\hat{z}}\cdot(\boldsymbol{r_0}+\boldsymbol{Rr'_{ij}})} \ \ ,
\label{energy map equation}
\end{split}
\end{equation}
where $\theta_{B}$ is the Bragg angle associated with pixel $(i,j)$, $h$ is the Planck constant, $c$ is the speed of light and $d$ is the spacing between the reflecting planes of the crystal. The geometrical quantities $\boldsymbol{r_0}$, $\boldsymbol{r'_{ij}}$, $\boldsymbol{R}$, are given by the following expressions (see Fig. \ref{Flat crystal spec}):

\begin{align}
& \boldsymbol{r_0} = D \ (\cos\bar{\theta}_B \ , \ 0 \ , \ \sin\bar{\theta}_B) \ , \\
& \boldsymbol{r'_{ij}} =  (x'_{ij} \ , \ y'_{ij} \ , \ 0) \\
& \boldsymbol{R} = \boldsymbol{R}_x(\alpha_x)\boldsymbol{R}_y(\alpha_y)\boldsymbol{R}_z(\alpha_z) \ .
\label{Geometrical quantities}
\end{align}
Here, $\boldsymbol{R}$ is the rotation matrix that specifies the orientation of the camera relative to the laboratory frame, parametrized with Euler's angles $(\alpha_x, \alpha_y, \alpha_z)$~\cite{goldstein:mechanics}. Such angles, along with the distance $D$ from the source to the camera origin $O'$ and the Bragg angle $\bar{\theta}_B$ associated with $O'$, constitute the geometrical parameters $\boldsymbol{\Lambda}$ for fitting and determining the energy map. In order to do this, we first identify, in terms of pixels coordinates, a bright curve on the experimental image associated with an emission line of known energy $\bar{E}$:

\begin{equation}
\begin{split}
& S_{exp} = \biggl\{(i, f(i)) \ , \ i = 1, \dots, N_y \biggr\} \\ 
& \mathrm{with}  \ f(i) = \max_{N_1 \leq j \leq N_2} A_{ij} \ , 
\label{experimental line1}  
\end{split}
\end{equation}
where $N_1$ and $N_2$ are appropriately chosen to isolate the emission line of energy $\bar{E}$. A more sophisticated and accurate method is to fit, for each row, the emission line to an analytic function and to take the peak of the fit as the line position. This approach is less susceptible to noise and allows for sub-pixel resolution, by constructing the energy map on a finer sub-pixel grid. However, for the purposes of this paper, we adopt the method described by Eq. (\ref{experimental line1}). As a second step, we compute the relative theoretical contour $\{(i,j)| \ E_{ij} = \bar{E}\}$ using Eq. (\ref{energy map equation}) with a guess for $\boldsymbol{\Lambda}$:

\begin{equation}
\begin{split}
& S_{th}(\boldsymbol{\Lambda}) = \biggl\{(i, g(i;\boldsymbol{\Lambda})) \ , \ i = 1, \dots, N_y \biggr\} \\ 
& \mathrm{with}  \ g(i;\boldsymbol{\Lambda}) = \min_{j} |E_{ij}(\boldsymbol{\Lambda})-\bar{E}| \ . 
\label{experimental line}  
\end{split}
\end{equation}
$S_{exp}$ and $S_{th}$ are parametrized with $i$ as the independent variable because of the expected shape for these curves. Finally, we calculate a loss function $\mathcal{L}$, dependent on the geometrical parameters $\boldsymbol{\Lambda}$, measuring a distance between the experimental line $S_{exp}$ and the theoretically-computed contour $S_{th}$:

\begin{equation}
\mathcal{L}(\boldsymbol{\Lambda}) = \frac{1}{N_y}\sum_{i=1}^{N_y} (f(i)-g(i;\boldsymbol{\Lambda}))^2
    \label{loss function}
\end{equation}

\begin{figure}
    \centering
    \includegraphics[width=\columnwidth,keepaspectratio]{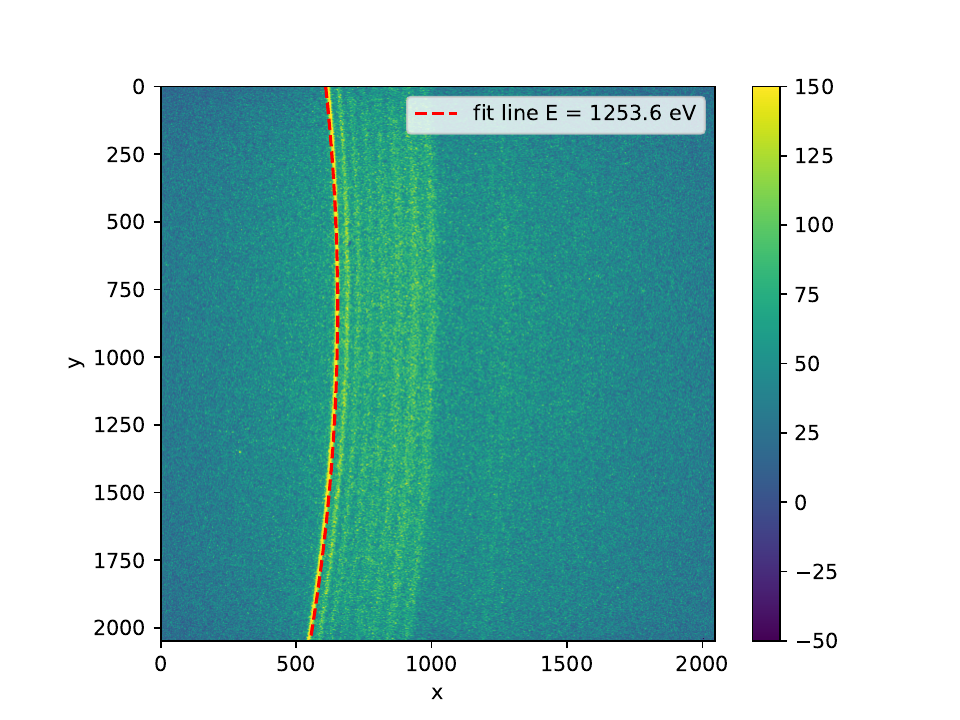}
    \caption{Detector image of MgF$_2$ emission spectrum with the theoretically-computed energy contour (red) fitted to a Mg $K\alpha$ line.
    }
    \label{FC_calibration}
\end{figure}

At this point, we can minimize $\mathcal{L}$ over the space of the geometrical parameters with standard optimization routines~\cite{gao2012implementing}. The loss function can include multiple emission lines by summing the loss functions of each individual line, as previously defined. This summation can include different weights for each line, proportional to their intensities. Figure \ref{FC_calibration} illustrates an example of this calibration for the Mg spectrum discussed earlier. Here, the geometrical parameters of the setup are determined by fitting the Mg $K\alpha$ line, and are found to be:

\begin{equation}
\begin{split}
& (\alpha_x, \alpha_y, \alpha_z) = (-1.9^\circ,55.8^\circ,0^\circ) \\
& D =  79.5 \ \mathrm{mm} \\
& \bar\theta_B = 34.7^\circ\ \ .
\label{Mg parameters}
\end{split}
\end{equation}
As we see from these values, it is common practice to work with the camera orthogonal to $\boldsymbol{r_0}$ in order to maximize the solid angle coverage of the detector, as well as to minimize any grazing incidence broadening in the sensor layer (see Section \ref{sec:Resolution}).

\subsubsection{\label{Von Hamos spectrometer}Von H\'amos spectrometer}

\begin{figure*}
    \centering
    \includegraphics[width=\textwidth,keepaspectratio]{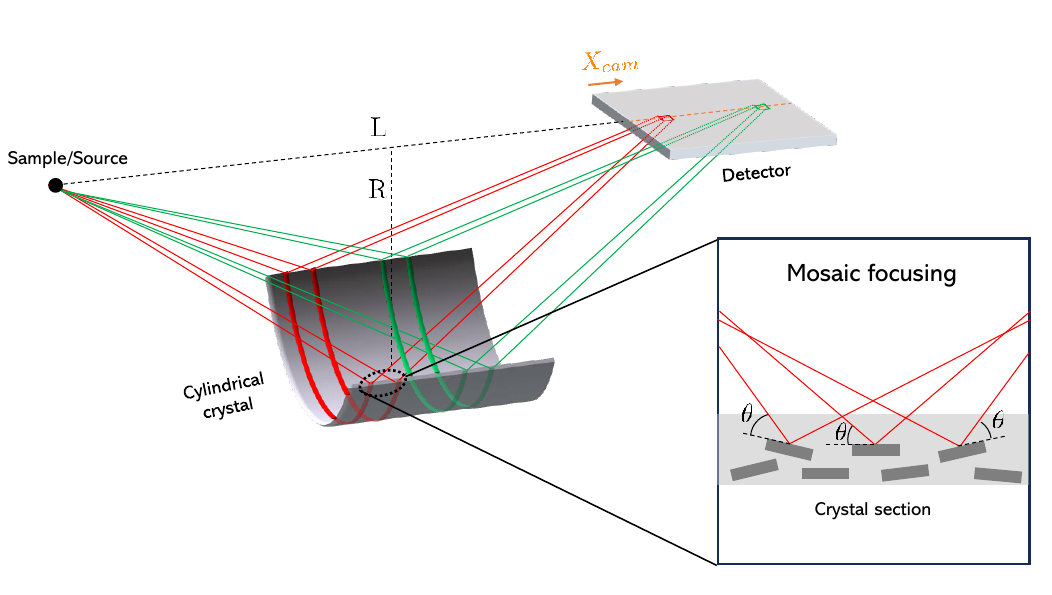}
    \caption{Von H\'amos spectrometer setup with an illustration of mosaic focusing (inset). The mosaicity of the crystal increases the area that reflects a certain photon energy. For the correct functioning of the spectrometer, the crystal must be placed halfway between the source and the centre of the detector. As in the flat crystal case, $\bar{\theta}_B$ is defined as the Bragg angle associated with the centre of the detector.
    }
    \label{VH_scheme}
\end{figure*}

\begin{figure}
    \centering
    \includegraphics[width=\columnwidth,keepaspectratio]{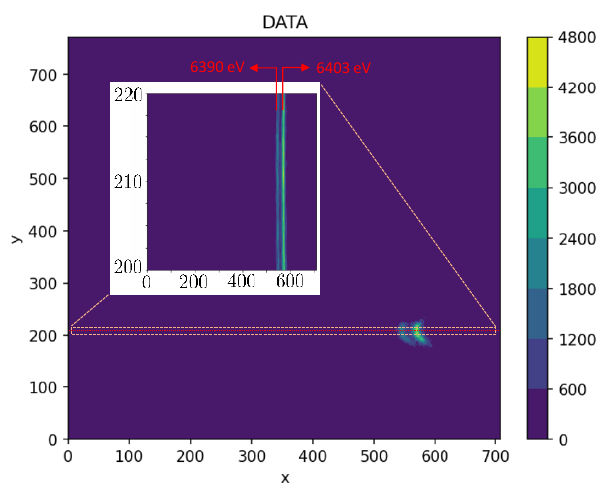}
    \caption{Detector image of Fe emission using a von H\'amos spectrometer from the experiment reported in Ref.~\cite{forte2024resonant}, with a zoomed inset around the central dispersion axis, indicated by the red dotted line. The two vertical bright lines in the inset correspond to the isoenergy contours of the Fe K-alpha lines. Generally, the central dispersion axis may be slightly tilted relative to the detector's edge. This tilt is accounted for by adjusting the region of interest while moving along the dispersive direction. However, Eq. (\ref{VH dispersion}) remains valid, as corrections due to the tilt are typically negligible.}
    \label{VH_calibration}
\end{figure}

In the von H\'amos (VH) geometry~\cite{vonHamos_1932}, an axis (the dispersion axis) is drawn between the source and the detector, and the detector surface lies along this line (see Fig.~\ref{VH_scheme}). A cylindrically-bent crystal is then placed parallel to this axis, with the crystal positioned at its radius of curvature below this axis. The crystal will therefore focus the reflected rays back on to the dispersion axis at a distance twice as a far away in the dispersive direction. This results in a simple dispersion relationship between the photon energy $E$, the position from the source to the front of the camera along the dispersion axis $L$, and the position on camera along the dispersion axis $X_{\rm cam}$:
\begin{equation}
    E = \frac{hc}{2d}\sqrt{1 + \frac{(L+X_{\rm cam})^2}{4R^2} } \, ,
\label{VH dispersion}
\end{equation}
where $R$ is the radius of curvature of the crystal, and the remaining symbols have the same meaning as before.
The calibration is straightforward and involves the optimization of $R$ and $L$ using analogous techniques to those described for the flat crystal geometry. Bright emission lines with known energies are well-suited to this task. Two sets of lines that cover the detector range are ideal for the calibration and should result in a very well-constrained dispersion relationship. We show an example of an experimental calibration in Fig.~\ref{VH_calibration}.

The primary benefit of a von H\'amos spectrometer is its collection efficiency. By focusing all the rays onto a single axis instead of arcs as with the flat crystal spectrometer, it is easier to detect events above the noise.
As the bending process deforms the lattice of a perfect crystal, its rocking curve becomes extended, which degrades the intrinsic resolution of the crystal~\cite{taupin1964theorie,uschmann1993x}. It is therefore common to use mosaic crystals as the dispersive crystal to further increase the collection efficiency of the spectrometer.
Common examples of such crystals are highly annealed pyrolytic graphite (HAPG), highly oriented pyrolytic graphite (HOPG) and Lithium Fluoride (LiF). A mosaic crystal is one that consists of small perfect crystallites with their normal vectors randomly distributed around the crystal surface normal. As a result, in contrast to perfect crystals where reflections only take place around the Bragg angle within the rocking curve, the Bragg condition can be satisfied anywhere on a mosaic crystal by finding a crystallite appropriately angled to the normal (see Fig.~\ref{VH_scheme}). This effect increases the area of the crystal reflecting a certain photon energy on a specific spot of the detector (mosaic focusing). The end result is that von H\'amos spectrometers have very high collection efficiencies, and so are often employed in situations where photon numbers are very low, such as in x-ray Thomson scattering and resonant inelastic x-ray scattering experiments~\cite{gawne2024effectsmosaiccrystalinstrument,forte2024resonant}.

As the x-rays are focused onto the dispersion axis, the spectrum is produced by integrating over a few pixels along the non-dispersive direction. The number of pixels used is chosen to maximize the signal-to-resolution ratio (see~\cite{preston}). Outside this region of interest (RoI), unfocused x-rays are disregarded, as different photon energies overlap on the same pixels.

The drawback to mosaic crystal VH spectrometers is that their resolution tends to be worse than for flat crystal spectrometers. The bulk of this is due to the mosaicity of the crystal, which both enhances depth broadening and itself contributes directly to the broadening. As a result, photons of a specific energy are sent to different locations on the detector rather than to a precise, nominal point, producing a non-trivial broad and extended instrument function~\cite{gawne2024effectsmosaiccrystalinstrument}. Furthermore, the mosaicity causes broadening in the non-dispersive direction, resulting in a characteristic X-shaped instrument function across the detector, with wider crystals producing a more extreme shape~\cite{2012_Zastrau_Focal,zastrau2013characterization}. This broadening leads to the aforementioned energy mixing outside the central RoI. If the spectrometer lies slightly out of focus, the broadening in the non-dispersive direction can become curved, in either direction, depending on the vertical position of the crystal and its various physical properties such as mosaicity. Indeed, we observe this sort of defocusing in the energy contours Fig. \ref{VH_calibration}), where emission lines show a backwards sweep.

Focal aberrations are in fact an additional factor that can degrades resolution~\cite{2012_Zastrau_Focal}, but this effect can be reduced by using a relatively narrow crystal~\cite{2012_Zastrau_Focal,preston}. Typical resolutions of von H\'amos spectrometers range from 1 to 10s eV~\cite{preston,glenzer2016matter,MacDonald_PoP_2021}.


\subsubsection{\label{sec:Resolution}Broadening Effects}

The energy of each photon should map precisely to a corresponding position on the detector. In practice, however, various broadening mechanisms can cause photons of the same nominal energy to be recorded at different locations on the detector.
As the spectrometer is calibrated so that spatial points on the detector correspond to specific energies, there is no way to distinguish between a photon arriving at its nominal position, and a photon with a different energy landing at the same point due to some broadening effect. This effect is described by a point spread function, also known as an instrument function (IF), and results in the observation of broadened features and a loss of spectral resolution.

While high resolution spectra are generally desirable, there is a trade-off between resolution and signal intensity as the very effects that result in broadening also tend to result in higher spectrometer efficiencies. 

Broadening effects are typically non-trivial to account for. The IF can be strongly geometry-dependent and varies between experiments and setups, and the IF also depends on the photon energies incident on the crystal~\cite{gawne2024effectsmosaiccrystalinstrument}. In other words, rather than the IF being applied to the spectrum as a simple convolution, it is in general a complicated kernel problem~\cite{gawne2024effectsmosaiccrystalinstrument}. Even if the IF was a convolution problem, finite spectral windows and experimental noise results in the deconvolution being notoriously unstable.
We list the main effects that deteriorate the resolution of the measured spectra.

\begin{itemize}
    \item \textbf{Finite pixel size and quantization error:} Pixels are rectangles with sides of finite length $L_p$, causing each pixel to collect photons within an energy range of order $O(\frac{2hc}{d}\frac{\cos\bar\theta_B}{\sin^2\bar\theta_B}\frac{L_p}{D})$.
    The energy calibration of a pixel $E_{ij}$ thus represents an average energy.
    For the cameras used in the experiments, typical $L_p$ values are in the range $\sim$10-100~$\mu$m~\cite{mozzanica2018jungfrau}, which is generally sufficiently small to measure a smooth spectrum with energy bins $\lesssim 2$~eV. The energy bin width $\Delta E$ must be chosen as small as possible to reduce the quantization error ($\sim\Delta E/\sqrt{12}$), but large enough to contain several $E_{ij}$ in order to obtain smooth spectra. As an aside, there is a trade-off to be considered in pixel size -- while smaller pixels will have less pixel broadening, they also suffer from increased charge spreading between pixels. If the charge of a photon is spread between too many pixels, it may become difficult to detect signal above the noise. As a result, there is a balance between the possible energy bin size and the ability to reliably detect photons above the noise.
    \item \textbf{Source broadening (SB):} As the incident beam has a finite size, photons leave the target from a region on the sample with an effective size $D_s$, typically ranging from several nm to dozens of $\mu$m (this region is often approximated as a circle, although imprints of the beam show that the laser spatial profile can be more complex~\cite{Chalupsky_Opt_2010}). Additionally, as samples have finite thickness, photons also emerge from different points from within the sample.
    Consequently, photons of the same energy emitted from different points in the sample will reach the detector at different points, representative of the shape of the source, which results in photons of different energies landing on the same point, broadening the measured spectrum (see Fig.~\ref{Flat crystal spec}). The energy uncertainty on each pixel, due to this effect, is of order $O(\frac{2hc}{d}\frac{D_s \cos\bar\theta_B}{D})$.
    \item \textbf{Depth Broadening (DB):} A photon can travel a finite depth into a crystal before being reflected out again. As a result, it reflects at a point displaced from the nominal position it would reflect from on the surface, and will therefore hit the detector at a different point. The DB resembles an exponential function as it represents the probability a photon reflects from a certain depth in the crystal. For perfect crystals, this is typically a very narrow function and does not contribute significantly to the broadening. However, for mosaic crystals the broadening can be substantial (several eV) due to the distribution of the crystallites~\cite{gawne2024effectsmosaiccrystalinstrument}, and is made even broader by non-negligible multi-reflection paths within the crystal~\cite{Schlesiger_JAC_2017}. When the path travelled inside the crystal is much smaller than its attenuation length -- a necessary condition to maintain the crystal's reflectivity -- the SB is proportional to $t/sin(\theta_B)$, with $t$ representing the crystal thickness. Since $t$ cannot be reduced excessively without compromising reflectivity, it is important to avoid working at overly shallow incidence angles. The same consideration applies to the detector, where small incidence angles influence the depth and, depending on $L_p$, the pixel position of the photoabsorption, thereby impacting pixel detection and increasing charge spreading~\cite{dusi2003study}.
    \item \textbf{Finite width of the rocking curve:} Another source of energy broadening is the finite width of the crystal rocking curve (RC), $D_W$. The RC allows for a photon that does not satisfy the Bragg condition to nevertheless be reflected from the crystal~\cite{zachariasen1967general}, albeit with a lower probability. For perfect crystals, their intrinsic RC can be calculated using dynamic diffraction theory, and it is an extremely narrow function~\cite{zachariasen1967general}. Typical values for $D_W$ are dozens of $\mu$rad. For mosaic crystals, the mosaic distribution of the crystallites also contributes to the RC, and is about an order of magnitude larger than the intrinsic RC of the crystallites~\cite{Gerlach_JAC_2015}.
    The energy uncertainty per pixel from this effect is then simply of order $O(\frac{2hc}{d}D_W)$. $D_W$ is generally weakly dependent on the photon energy~\cite{Gerlach_JAC_2015}, and is often assumed to be constant for typical spectrometer energy windows.
\end{itemize}

The first effect, unlike the others, is not technically a broadening effect, as it degrades resolution without affecting the correspondence between photon energy and its position on the detector. For the geometrical setup of the Mg experiment, summarised by Parameters~(\ref{Mg parameters}), with typical values for $L_p$, $D_s$ and $D_W$, the energy resolution losses produced by these effects are of around $0.5-1$ eV. In many scientific fields, these resolution losses are of secondary importance, as other factors such as source bandwidth dominate~\cite{feldhaus1997possible}. Nevertheless, they must be considered if one wants to reach high resolutions (sub-eV). These resolutions can be achieved by means of specific experimental setups, characterised by large source-detector distances and high resolution crystals with very narrow RC~\cite{wollenweber2021high}. However, these features also tend to decrease the SNR of the images. It is therefore useful to contrive alternative methods that mitigate these resolution losses, without compromising the signal intensity. One potential solution to overcome the resolution limitations imposed by the finite pixel size is to analyze the charge distribution between neighbouring pixels. Techniques used to manage charge spreading, based primarily on centroid calculations, have been successfully employed in the past to handle such correlations and achieve sub-pixel resolutions~\cite{abboud2013sub, hagino2019measurement}. Nevertheless, it is important to note that these methods are applicable only when the pixel dimensions are sufficiently small compared to $R_s$ and in low photon-density regimes. In Section~\ref{Estimate of the photon hits distribution} we will introduce an alternative approach to treat these correlations, based on deconvolution algorithms capable of also handling cases with large photon densities on the detector.

The remaining three effects can be mitigated by studying the instrument function of the experimental setup and solving the complicated kernel problem mentioned earlier. Assuming the pixel size effect to be negligible or already accounted for, we can study the broadening due to the spectrometer instrument function by irradiating the target with a monochromatic source and compare the elastic scattering measured on the spectrometer to a high-resolution beam spectrometer. For this method to work, a material with very weak inelastic scattering in the vicinity of the elastic scattering must be chosen. Good choices include plastics such as polymethyl methacrylate and Polypropylene, or metallic glasses~\cite{Voigt_PoP_20201,gawne2024effectsmosaiccrystalinstrument}.
Another common way to determine the resolution of a setup is from emission lines with known lineshapes; e.g. H\"olzer \textit{et al.}'s $3d$ metal transition lines~\cite{Hoelzer_PRA_1997}. In any case, the FWHM of the instrument function used provides an estimate of the spectrometer resolution.





\subsection{\label{Experimental estimation of camera parameters}Experimental estimation of camera parameters}

A fundamental quantity for converting the ADU map to the distribution of photon hits is the RMS of the camera read noise, $\sigma_{N}$, which measures the level of thermal noise in the detector. This can be measured pixel-wise by running the camera without a source and calculating the variance of the pixel values during these dark runs. However, for simplicity we will consider this quantity to be constant across the camera, which is a reasonable approximation for our detectors. In these cases, $\sigma_{N}$ is directly inferable from the histogram of the camera images with signal present, the shape of which we will now discuss.

We will assume the detector images have undergone dark image subtraction, so that the expected value of the ADU count for a pixel not hit by any photon is zero. Additionally, for our cases, the analysis of the histogram is simplified because, in Bragg spectroscopy, photons energies are determined by their position on the detector. Consequently, we can subdivide the camera into regions sufficiently small to approximate $ADU_{sp}(E)$, the ADU counts produced by a photon of energy $E$, as constant ($ADU_{sp}$) in each region. The shape of the histogram for a single region is in general a series of peaks centered approximately in the following sequence of values:
\begin{eqnarray}
P_{n} = \zeta \cdot n \cdot ADU_{sp}, \ \ \ \ \ n = 0,1,2,\dots
\label{ADU sequence}
\end{eqnarray}
where $\zeta$ is a constant smaller than 1 (typically around 0.8-0.9), whose origin will be discussed below. The n-th peak contains the pixels which are likely to have been hit by $n$ photons. In reality, only the initial peaks of this sequence will appear in the histogram, depending on the density of photon hits, which is measured by the so-called fill fraction $\lambda$. This important parameter is defined as:
\begin{eqnarray}
 \lambda = \frac{\# \ \textrm{photons on the camera}}{\# \ \textrm{pixels in the camera}} \ \ .
\label{fill fraction}
\end{eqnarray}

\begin{figure}
    \centering
    \includegraphics[width=\columnwidth,keepaspectratio]{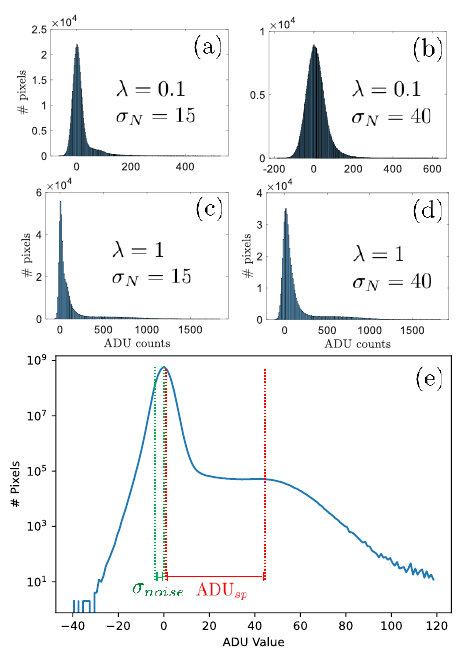}
    \caption{(a-d) Synthetic histograms of 700x700 images for various levels of noise and fill fractions. $ADU_{sp}$ is taken equal to 120. (e) Histogram of the camera image reported in Fig.~\ref{fig:general process}.
    }
    \label{histogram}
\end{figure}

Each peak is broadened by the intrinsic noise of the detector and the charge spreading, quantified by $\sigma_{N}$ and $R_s$, respectively. The magnitude of these broadening effects determines whether the peaks will be merged together or remain distinct. Additionally, charge spreading systematically shifts the peaks, initially centred at $n ADU_{sp}$, towards lower ADU values, as some of the photon energy is deposited across multiple pixels. The constant $\zeta$ in Eq. (\ref{ADU sequence}) accounts for this shift, decreasing as $R_s$ increases. To illustrate these effects, in Figs.~\ref{histogram}~(a-d) we present histograms generated from synthetic camera images with various levels of noise and fill fractions. More details on the fabrication of these synthetic images are provided in section~\ref{Synthetic data}. Alongside these synthetic data, we include an experimental histogram for MgF$_2$ in Fig.~\ref{histogram}~(e) for data from Ref.~\cite{Gawne_PRE_2023}.

If $ADU_{sp}(E)$ is not provided by the facility, it can be directly inferred from the position of the first peak in the series given by Eq. (\ref{ADU sequence}), which corresponds to single photons hitting pixels on the camera.
Similarly, from the width of the peak centred around 0, we can estimate $\sigma_{N}$. Intuitively, the more the peaks merge together, the more challenging it becomes to accurately determine these quantities. In such cases, one approach to extract these two parameters is to fit the experimental histograms to a theoretical model. This model can be constructed (see sections \ref{Synthetic data}) given certain values for $ADU_{sp}$, $\sigma_N$, and $\lambda$, which can be estimated from the expected number of photons, and $R_s$, usually provided in the camera documentation. 


\subsection{\label{Estimate of the photon hits distribution}Estimate of the photon hits distribution}

The objective of this section is to describe how to compute the distribution of photon hits, $N^{\gamma}_{ij}$, from the camera image ADU$_{ij}$, thereby eliminating the distortions due to thermal noise and charge spreading. Depending on the spectrum of photons hitting the camera, the image will have regions with different local fill fractions (see Fig.~\ref{synthetic images}), which must be treated with distinct methods. Therefore, to process the image, we apply the following two techniques:

\begin{enumerate}
    \item \textbf{Photon counting algorithms:} To address regions with small fill fractions ($\lambda \lesssim 0.1$), where single photon clusters are present, we use photon counting algorithms. These algorithms can identify single photon clusters, exploiting the correlation between neighbouring pixels to reduce the uncertainty on the photon detection.
    \item \textbf{Image deconvolution techniques:} To treat the regions with large fill fractions ($\lambda \gtrsim 1$), where agglomerated clusters are present (see next section), we process the image with the Richardson-Lucy deconvolution method~\cite{richardson1972bayesian}.
\end{enumerate}
As we will see, these two operations can be carried out sequentially. Considering the discussion in section~\ref{Experimental estimation of camera parameters}, we will assume in the following that the camera parameters $ADU_{sp}$, $\sigma_N$, and $R_s$ are known.

\subsubsection{\label{Photon counting algorithms}Clustering Algorithms}

\begin{figure}
    \includegraphics[width=\columnwidth,keepaspectratio]{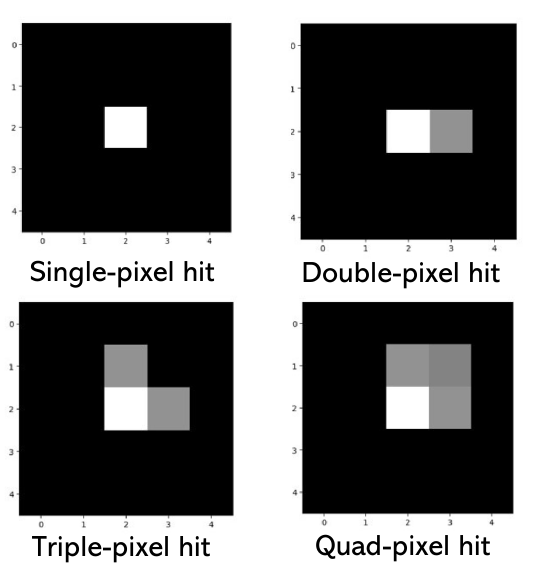}
    \caption{The 4 layouts for a single photon cluster. The brighter pixel represents the pixel hit by the photon. Each of the last 3 can rotate by $90^{\circ}$, $180^{\circ}$ and $270^{\circ}$ to produce 13 layouts in total.
    }
\label{clusters} 
\end{figure}

If we suppose, as is often the case, that the charge produced by a photon does not spread further than a pixel side length, then the shapes of single-photon clusters (i.e. the pixels in which the charged produced by the photon has leaked) are few, and given in Fig.~\ref{clusters}.
As there are only 13 possible shapes, a simple clustering algorithm can be employed to identify these clusters and re-accumulate the intensity back on to the central pixel. The benefit of doing so is that this makes the pixel brighter relative to the background, allowing for more of these events to be included once the final thresholding is performed.

The first step to identify the photon hits from the ADU map is thresholding the camera image. Although different choices are possible for the thresholding value, here, we chose 1.5 $\sigma_{N}$, as a good compromise to cut the noise and retain the signal:
\begin{equation}
A^{th}_{ij} = 
\begin{cases}
A_{ij} \ \ \ \ \ A_{ij}\ge 1.5 \ \sigma_{N} \\
0 \ \ \ \ \ \ \  \ A_{ij}< 1.5 \ \sigma_{N}
\end{cases}
\label{thresholding}
\end{equation}
The thresholding values can be chosen in a systematic way by analysing their performance over multiple synthetic images for a given set of parameters ($ADU_{sp}$, $\sigma_N$ and $R_s$). This method is briefly described in Appendix~\ref{Determination of thresholding values}. After the thresholding, the pixels with an ADU value different from zero are the ones that most likely received a contribution to their ADU by a photon hit. 
One may also set pixels exceeding some upper threshold to zero to mask out possible faulty pixels or high intensity events that do not correspond to real signal. 

Next, we locate any clusters and determine whether they fall into one of two cases:
\begin{enumerate}
    \item \textbf{Single-photon clusters.}
    If the shape of the cluster $C$ matches that in Fig.~\ref{clusters}, then the charge spread is considered to originate from the brightest pixel, and we accumulate all the ADU values of the cluster onto its brightest pixel ($\bar{i},\bar{j}$):
    \begin{equation}
        \begin{split}
            & A_{\bar{i},\bar{j}}' = \sum_{ij\in \textrm{C}}A_{ij}^{th} \\
            & A_{ij}'=0 \ \ \textrm{for } (i,j)\in \textrm{C and } (i,j) \ne (\bar{i},\bar{j}) \ .
        \end{split}
        \label{cluster sum}
    \end{equation}
    \item\textbf{Agglomerated clusters.}
    In this case we do not modify the ADU values:
    \begin{equation}
        A_{ij}' = A_{ij}^{th}
        \label{cluster sum 2}
    \end{equation}
\end{enumerate}
After the clustering algorithm is applied, the image is thresholded once more with a higher threshold, assuming that any events that fall below the threshold did not originate from a single photon event. For this work, we choose $(ADU_{sp} - \ \sigma_N)$ as the second threshold, which results in discarding roughly 20\% of the single photons ($A'_{ij}$ has a larger variance than $A_{ij}$), but eliminates almost entirely noise fluctuations. This second threshold can also be systematically determined using the method outlined in Appendix \ref{Determination of thresholding values}. Charge accumulation over the brightest pixel can also be carried out by considering the centroid of the single-photon cluster, enabling subpixel resolution~\cite{abboud2013sub}. However, this more advanced process is beyond the scope of our discussion here.

\begin{figure*}
    \centering
    \includegraphics[width=\textwidth,keepaspectratio]{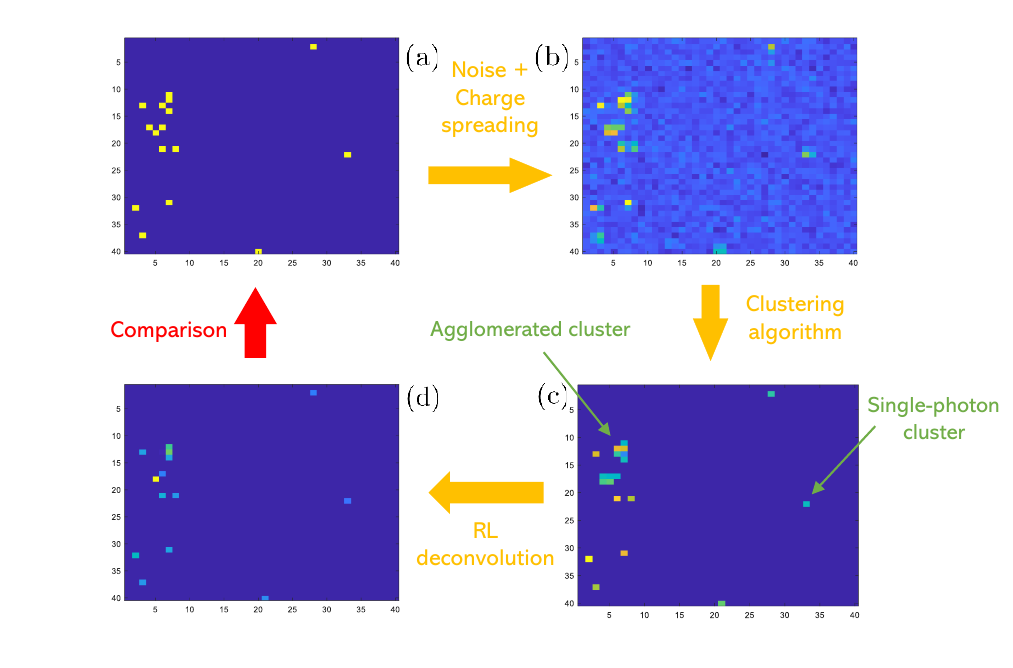}
    \caption{Illustration of the various steps in the camera image processing. (a) Map of photon hits on the camera for a small $\lambda$, to which thermal noise and charge spreading have been applied to obtain the camera image (b). The same image, processed with the clustering counting algorithm (c), and then with the deconvolution algorithm (d). In the figure, we indicate an example of a single-photon cluster and of an agglomerated cluster. All camera images, except the first initial photon map, are in ADU units.}
    \label{algorithms}
\end{figure*}

In Fig.~\ref{algorithms} we report an example of this operation using a synthetic camera image (see section~\ref{Synthetic data} for more details). As is observed, agglomerated clusters remain unchanged by this algorithm and will be addressed in the following section.

Exploiting the correlation of ADU values across neighbouring pixels enables us to boost the signal. As an example, consider three adjacent pixels arranged in an L-shape, with ADU values between the first and second thresholds. If the correlation is ignored, and consequently, we do not accumulate the ADU values on the brightest pixel, all three pixels will be zeroed after the second thresholding, resulting in a loss of signal that likely corresponds to a photon. In fact, a simple analysis shows that it is significantly more likely that three L-shaped pixels with values above 1.5 $\sigma_N$ are caused by a photon compared with noise. 

In addition to this, thresholding mitigates noise-related fluctuations in the reconstructed spectrum. The noise reduction is approximately proportional to $\mathcal{P}(N^{\gamma}_{ij} \geq 1) < 1$, where $\mathcal{P}(N^{\gamma}_{ij} \geq 1)$ is the probability that at least one photon lands on pixel $(i,j)$. This result can be derived by considering the variance of the random variables $A_{ij}\approx N^{\gamma}_{ij}ADU_{sp} + \eta_{ij}$ (see Appendix \ref{Probability distribution of the pixel values}) and $A^{th}_{ij}$ (as defined earlier), where $\eta_{ij}$ represents the random noise on pixel $i$,$j$. Firstly, fixing the number of photons $N^{\gamma}_{ij}$, we find:

\begin{equation}
    \begin{split}
            & \mathbb{V}(A_{ij}|N^{\gamma}_{ij}) = \mathbb{V}(\eta_{ij}) = \sigma_N^2 \ \ \ \ \ \ \ \forall N^{\gamma}_{ij}, \\
            & \mathbb{V}(A^{th}_{ij}|N^{\gamma}_{ij}) =  
            \begin{cases}
            \sigma_N^2 \ \ \ \ \ A_{ij}\ge 1.5 \ \sigma_{N} \\
            0 \ \ \ \ \ \ \ \ A_{ij}< 1.5 \ \sigma_{N}
             \end{cases} .
    \end{split}
    \label{bonus formula}
\end{equation}
Now, passing to the mean values, we obtain:

\begin{equation}
    \begin{split}
            & \mathbb{E}(\mathbb{V}(A_{ij})) =  \sigma_N^2 \mathcal{P}(N^{\gamma}_{ij}=0) + \sigma_N^2 \mathcal{P}(N^{\gamma}_{ij}=1) + \cdots = \sigma_N^2, \\
            & \mathbb{E}(\mathbb{V}(A^{th}_{ij})) = 0 \cdot\mathcal{P}(N^{\gamma}_{ij}=0) + \sigma_N^2 \mathcal{P}(N^{\gamma}_{ij}\ge1) < \sigma_N^2,
    \end{split}
    \label{bonus formula 1}
\end{equation}
having taken as an approximation that $\mathcal{P}(A_{ij}<1.5 \ \sigma_N|N^{\gamma}_{ij}>1) = 0$ and $\mathcal{P}(A_{ij}>1.5 \ \sigma_N|N^{\gamma}_{ij}=0) = 0$. Summing the ADU values over the energy contours preserves the ratio between the variance of the raw and thresholded values.

\subsubsection{\label{Images devonvolution}Images deconvolution}

To address charge spreading and noise contributions in regions with large $\lambda$, we apply the Richardson-Lucy (RL) deconvolution algorithm~\cite{richardson1972bayesian} to the matrix $A'_{ij}$ . This method is an iterative procedure, based on Bayesian inference, that computes the most likely expected values $V_{ij}$ of the Poissonian and independent variables $X_{ij}$, given a realization $R_{ij}$, which has been blurred out by a known kernel function $H$ and corrupted by random noise. In formulas:

\begin{equation}
\begin{split}
    & R_{ij} = (X \ast H)_{ij} + \eta_{ij}, \\
    & X_{ij} \sim \textrm{Poisson}(V_{ij}), \ \ \eta_{ij} \sim \mathcal{N}(0,\sigma^2_N) \\
    & R_{ij} \xrightarrow{\text{RL algorithm}} V_{ij}
    \label{RL formulas}    
\end{split}
\end{equation}
where $\ast$ denotes the convolution symbol, and $X_{ij}$ and $\eta_{ij}$ are uncorrelated random variables for each $i$ and $j$. It is intuitive to see that this problem is analogous to our case, where $X_{ij}$ is the number of photons landing on pixel $ij$, $V_{ij}$ is the intensity of the spectrum at that pixel energy $I(E_{ij})$, $H$ is an effective charge spreading function (defined in Appendix \ref{Probability distribution of the pixel values}), $\eta_{ij}$ is the thermal noise on pixel $ij$ and $R_{ij}$ is the ADU map. Therefore, defining $A_{ij}^{proc}$ as:

\begin{equation}
A'_{ij} \xrightarrow{\text{RL algorithm}} A_{ij}^{proc} \ ,
\label{RL deconvolution}
\end{equation}
it will constitute an estimate for $I(E_{ij})$, which is the quantity we are interested in. A more rigorous justification for the application of this method is presented in Appendix~\ref{Probability distribution of the pixel values}.

In Fig.~\ref{algorithms} (d), we illustrate the result of applying the RL algorithm to the camera image already treated with the photon counting algorithms.
As observed, the algorithm decomposes the agglomerated clusters into the most likely distribution of photon hits (still in ADU units), leaving the single-photon clusters unchanged. The comparison between the reconstructed map $A_{ij}^{proc}$ and the original photon hits is excellent, with only very few mismatches.


Finally, we convert the camera units to photon numbers, which provides the intensity in the measured spectrum:
\begin{equation}
    N^{\gamma}_{ij} = \frac{A_{ij}^{proc}}{ADU_{sp}} \ \ . 
    \label{photon number}
\end{equation}


\subsection{\label{Construction of the spectrum and solid angle corrections}Construction of the spectrum and solid angle corrections}

\begin{figure}
    \centering
    \includegraphics[width=\columnwidth,keepaspectratio]{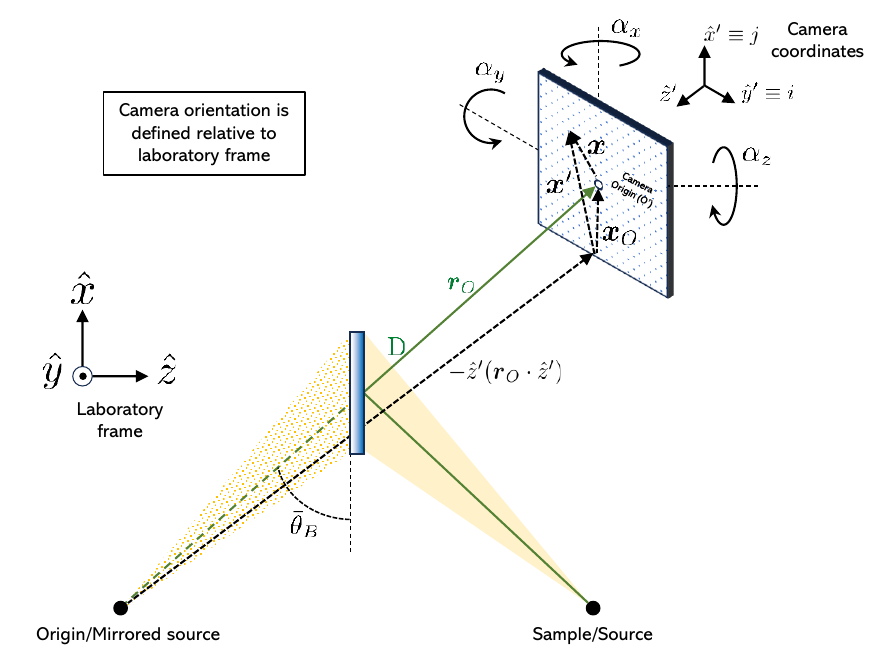}
    \caption{Scheme for the solid angles calculations in the flat crystal spectrometer case.
    }
    \label{solid angles}
\end{figure}

The final step is to construct an estimate of the photon emission spectrum, $S(E)$, in the direction of the detector, based on the photon hit distribution $N^{\gamma}_{ij}$ obtained in the previous section. In the following, we will assume the emission spectrum to be independent of the unit direction $\boldsymbol{\hat{k}}$ in the solid angle detected by the camera $\Omega_{det}$. If this assumption does not hold, the following discussion remains largely unaffected, except that the calculated spectrum will represent values averaged over all detected directions for a given energy. The measured spectrum, $I(E)$, can be expressed as follows:

\begin{equation}
\begin{split}
& I(E_l) = \int_{\Omega_{det}}d\Omega(\boldsymbol{\hat{k}}) \int_{E_l-\Delta E/2}^{E_l+\Delta E/2} dE \ S'(E,\boldsymbol{\hat{k}}) \\
& l=1,\dots,N_b \ ,  
\end{split}
\label{measured spectrum} 
\end{equation}
where $E$ and $\boldsymbol{\hat{k}}$ are the energy and unit wave vector of the photons, $\Delta E$ is the energy resolution, and $N_b$ is the number of energy bins. Here, $S'(E,\boldsymbol{\hat{k}})$ is the emission spectrum $S(E)$ modified by the action of the Bragg crystal, summarized in the shape function $F(E,\boldsymbol{\hat{k}})$:

\begin{equation}
S'(E, \boldsymbol{\hat{k}}) = S(E)F(E,\boldsymbol{\hat{k}}) \ .
\label{emissionspectrum}
\end{equation}
Integrating Eq. (\ref{measured spectrum}) over the energies, we obtain:

\begin{equation}
I(E_l) = \int_{\Omega_{det}}d\Omega(\boldsymbol{\hat{k}}) \ \bar{S}'(\boldsymbol{\hat{k}}; E_l) \ .
\label{passage 1}
\end{equation}
Given the energy-dispersive effect of the crystal on the radiation, $\bar{S}'(\boldsymbol{\hat{k}}; E_l)$ can be approximated in the following way:

\begin{equation}
\begin{split}
& \bar{S}'(\boldsymbol{\hat{k}}; E_l) = \bar{S}(E_l)=\int_{E_l-\Delta E/2}^{E_l+\Delta E/2} dE \ S(E) \\ 
& \forall k \ | \ E_l = \argmin_{E_{ij}}|E(\boldsymbol{\hat{k}})-E_{ij}| \ ,
\end{split}
\label{passage 2}
\end{equation}
and equal to zero for all the other values of $\boldsymbol{\hat{k}}$. Here, $E(\boldsymbol{\hat{k}})$ is given by Eq. (\ref{energy map equation}), with:

\begin{equation}
\boldsymbol{\hat{k}} = \frac{\boldsymbol{r_0}+\boldsymbol{Rr'}}{\|\boldsymbol{r_0}+\boldsymbol{Rr'}\|} \ , \ \boldsymbol{r'} = (x',y',0) \ .
\label{passage 3}
\end{equation}
Now, the integration over the solid angle in Eq. (\ref{passage 1}) becomes:

\begin{equation}
I(E_l) = S(E_l)\cdot \sum_{ij|E_{ij}\in \Delta (E_l)} \Omega_{ij} \ \ . 
\label{passage 4}
\end{equation}
Having denoted the interval $(E_l - \Delta E/2, E_l - \Delta E/2)$ as $\Delta (E_l)$, and $\Omega_{ij}$ the solid angle associated with pixel $ij$. This solid angle  for the flat crystal spectrometer can be computed through the following formulas:
\begin{eqnarray}
\hat{z}' &=& \boldsymbol{R}\hat{z} \nonumber \\
\boldsymbol{x}_0 &=& \hat{z}'(\hat{z}'\cdot \boldsymbol{r}_0) \nonumber \\
\boldsymbol{x}'_0 &=& \boldsymbol{r}_0 - \boldsymbol{x}_0 \\
\boldsymbol{x}'' &=& (\boldsymbol{x}' + \boldsymbol{R}^{-1}\boldsymbol{x}) \nonumber \\
\Omega(x,y) &=& \frac{dx dy \cos(\beta)}{d(\boldsymbol{x}')} = \frac{D'}{(x''^{2}+y''^{2}+ D'^{2})^{3/2}}, \nonumber
\label{passage 5}
\end{eqnarray}
where the relative quantities are defined in Fig. \ref{solid angles}. Similar reasoning can be carried out for computing the solid angles in the von H\'amos geometry.

The intensity $I(E_l)$ can be estimated (see Appendix~\ref{error calculations}) from the distribution of photon hits $N^{\gamma}_{ij}$ as:

\begin{equation}
\tilde{I}(E_l) = \sum_{ij|E_{ij}\in \Delta E_l} N^{\gamma}_{ij} \ \ .
\label{passage 6}
\end{equation}
Putting Eq.~(\ref{passage 4}) and~(\ref{passage 6}) together, we obtain an estimate $\tilde{S}(E_l)$ for the emission spectrum:

\begin{equation}
\tilde{S}(E_l) = \frac{\sum_{ij|E_{ij}\in \Delta E_l} N^{\gamma}_{ij}}{\sum_{ij|E_{ij}\in \Delta (E_l)} \Omega_{ij}} = \frac{\tilde{I}(E_l)}{\Omega(E_l)} \ \ .
\label{passage 7}
\end{equation}
When summing the number of photons over the energy contours associated to these bins, we will reduce the fluctuations due to the thermal noise and Poissonian statistics, thanks to the large numbers of pixels.

\section{\label{sec3} Results}

In this section, we apply the discussed methods to both synthetic and experimental data. The synthetic data will be utilized to evaluate the algorithms' performance in enhancing resolution and SNR for different conditions. The experimental data will instead demonstrate how these methods can amplify specific spectral features that would otherwise be weak, showcasing the efficacy of these techniques in real situations.

\subsection{\label{Synthetic data}Synthetic data}

\begin{figure*}
    \centering
    \includegraphics[width=\textwidth,keepaspectratio]{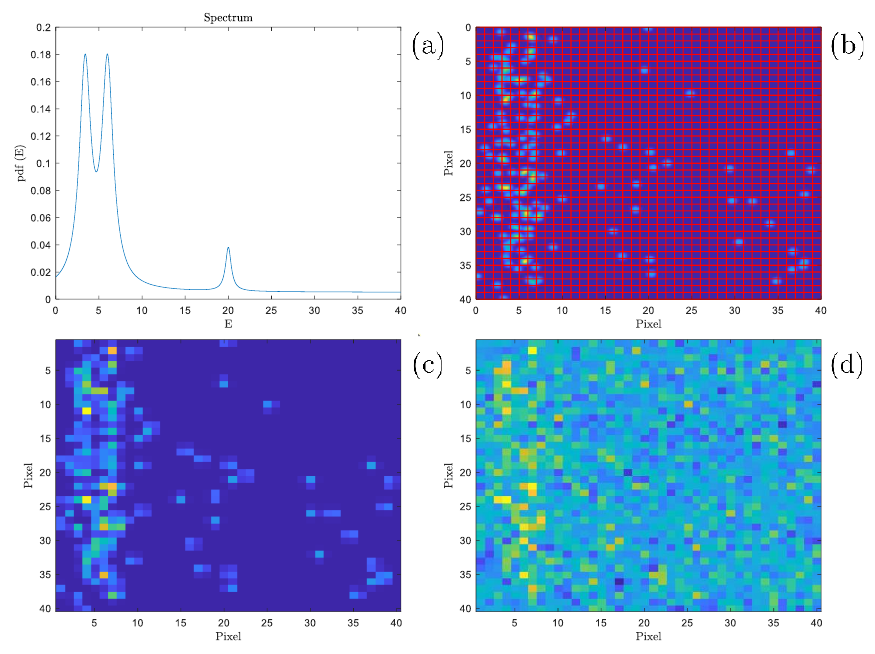}
    \caption{(a) The normalised spectrum used for producing the synthetic camera images.(b) The charge deposited on the camera surface, with the pixel grid depicted.(c) The ADU map on the pixel grid.(d) The final camera image with the noise applied.
    }
    \label{synthetic images}
\end{figure*}

In order to test our techniques, we generated some artificial camera images through the following steps, further illustrated in Fig. \ref{synthetic images}:

\begin{enumerate}
    \item We create a 40x40 array to represent the camera pixels and a finer 4000x4000 subgrid to depict its surface. For simplicity, the energy contours on the camera correspond to the columns of the arrays. 
    \item A number of photons, specified by the selected $\lambda$, are sent onto the subgrid. Their energy, which determines the pixel column they land on, is sampled from the spectrum $S(E_l)$ shown in Fig. \ref{synthetic images}a and then converted to an ADU value, taking ADU$_{sp}=120$.
    \item Charge spreading is applied by convolving the subgrid image with a gaussian kernel of width $R_s$. The ADU values of the subgrid belonging to the same pixel are then summed together to obtain the actual camera image (see Fig \ref{synthetic images}c)
    \item Finally, gaussian random noise with standard deviation $\sigma_N$ is added to the image (Fig. \ref{synthetic images}d).
\end{enumerate}

\begin{figure*}
    \centering
    \includegraphics[width=\textwidth,keepaspectratio]{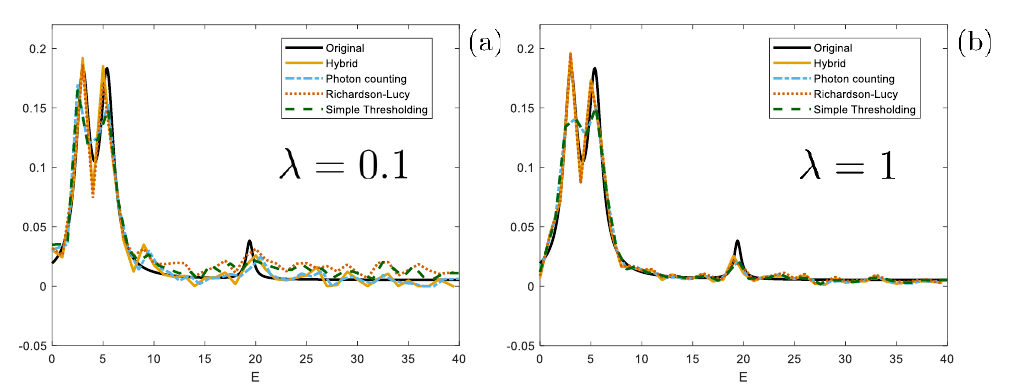}
    \caption{The original spectrum compared, for $\lambda = 0.1$ (a) and $\lambda = 1$ (b), to the spectra reconstructed with different methods: only the photon counting algorithm (`Photon counting'), only the deconvolution algorithm (`Richardson-Lucy'), both the techniques combined (`Hybrid'), none of them (`Simple Thresholding'). The camera parameters are $R_s=0.4$, $\sigma_N=30$.
    }
    \label{performances}
\end{figure*}

We evaluated the performances of our techniques for different values of $\lambda$, $R_s$ and $\sigma_N$ by computing the $\mathcal{L}_2$ distance between $S(E_l)$ and the reconstructed spectrum $\tilde{S}(E_l)$. The latter is obtained by integrating the processed camera images along its columns and then normalising it for comparison with $S(E_l)$. In Fig. \ref{performances}, we compare, for $\lambda = 0.1$ and $\lambda = 1$, the original spectrum with the spectra reconstructed using different methods to treat the camera image: only simple thresholding, only the photon counting algorithm, only the deconvolution algorithm, and the photon counting and deconvolution techniques combined (`Hybrid' mode). Panel (a) shows that, for small $\lambda$ values, the photon counting algorithm alone performs better than the deconvolution algorithm alone and the simple thresholding, especially in the regions with very few photons where ADU counts are mostly produced by noise. In contrast, for large $\lambda$ values, the photon counting algorithm and the simple thresholding offer little improvement over the thresholded image, whereas the deconvolution algorithm is able better to treat the regions with high $\lambda$, enhancing and better resolving the spectral features. Sequential application of the two methods combines their benefits, yielding the best performances for both values of $\lambda$. We note that the applied method can have a substantial effect on the extracted spectral features, such as spectral linewidths or line ratios, which are often used as key diagnostics in experiments.

\begin{figure}
    \centering
    \includegraphics[width=\columnwidth,keepaspectratio]{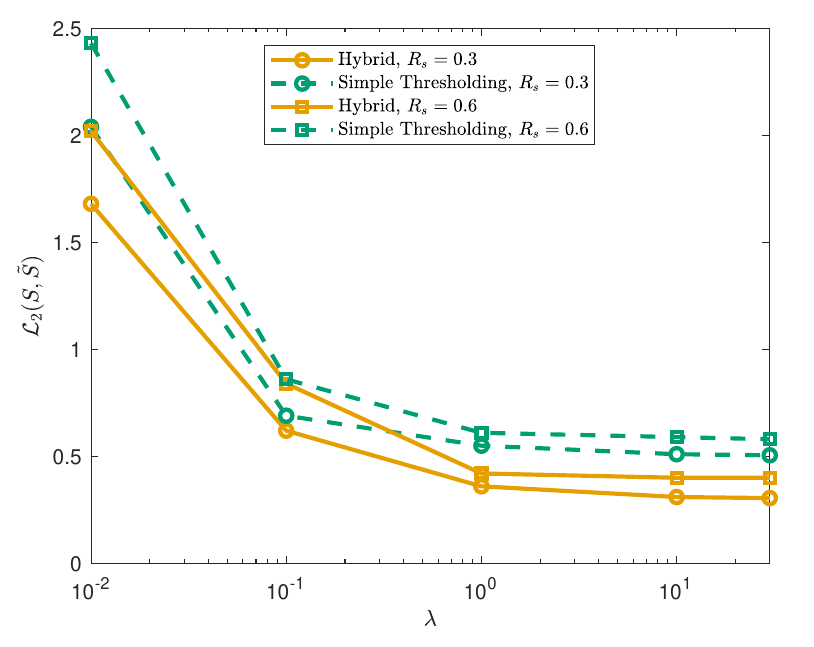}
    \caption{$\mathcal{L}_2$ distance between the original spectrum and the one reconstructed from the processed and thresholded image as a function of the filling fraction, for two values of $R_s$.
    }
    \label{performances3}
\end{figure}

To evaluate performance, we study the quality of the spectrum reconstructed from the `Hybrid'-processed image compared to that obtained by `Simple thresholding' of the raw image as a function of the filling fraction. The effect of changing the noise level is included in this study, as the ratio $\lambda/\sigma_N$ is the dominant parameter determining the quality of the reconstruction. This analysis was conducted for two values of the charge spreading, $R_s=0.3$ and $R_s=0.6$ in pixel units, and its results are reported in Fig. \ref{performances3}.
As anticipated, the `Hybrid'-processed image consistently yields a higher quality spectrum than the thresholded image across all $\lambda$ values and for each $R_s$. The smallest difference occurs at intermediate $\lambda$ values, where neither the photon counting algorithm nor the deconvolution algorithm are particularly effective. Additionally, the quality of the reconstruction improves for smaller $R_s$ and larger $\lambda$, reaching an asymptote as $\lambda\to\infty$. This behaviour is expected, as a larger number of photons on the camera reduces the Poissonian error and the relative strength of noise, while a smaller $R_s$ facilitates the deconvolution process and the identification of the thresholding value, better separating the 0th and 1st peak in the histogram of photon counts. 

\subsection{\label{Real data}Experimental data}

\begin{figure*}
    \includegraphics[width=\textwidth,keepaspectratio]{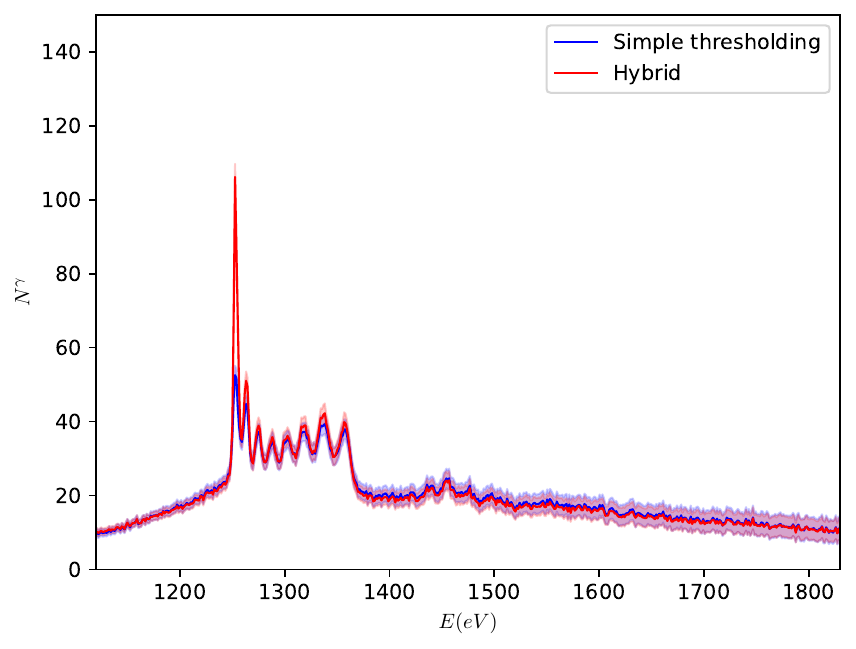}
    \caption{Emission spectrum of MgF$_2$ obtained with simple thresholding (red), and processing the images with the photon counting and deconvolution methods (blue). Error bounds are indicated by shaded regions. The peaks in the spectrum correspond to the $K\alpha$ (from 1250 to 1370 eV) and $K\beta$ (from 1430 to 1500 eV) emission lines of the different ionization states of the Mg ions.}
    \label{real_data}
\end{figure*}

In this section we show the application of our analysis approach to a real camera image and the resulting improvements in the spectrum obtained. Figure \ref{real_data} reports the MgF$_2$ emission spectrum computed either with simple thresholding or with the image processing described in section \ref{Estimate of the photon hits distribution}. The parameters used in these operations ($ADU_{sp}$, $\sigma_N$, $\lambda$) were determined by fitting a synthetic histogram to the experimental one (see Fig. \ref{histogram}e). To generate the synthetic histogram, we used the parameters of the PIXIS-XF camera (the detector employed for this experiment), which were provided in the relevant documentation~\cite{yaffe1997x}. From this documentation and the fitting process, we obtained $R_s = 1.7$ in pixel unit, $ADU_{sp} = 43$ ADU and $\sigma_N=4$ ADU, which were applied in the processing of the experimental images. For the thresholding we used the values specified in Section \ref{Photon counting algorithms}. The error bars for the two spectra were computed using Poissonian inference, and neglecting noise-related errors, drastically reduced by thresholding and integration over the energy contours. Details of these error calculations are provided in Appendix \ref{error calculations}.

The enhancement of the features is visible, especially for the brightest ones. We believe that such improvements are mainly due to the reduction of the background noise, and they could therefore be essential when the physical information of the probed system is contained in weaker features (e.g. Raman scattering peaks). For the photon counting algorithm to be effective, it must be applied to single-shot images rather than averaged ones, where single photon clusters are reduced by summation. This requirement, combined with the typical size of a camera image (here 2048x2048 pixels), makes the process time-consuming for real-time analysis. However, this procedure can be performed during post-analysis, where time constraints are less demanding.

\section{\label{sec4} Conclusions}

In summary, we have provided a detailed guide for constructing spectra from the relative camera images, for two of the most common geometries employed in x-ray dispersion spectroscopy. Specifically, we examined the computation of the energy map and the solid angle correction for the pixel array, the estimation of important camera parameters such as the thermal noise intensity, and presented techniques to estimate the distribution of photon hits from the raw camera data, with the aim of increasing the SNR. These image processing techniques are based on the sequential combination of clustering algorithms and deconvolution methods for simultaneously treating regions with different local fill fractions. Application of these methods to both synthetic and experimental data demonstrated notable improvements in SNR and feature clarity, proving their potential in contexts where important physical features are weak, such as HED science. Future work could explore introducing in this framework the capacity of multiple photons-cluster identification, or expanding the variety of allowed cluster shapes. Additionally, it would be useful to integrate this work with studies on the instrument function to combine the improvement of SNR with resolution enhancement.

\appendix

\section{\label{Determination of thresholding values} Determination of thresholding values}

The two thresholding values can be determined using the following procedure. First, we generate $N_s$ synthetic images (see also Section \ref{Synthetic data}) for specified parameters ($ADU_{sp}$, $\sigma_N$, $R_s$), and for a small fill fraction (e.g. $\lambda=0.1$), so that single-photon clusters are predominant. Each of these synthetic images is then processed with the photon clustering algorithm, applying a trial pair of thresholding values, and divided by $ADU_{sp}$ to reconstruct the photon hits map $\{N^{rec}\}$. The reconstructed photon maps are then compared to the original photon map $\{N^{orig}\}$, and a global loss function is computed:

\begin{equation}
\mathcal{L} = \frac{1}{N_s} \sum_{l = 1}^{N_s} \mathcal{L}_F(\{N^{rec}\}_l - \{N^{orig}\}) \ \ ,
\label{appendix 0}
\end{equation}
where $\{N^{rec}\}_l$ and $\{N^{orig}\}$ are two matrices and $\mathcal{L}_F$ is the Frobenius norm. The optimal threshold pair is obtained as the one that minimizes this loss function.

\section{\label{Probability distribution of the pixel values}Probability distribution of the ADU counts}

Let us focus on regions with high local $\lambda$, where the photon counting algorithm leaves pixel values unchanged. While this discussion can be extended to include thresholding and charge accumulation operations, doing so adds complexity without altering the fundamental concepts. First, consider the signal on a sufficiently fine subgrid $k,l = 1,2, ...$ (see also Section \ref{Synthetic data}):

\begin{equation}
    A_{kl} = \sum_{k'l'} ADU_{sp} (N^{\gamma}_{k'l'} + \beta_{k' l'}) H_{k-k' \ l-l'} \ \ \ \ \ \ ,
    \label{extra1}
\end{equation}
where $H$ is the charge spreading function, and $N^{\gamma}_{k'l'} \sim {\rm Poisson}(I_{k'l'})$, with $I_{k'l'}$ denoting the spectral intensity at subpixel $(k'l')$. The term $\beta_{k' l'} \sim \mathcal{N}(0,f N^{\gamma}_{k'l'})$ is a compound probability distribution accounting for random fluctuations in e-h pairs production, quantified by the Fano factor $f$. This factor is usually small (e.g. $f=0.12$ in Silicon~\cite{humphries2020isochoric}), and therefore this noise source is often neglected. The value $A_{ij}$ is obtained by summing $A_{kl}$ over the subpixels within each pixel, and then adding the thermal noise specific to the pixel:

\begin{equation}
    A_{ij} = \sum_{k,l \in (i,j)} A_{kl} + \eta_{ij} \ \ ,
    \label{extra2}
\end{equation}
where $k,l\in(i,j)$ indicates that subpixel $(k,l)$ lies in pixel $(i,j)$. Using the properties of the Poisson distribution and neglecting the e-h pairs production noise, we can now write:

\begin{equation}
\begin{split}
      & A_{ij} = \sum_{k,l \in (i,j)} \sum_{k'l'} ADU_{sp}{\rm Poisson}(I_{k'l'}) H_{k-k' \ l-l'} + \eta_{ij} \\
      & = ADU_{sp} \ {\rm Poisson}\biggl(\sum_{k,l \in (i,j)} \sum_{k'l'} I_{k'l'} H_{k-k' \ l-l'}\biggr) + \eta_{ij} \\
      & = ADU_{sp} \ {\rm Poisson}\biggl(\sum_{i' j'} \sum_{k'l' \in (i',j')} I_{k'l'} \sum_{k,l \in (i,j)} H_{k-k' \ l-l'}\biggr) + \eta_{ij} \\
      & \approx ADU_{sp} \ {\rm Poisson}\biggl(\sum_{i' j'} \bar{I}_{i'j'} \sum_{k'l' \in (i',j')} \sum_{k,l \in (i,j)} H_{k-k' \ l-l'}\biggr) + \eta_{ij} \\
      & = ADU_{sp} \ {\rm Poisson}\biggl(\sum_{i' j'} I_{i'j'} \tilde{H}_{i-i' \ j-j'}\biggr) + \eta_{ij} \\
      & = ADU_{sp} \ \sum_{i' j'} {\rm Poisson} (I_{i'j'}) \tilde{H}_{i-i' \ j-j'} + \eta_{ij} \ \ ,
\end{split}
    \label{extra3}
\end{equation}
where we reasonably assumed $I_{k'l'}$ to be constant within each pixel and equal to $\bar{I}_{i'j'}$. Here, we defined $I_{i'j'} = \sum_{k'l' \in (i',j')} I_{k'l'}\approx N_p \bar{I}_{i' j'}$, with $N_p$ representing the number of subpixels per pixel, and an effective charge spreading function as:

\begin{equation}
    \tilde{H}_{i-i' \ j-j'} = \frac{1}{N_p}\sum_{i' k' \in (i' j')} \sum_{i k \in (i j)} H_{k-k' \ l-l'} \ \ . 
    \label{extra4}
\end{equation}
The translation invariance of $\tilde{H}$ follows directly from the inherent translation invariance of $H$. Equation (\ref{extra3}) proves that the operation defined by Eq. (\ref{RL deconvolution}), combined with the division in Eq. (\ref{photon number}), yields an estimate for $I_{ij}$.

\section{\label{error calculations} Error calculations}

We begin by computing the best estimate of the parameter $I$ (also the mean) of a Poisson distribution, given an observation $n$:

\begin{equation}
N \sim Poisson(I) \  \Rightarrow \ \mathcal{P}(N = n|I) = \frac{I^n e^{-I}}{n!}.
\label{appendix 1}
\end{equation}
Assuming a flat prior distribution for $I$ over a large interval $[0,M]$ (since $I$ cannot be infinitely large), and noting that $\mathcal{P}(n)=\int_0^M \mathcal{P}(n|I)  \mathcal{P}(I) = \int_0^M dI \frac{I^n e^{-I}}{n!} \frac{1}{M} \approx \frac{1}{M}$ if $M$ is sufficiently large, from Bayes' theorem we find:

\begin{equation}
\mathcal{P}(I|n) = \frac{\mathcal{P}(n|I)\mathcal{P}(I)}{\mathcal{P}(n)} = \mathcal{P}(n|I) \ \ .
\label{appendix 2}
\end{equation}
From this equation, we can calculate the best estimate for $I$ as:

\begin{equation}
\langle I \rangle (n) = \int_0^M dI I \mathcal{P}(I|n) \int_0^M dI I \frac{I^n e^{-I}}{n!} \approx n+1 \ \ .
\label{appendix 3}
\end{equation}
Similarly, we determine the error on this estimate as the standard deviation of $\mathcal{P}(I|n)$:

\begin{equation}
\sqrt{\langle I^2 - \langle I \rangle^2 \rangle} (n) \approx \sqrt{n+1} \ \ .
\label{appendix 4}
\end{equation}
To analyse our case, we consider $\tilde{I}(E_l)= \sum_{ij|E_{ij}\in \Delta E_l} N^{\gamma}_{ij}$ as a realization of the random variable $I(E_l) \sim {\rm Poisson}(\Omega(E_l) S(E_l))$. Here, we are disregarding the contribution to $I(E_l)$ from noise, which is generally small due to the thresholding operations and the summation over many pixels. Additionally, we are not considering the error propagation through processing steps, taking instead directly $N^{\gamma}_{ij}$ as the estimate of the photon hits map. Under these assumptions, the spectrum and its error, for each energy bin, are given by:
\begin{equation}
\langle S(E_l) \rangle = \frac{\tilde{I}(E_l)+1}{\Omega(E_l)} \ \  \sigma(S(E_l)) = \frac{\sqrt{I(E_l)+1}}{\Omega(E_l)} \ \ .
\label{appendix 5}
\end{equation}
As a final consideration, we observe that the error normalized to the signal, $\sigma(S(E_l))/\langle S(E_l) \rangle = 1/\sqrt{\tilde{I}(E_l)+1}$, correctly decreases as the number of photons per bin increases.

\section*{Data availability}
The data that support the findings of this study are available from the corresponding author upon reasonable request.

\section*{Acknowledgements}
A.F. and S.M.V. acknowledge support for the STFC UK Hub for the Physical Sciences on XFELS.
T.G. and S.M.V. acknowledge support from AWE via the Oxford Centre for High Energy Density Science (OxCHEDS).
T. C. and S.M.V. acknowledge support from the Royal Society. 
Y.S. and S.M.V. acknowledge support from the UK EPSRC under grants EP/P015794/1 and EP/W010097/1. T.G. acknowledges support by the Center for Advanced Systems Understanding (CASUS), financed by Germany’s Federal Ministry of Education and Research (BMBF) and the Saxon state government out of the State budget approved by the Saxon State Parliament. This work has received funding from the European Union’s Just Transition Fund (JTF) within the project Röntgenlaser-Optimierung der Laserfusion (ROLF), Contract No. 5086999001, co-financed by the Saxon state government out of the State budget approved by the Saxon State Parliament.

\section*{Contributions}
A.F., O.S.H., T.G. and S.M.V developed the methods. A.F., T.G implemented the codes, with contributions from O.S.H., T.C. and Y.S. A.F. wrote the manuscript with contributions from T.G. and O.S.H, and all authors reviewed and edited the final version. S.M.V. supervised the project.

\section*{Competing interests}
The authors declare no competing interests.

\bibliography{bibliography}

\providecommand{\noopsort}[1]{}\providecommand{\singleletter}[1]{#1}%
\begin{thebibliography}{60}%
\makeatletter
\providecommand \@ifxundefined [1]{%
 \@ifx{#1\undefined}
}%
\providecommand \@ifnum [1]{%
 \ifnum #1\expandafter \@firstoftwo
 \else \expandafter \@secondoftwo
 \fi
}%
\providecommand \@ifx [1]{%
 \ifx #1\expandafter \@firstoftwo
 \else \expandafter \@secondoftwo
 \fi
}%
\providecommand \natexlab [1]{#1}%
\providecommand \enquote  [1]{``#1''}%
\providecommand \bibnamefont  [1]{#1}%
\providecommand \bibfnamefont [1]{#1}%
\providecommand \citenamefont [1]{#1}%
\providecommand \href@noop [0]{\@secondoftwo}%
\providecommand \href [0]{\begingroup \@sanitize@url \@href}%
\providecommand \@href[1]{\@@startlink{#1}\@@href}%
\providecommand \@@href[1]{\endgroup#1\@@endlink}%
\providecommand \@sanitize@url [0]{\catcode `\\12\catcode `\$12\catcode
  `\&12\catcode `\#12\catcode `\^12\catcode `\_12\catcode `\%12\relax}%
\providecommand \@@startlink[1]{}%
\providecommand \@@endlink[0]{}%
\providecommand \url  [0]{\begingroup\@sanitize@url \@url }%
\providecommand \@url [1]{\endgroup\@href {#1}{\urlprefix }}%
\providecommand \urlprefix  [0]{URL }%
\providecommand \Eprint [0]{\href }%
\providecommand \doibase [0]{https://doi.org/}%
\providecommand \selectlanguage [0]{\@gobble}%
\providecommand \bibinfo  [0]{\@secondoftwo}%
\providecommand \bibfield  [0]{\@secondoftwo}%
\providecommand \translation [1]{[#1]}%
\providecommand \BibitemOpen [0]{}%
\providecommand \bibitemStop [0]{}%
\providecommand \bibitemNoStop [0]{.\EOS\space}%
\providecommand \EOS [0]{\spacefactor3000\relax}%
\providecommand \BibitemShut  [1]{\csname bibitem#1\endcsname}%
\let\auto@bib@innerbib\@empty
\bibitem [{\citenamefont {Born}\ and\ \citenamefont
  {Wolf}(2013)}]{born2013principles}%
  \BibitemOpen
  \bibfield  {author} {\bibinfo {author} {\bibfnamefont {M.}~\bibnamefont
  {Born}}\ and\ \bibinfo {author} {\bibfnamefont {E.}~\bibnamefont {Wolf}},\
  }\href@noop {} {\emph {\bibinfo {title} {Principles of optics:
  electromagnetic theory of propagation, interference and diffraction of
  light}}}\ (\bibinfo  {publisher} {Elsevier},\ \bibinfo {year}
  {2013})\BibitemShut {NoStop}%
\bibitem [{\citenamefont {Foord}\ \emph {et~al.}(1969)\citenamefont {Foord},
  \citenamefont {Jones}, \citenamefont {Oliver},\ and\ \citenamefont
  {Pike}}]{foord1969use}%
  \BibitemOpen
  \bibfield  {author} {\bibinfo {author} {\bibfnamefont {R.}~\bibnamefont
  {Foord}}, \bibinfo {author} {\bibfnamefont {R.}~\bibnamefont {Jones}},
  \bibinfo {author} {\bibfnamefont {C.}~\bibnamefont {Oliver}},\ and\ \bibinfo
  {author} {\bibfnamefont {E.}~\bibnamefont {Pike}},\ }\bibfield  {title}
  {\bibinfo {title} {The use of photomultiplier tubes for photon counting},\
  }\href@noop {} {\bibfield  {journal} {\bibinfo  {journal} {Applied optics}\
  }\textbf {\bibinfo {volume} {8}},\ \bibinfo {pages} {1975} (\bibinfo {year}
  {1969})}\BibitemShut {NoStop}%
\bibitem [{\citenamefont {Peterson}(2001)}]{peterson2001works}%
  \BibitemOpen
  \bibfield  {author} {\bibinfo {author} {\bibfnamefont {C.}~\bibnamefont
  {Peterson}},\ }\bibfield  {title} {\bibinfo {title} {How it works: the
  charged-coupled device, or ccd},\ }\href@noop {} {\bibfield  {journal}
  {\bibinfo  {journal} {Journal of young investigators}\ }\textbf {\bibinfo
  {volume} {3}} (\bibinfo {year} {2001})}\BibitemShut {NoStop}%
\bibitem [{\citenamefont {Svanberg}(2012)}]{svanberg2012atomic}%
  \BibitemOpen
  \bibfield  {author} {\bibinfo {author} {\bibfnamefont {S.}~\bibnamefont
  {Svanberg}},\ }\href@noop {} {\emph {\bibinfo {title} {Atomic and molecular
  spectroscopy: basic aspects and practical applications}}},\ Vol.~\bibinfo
  {volume} {6}\ (\bibinfo  {publisher} {Springer Science \& Business Media},\
  \bibinfo {year} {2012})\BibitemShut {NoStop}%
\bibitem [{\citenamefont {Willmott}(2019)}]{willmott2019introduction}%
  \BibitemOpen
  \bibfield  {author} {\bibinfo {author} {\bibfnamefont {P.}~\bibnamefont
  {Willmott}},\ }\href@noop {} {\emph {\bibinfo {title} {An introduction to
  synchrotron radiation: techniques and applications}}}\ (\bibinfo  {publisher}
  {John Wiley \& Sons},\ \bibinfo {year} {2019})\BibitemShut {NoStop}%
\bibitem [{\citenamefont {Jaroszynski}\ \emph {et~al.}(2006)\citenamefont
  {Jaroszynski}, \citenamefont {Bingham}, \citenamefont {Brunetti},
  \citenamefont {Ersfeld}, \citenamefont {Gallacher}, \citenamefont {van
  Der~Geer}, \citenamefont {Issac}, \citenamefont {Jamison}, \citenamefont
  {Jones}, \citenamefont {De~Loos} \emph {et~al.}}]{jaroszynski2006radiation}%
  \BibitemOpen
  \bibfield  {author} {\bibinfo {author} {\bibfnamefont {D.}~\bibnamefont
  {Jaroszynski}}, \bibinfo {author} {\bibfnamefont {R.}~\bibnamefont
  {Bingham}}, \bibinfo {author} {\bibfnamefont {E.}~\bibnamefont {Brunetti}},
  \bibinfo {author} {\bibfnamefont {B.}~\bibnamefont {Ersfeld}}, \bibinfo
  {author} {\bibfnamefont {J.}~\bibnamefont {Gallacher}}, \bibinfo {author}
  {\bibfnamefont {B.}~\bibnamefont {van Der~Geer}}, \bibinfo {author}
  {\bibfnamefont {R.}~\bibnamefont {Issac}}, \bibinfo {author} {\bibfnamefont
  {S.}~\bibnamefont {Jamison}}, \bibinfo {author} {\bibfnamefont
  {D.}~\bibnamefont {Jones}}, \bibinfo {author} {\bibfnamefont
  {M.}~\bibnamefont {De~Loos}}, \emph {et~al.},\ }\bibfield  {title} {\bibinfo
  {title} {Radiation sources based on laser--plasma interactions},\ }\href@noop
  {} {\bibfield  {journal} {\bibinfo  {journal} {Philosophical Transactions of
  the Royal Society A: Mathematical, Physical and Engineering Sciences}\
  }\textbf {\bibinfo {volume} {364}},\ \bibinfo {pages} {689} (\bibinfo {year}
  {2006})}\BibitemShut {NoStop}%
\bibitem [{\citenamefont {Schm{\"u}ser}\ \emph {et~al.}(2014)\citenamefont
  {Schm{\"u}ser}, \citenamefont {Dohlus}, \citenamefont {Rossbach},\ and\
  \citenamefont {Behrens}}]{schmuser2014free}%
  \BibitemOpen
  \bibfield  {author} {\bibinfo {author} {\bibfnamefont {P.}~\bibnamefont
  {Schm{\"u}ser}}, \bibinfo {author} {\bibfnamefont {M.}~\bibnamefont
  {Dohlus}}, \bibinfo {author} {\bibfnamefont {J.}~\bibnamefont {Rossbach}},\
  and\ \bibinfo {author} {\bibfnamefont {C.}~\bibnamefont {Behrens}},\
  }\bibfield  {title} {\bibinfo {title} {Free-electron lasers in the
  ultraviolet and x-ray regime},\ }\href@noop {} {\bibfield  {journal}
  {\bibinfo  {journal} {Springer Tracts in Modern Physics}\ }\textbf {\bibinfo
  {volume} {258}} (\bibinfo {year} {2014})}\BibitemShut {NoStop}%
\bibitem [{\citenamefont {Yamada}\ \emph {et~al.}(2024)\citenamefont {Yamada},
  \citenamefont {Matsuyama}, \citenamefont {Inoue}, \citenamefont {Osaka},
  \citenamefont {Inoue}, \citenamefont {Nakamura}, \citenamefont {Tanaka},
  \citenamefont {Inubushi}, \citenamefont {Yabuuchi}, \citenamefont {Tono}
  \emph {et~al.}}]{yamada2024extreme}%
  \BibitemOpen
  \bibfield  {author} {\bibinfo {author} {\bibfnamefont {J.}~\bibnamefont
  {Yamada}}, \bibinfo {author} {\bibfnamefont {S.}~\bibnamefont {Matsuyama}},
  \bibinfo {author} {\bibfnamefont {I.}~\bibnamefont {Inoue}}, \bibinfo
  {author} {\bibfnamefont {T.}~\bibnamefont {Osaka}}, \bibinfo {author}
  {\bibfnamefont {T.}~\bibnamefont {Inoue}}, \bibinfo {author} {\bibfnamefont
  {N.}~\bibnamefont {Nakamura}}, \bibinfo {author} {\bibfnamefont
  {Y.}~\bibnamefont {Tanaka}}, \bibinfo {author} {\bibfnamefont
  {Y.}~\bibnamefont {Inubushi}}, \bibinfo {author} {\bibfnamefont
  {T.}~\bibnamefont {Yabuuchi}}, \bibinfo {author} {\bibfnamefont
  {K.}~\bibnamefont {Tono}}, \emph {et~al.},\ }\bibfield  {title} {\bibinfo
  {title} {Extreme focusing of hard x-ray free-electron laser pulses enables 7
  nm focus width and 1022 w cm- 2 intensity},\ }\href@noop {} {\bibfield
  {journal} {\bibinfo  {journal} {Nature Photonics}\ ,\ \bibinfo {pages} {1}}
  (\bibinfo {year} {2024})}\BibitemShut {NoStop}%
\bibitem [{\citenamefont {Basden}\ \emph {et~al.}(2003)\citenamefont {Basden},
  \citenamefont {Haniff},\ and\ \citenamefont {Mackay}}]{basden2003photon}%
  \BibitemOpen
  \bibfield  {author} {\bibinfo {author} {\bibfnamefont {A.~G.}\ \bibnamefont
  {Basden}}, \bibinfo {author} {\bibfnamefont {C.}~\bibnamefont {Haniff}},\
  and\ \bibinfo {author} {\bibfnamefont {C.}~\bibnamefont {Mackay}},\
  }\bibfield  {title} {\bibinfo {title} {Photon counting strategies with
  low-light-level ccds},\ }\href@noop {} {\bibfield  {journal} {\bibinfo
  {journal} {Monthly notices of the royal astronomical society}\ }\textbf
  {\bibinfo {volume} {345}},\ \bibinfo {pages} {985} (\bibinfo {year}
  {2003})}\BibitemShut {NoStop}%
\bibitem [{\citenamefont {Eggert}\ \emph {et~al.}(2010)\citenamefont {Eggert},
  \citenamefont {Hicks}, \citenamefont {Celliers}, \citenamefont {Bradley},
  \citenamefont {McWilliams}, \citenamefont {Jeanloz}, \citenamefont {Miller},
  \citenamefont {Boehly},\ and\ \citenamefont {Collins}}]{eggert2010melting}%
  \BibitemOpen
  \bibfield  {author} {\bibinfo {author} {\bibfnamefont {J.}~\bibnamefont
  {Eggert}}, \bibinfo {author} {\bibfnamefont {D.}~\bibnamefont {Hicks}},
  \bibinfo {author} {\bibfnamefont {P.}~\bibnamefont {Celliers}}, \bibinfo
  {author} {\bibfnamefont {D.}~\bibnamefont {Bradley}}, \bibinfo {author}
  {\bibfnamefont {R.}~\bibnamefont {McWilliams}}, \bibinfo {author}
  {\bibfnamefont {R.}~\bibnamefont {Jeanloz}}, \bibinfo {author} {\bibfnamefont
  {J.}~\bibnamefont {Miller}}, \bibinfo {author} {\bibfnamefont
  {T.}~\bibnamefont {Boehly}},\ and\ \bibinfo {author} {\bibfnamefont
  {G.}~\bibnamefont {Collins}},\ }\bibfield  {title} {\bibinfo {title} {Melting
  temperature of diamond at ultrahigh pressure},\ }\href@noop {} {\bibfield
  {journal} {\bibinfo  {journal} {Nature Physics}\ }\textbf {\bibinfo {volume}
  {6}},\ \bibinfo {pages} {40} (\bibinfo {year} {2010})}\BibitemShut {NoStop}%
\bibitem [{\citenamefont {Zylstra}\ \emph {et~al.}(2022)\citenamefont
  {Zylstra}, \citenamefont {Hurricane}, \citenamefont {Callahan}, \citenamefont
  {Kritcher}, \citenamefont {Ralph}, \citenamefont {Robey}, \citenamefont
  {Ross}, \citenamefont {Young}, \citenamefont {Baker}, \citenamefont {Casey}
  \emph {et~al.}}]{Zylstra2022-kp}%
  \BibitemOpen
  \bibfield  {author} {\bibinfo {author} {\bibfnamefont {A.~B.}\ \bibnamefont
  {Zylstra}}, \bibinfo {author} {\bibfnamefont {O.~A.}\ \bibnamefont
  {Hurricane}}, \bibinfo {author} {\bibfnamefont {D.~A.}\ \bibnamefont
  {Callahan}}, \bibinfo {author} {\bibfnamefont {A.~L.}\ \bibnamefont
  {Kritcher}}, \bibinfo {author} {\bibfnamefont {J.~E.}\ \bibnamefont {Ralph}},
  \bibinfo {author} {\bibfnamefont {H.~F.}\ \bibnamefont {Robey}}, \bibinfo
  {author} {\bibfnamefont {J.~S.}\ \bibnamefont {Ross}}, \bibinfo {author}
  {\bibfnamefont {C.~V.}\ \bibnamefont {Young}}, \bibinfo {author}
  {\bibfnamefont {K.~L.}\ \bibnamefont {Baker}}, \bibinfo {author}
  {\bibfnamefont {D.~T.}\ \bibnamefont {Casey}}, \emph {et~al.},\ }\bibfield
  {title} {\bibinfo {title} {Burning plasma achieved in inertial fusion},\
  }\href@noop {} {\bibfield  {journal} {\bibinfo  {journal} {Nature}\ }\textbf
  {\bibinfo {volume} {601}},\ \bibinfo {pages} {542} (\bibinfo {year}
  {2022})}\BibitemShut {NoStop}%
\bibitem [{\citenamefont {Abu-Shawareb}\ \emph {et~al.}(2022)\citenamefont
  {Abu-Shawareb}, \citenamefont {Acree}, \citenamefont {Adams}, \citenamefont
  {Adams}, \citenamefont {Addis}, \citenamefont {Aden}, \citenamefont {Adrian},
  \citenamefont {Afeyan}, \citenamefont {Aggleton}, \citenamefont {Aghaian}
  \emph {et~al.}}]{LawsonCriterion-2022}%
  \BibitemOpen
  \bibfield  {author} {\bibinfo {author} {\bibfnamefont {H.}~\bibnamefont
  {Abu-Shawareb}}, \bibinfo {author} {\bibfnamefont {R.}~\bibnamefont {Acree}},
  \bibinfo {author} {\bibfnamefont {P.}~\bibnamefont {Adams}}, \bibinfo
  {author} {\bibfnamefont {J.}~\bibnamefont {Adams}}, \bibinfo {author}
  {\bibfnamefont {B.}~\bibnamefont {Addis}}, \bibinfo {author} {\bibfnamefont
  {R.}~\bibnamefont {Aden}}, \bibinfo {author} {\bibfnamefont {P.}~\bibnamefont
  {Adrian}}, \bibinfo {author} {\bibfnamefont {B.~B.}\ \bibnamefont {Afeyan}},
  \bibinfo {author} {\bibfnamefont {M.}~\bibnamefont {Aggleton}}, \bibinfo
  {author} {\bibfnamefont {L.}~\bibnamefont {Aghaian}}, \emph {et~al.}
  (\bibinfo {collaboration} {Indirect Drive ICF Collaboration}),\ }\bibfield
  {title} {\bibinfo {title} {{Lawson Criterion for Ignition Exceeded in an
  Inertial Fusion Experiment}},\ }\href
  {https://doi.org/10.1103/PhysRevLett.129.075001} {\bibfield  {journal}
  {\bibinfo  {journal} {Phys. Rev. Lett.}\ }\textbf {\bibinfo {volume} {129}},\
  \bibinfo {pages} {075001} (\bibinfo {year} {2022})}\BibitemShut {NoStop}%
\bibitem [{\citenamefont {Abu-Shawareb}\ \emph {et~al.}(2024)\citenamefont
  {Abu-Shawareb}, \citenamefont {Acree}, \citenamefont {Adams}, \citenamefont
  {Adams}, \citenamefont {Addis}, \citenamefont {Aden}, \citenamefont {Adrian},
  \citenamefont {Afeyan}, \citenamefont {Aggleton}, \citenamefont {Aghaian}
  \emph {et~al.}}]{Gain-2024}%
  \BibitemOpen
  \bibfield  {author} {\bibinfo {author} {\bibfnamefont {H.}~\bibnamefont
  {Abu-Shawareb}}, \bibinfo {author} {\bibfnamefont {R.}~\bibnamefont {Acree}},
  \bibinfo {author} {\bibfnamefont {P.}~\bibnamefont {Adams}}, \bibinfo
  {author} {\bibfnamefont {J.}~\bibnamefont {Adams}}, \bibinfo {author}
  {\bibfnamefont {B.}~\bibnamefont {Addis}}, \bibinfo {author} {\bibfnamefont
  {R.}~\bibnamefont {Aden}}, \bibinfo {author} {\bibfnamefont {P.}~\bibnamefont
  {Adrian}}, \bibinfo {author} {\bibfnamefont {B.~B.}\ \bibnamefont {Afeyan}},
  \bibinfo {author} {\bibfnamefont {M.}~\bibnamefont {Aggleton}}, \bibinfo
  {author} {\bibfnamefont {L.}~\bibnamefont {Aghaian}}, \emph {et~al.}
  (\bibinfo {collaboration} {The Indirect Drive ICF Collaboration}),\
  }\bibfield  {title} {\bibinfo {title} {{Achievement of Target Gain Larger
  than Unity in an Inertial Fusion Experiment}},\ }\href
  {https://doi.org/10.1103/PhysRevLett.132.065102} {\bibfield  {journal}
  {\bibinfo  {journal} {Phys. Rev. Lett.}\ }\textbf {\bibinfo {volume} {132}},\
  \bibinfo {pages} {065102} (\bibinfo {year} {2024})}\BibitemShut {NoStop}%
\bibitem [{\citenamefont {Kraus}\ \emph {et~al.}(2016)\citenamefont {Kraus},
  \citenamefont {Ravasio}, \citenamefont {Gauthier}, \citenamefont {Gericke},
  \citenamefont {Vorberger}, \citenamefont {Frydrych}, \citenamefont
  {Helfrich}, \citenamefont {Fletcher}, \citenamefont {Schaumann},
  \citenamefont {Nagler}, \citenamefont {Barbrel}, \citenamefont {Bachmann},
  \citenamefont {Gamboa}, \citenamefont {G{\"o}de}, \citenamefont {Granados},
  \citenamefont {Gregori}, \citenamefont {Lee}, \citenamefont {Neumayer},
  \citenamefont {Schumaker}, \citenamefont {D{\"o}ppner}, \citenamefont
  {Falcone}, \citenamefont {Glenzer},\ and\ \citenamefont
  {Roth}}]{Kraus2016-hc}%
  \BibitemOpen
  \bibfield  {author} {\bibinfo {author} {\bibfnamefont {D.}~\bibnamefont
  {Kraus}}, \bibinfo {author} {\bibfnamefont {A.}~\bibnamefont {Ravasio}},
  \bibinfo {author} {\bibfnamefont {M.}~\bibnamefont {Gauthier}}, \bibinfo
  {author} {\bibfnamefont {D.~O.}\ \bibnamefont {Gericke}}, \bibinfo {author}
  {\bibfnamefont {J.}~\bibnamefont {Vorberger}}, \bibinfo {author}
  {\bibfnamefont {S.}~\bibnamefont {Frydrych}}, \bibinfo {author}
  {\bibfnamefont {J.}~\bibnamefont {Helfrich}}, \bibinfo {author}
  {\bibfnamefont {L.~B.}\ \bibnamefont {Fletcher}}, \bibinfo {author}
  {\bibfnamefont {G.}~\bibnamefont {Schaumann}}, \bibinfo {author}
  {\bibfnamefont {B.}~\bibnamefont {Nagler}}, \bibinfo {author} {\bibfnamefont
  {B.}~\bibnamefont {Barbrel}}, \bibinfo {author} {\bibfnamefont
  {B.}~\bibnamefont {Bachmann}}, \bibinfo {author} {\bibfnamefont {E.~J.}\
  \bibnamefont {Gamboa}}, \bibinfo {author} {\bibfnamefont {S.}~\bibnamefont
  {G{\"o}de}}, \bibinfo {author} {\bibfnamefont {E.}~\bibnamefont {Granados}},
  \bibinfo {author} {\bibfnamefont {G.}~\bibnamefont {Gregori}}, \bibinfo
  {author} {\bibfnamefont {H.~J.}\ \bibnamefont {Lee}}, \bibinfo {author}
  {\bibfnamefont {P.}~\bibnamefont {Neumayer}}, \bibinfo {author}
  {\bibfnamefont {W.}~\bibnamefont {Schumaker}}, \bibinfo {author}
  {\bibfnamefont {T.}~\bibnamefont {D{\"o}ppner}}, \bibinfo {author}
  {\bibfnamefont {R.~W.}\ \bibnamefont {Falcone}}, \bibinfo {author}
  {\bibfnamefont {S.~H.}\ \bibnamefont {Glenzer}},\ and\ \bibinfo {author}
  {\bibfnamefont {M.}~\bibnamefont {Roth}},\ }\bibfield  {title} {\bibinfo
  {title} {Nanosecond formation of diamond and lonsdaleite by shock compression
  of graphite},\ }\href@noop {} {\bibfield  {journal} {\bibinfo  {journal}
  {Nature Communications}\ }\textbf {\bibinfo {volume} {7}},\ \bibinfo {pages}
  {10970} (\bibinfo {year} {2016})}\BibitemShut {NoStop}%
\bibitem [{\citenamefont {Miao}\ \emph {et~al.}(2020)\citenamefont {Miao},
  \citenamefont {Sun}, \citenamefont {Zurek},\ and\ \citenamefont
  {Lin}}]{Miao2020-ew}%
  \BibitemOpen
  \bibfield  {author} {\bibinfo {author} {\bibfnamefont {M.}~\bibnamefont
  {Miao}}, \bibinfo {author} {\bibfnamefont {Y.}~\bibnamefont {Sun}}, \bibinfo
  {author} {\bibfnamefont {E.}~\bibnamefont {Zurek}},\ and\ \bibinfo {author}
  {\bibfnamefont {H.}~\bibnamefont {Lin}},\ }\bibfield  {title} {\bibinfo
  {title} {Chemistry under high pressure},\ }\href@noop {} {\bibfield
  {journal} {\bibinfo  {journal} {Nature Reviews Chemistry}\ }\textbf {\bibinfo
  {volume} {4}},\ \bibinfo {pages} {508} (\bibinfo {year} {2020})}\BibitemShut
  {NoStop}%
\bibitem [{\citenamefont {Toleikis}\ \emph {et~al.}(2010)\citenamefont
  {Toleikis}, \citenamefont {Bornath}, \citenamefont {Döppner}, \citenamefont
  {Düsterer}, \citenamefont {Fäustlin}, \citenamefont {Förster},
  \citenamefont {Fortmann}, \citenamefont {Glenzer}, \citenamefont {Göde},
  \citenamefont {Gregori}, \citenamefont {Irsig}, \citenamefont {Laarmann},
  \citenamefont {Lee}, \citenamefont {Li}, \citenamefont {Meiwes-Broer},
  \citenamefont {Mithen}, \citenamefont {Nagler}, \citenamefont {Przystawik},
  \citenamefont {Radcliffe}, \citenamefont {Redlin}, \citenamefont {Redmer},
  \citenamefont {Reinholz}, \citenamefont {Röpke}, \citenamefont {Tavella},
  \citenamefont {Thiele}, \citenamefont {Tiggesbäumker}, \citenamefont
  {Uschmann}, \citenamefont {Vinko}, \citenamefont {Whitcher}, \citenamefont
  {Zastrau}, \citenamefont {Ziaja},\ and\ \citenamefont
  {Tschentscher}}]{Toleikis_2010}%
  \BibitemOpen
  \bibfield  {author} {\bibinfo {author} {\bibfnamefont {S.}~\bibnamefont
  {Toleikis}}, \bibinfo {author} {\bibfnamefont {T.}~\bibnamefont {Bornath}},
  \bibinfo {author} {\bibfnamefont {T.}~\bibnamefont {Döppner}}, \bibinfo
  {author} {\bibfnamefont {S.}~\bibnamefont {Düsterer}}, \bibinfo {author}
  {\bibfnamefont {R.~R.}\ \bibnamefont {Fäustlin}}, \bibinfo {author}
  {\bibfnamefont {E.}~\bibnamefont {Förster}}, \bibinfo {author}
  {\bibfnamefont {C.}~\bibnamefont {Fortmann}}, \bibinfo {author}
  {\bibfnamefont {S.~H.}\ \bibnamefont {Glenzer}}, \bibinfo {author}
  {\bibfnamefont {S.}~\bibnamefont {Göde}}, \bibinfo {author} {\bibfnamefont
  {G.}~\bibnamefont {Gregori}}, \bibinfo {author} {\bibfnamefont
  {R.}~\bibnamefont {Irsig}}, \bibinfo {author} {\bibfnamefont
  {T.}~\bibnamefont {Laarmann}}, \bibinfo {author} {\bibfnamefont {H.~J.}\
  \bibnamefont {Lee}}, \bibinfo {author} {\bibfnamefont {B.}~\bibnamefont
  {Li}}, \bibinfo {author} {\bibfnamefont {K.-H.}\ \bibnamefont
  {Meiwes-Broer}}, \bibinfo {author} {\bibfnamefont {J.}~\bibnamefont
  {Mithen}}, \bibinfo {author} {\bibfnamefont {B.}~\bibnamefont {Nagler}},
  \bibinfo {author} {\bibfnamefont {A.}~\bibnamefont {Przystawik}}, \bibinfo
  {author} {\bibfnamefont {P.}~\bibnamefont {Radcliffe}}, \bibinfo {author}
  {\bibfnamefont {H.}~\bibnamefont {Redlin}}, \bibinfo {author} {\bibfnamefont
  {R.}~\bibnamefont {Redmer}}, \bibinfo {author} {\bibfnamefont
  {H.}~\bibnamefont {Reinholz}}, \bibinfo {author} {\bibfnamefont
  {G.}~\bibnamefont {Röpke}}, \bibinfo {author} {\bibfnamefont
  {F.}~\bibnamefont {Tavella}}, \bibinfo {author} {\bibfnamefont
  {R.}~\bibnamefont {Thiele}}, \bibinfo {author} {\bibfnamefont
  {J.}~\bibnamefont {Tiggesbäumker}}, \bibinfo {author} {\bibfnamefont
  {I.}~\bibnamefont {Uschmann}}, \bibinfo {author} {\bibfnamefont {S.~M.}\
  \bibnamefont {Vinko}}, \bibinfo {author} {\bibfnamefont {T.}~\bibnamefont
  {Whitcher}}, \bibinfo {author} {\bibfnamefont {U.}~\bibnamefont {Zastrau}},
  \bibinfo {author} {\bibfnamefont {B.}~\bibnamefont {Ziaja}},\ and\ \bibinfo
  {author} {\bibfnamefont {T.}~\bibnamefont {Tschentscher}},\ }\bibfield
  {title} {\bibinfo {title} {Probing near-solid density plasmas using soft
  x-ray scattering},\ }\href@noop {} {\bibfield  {journal} {\bibinfo  {journal}
  {Journal of Physics B: Atomic, Molecular and Optical Physics}\ }\textbf
  {\bibinfo {volume} {43}},\ \bibinfo {pages} {194017} (\bibinfo {year}
  {2010})}\BibitemShut {NoStop}%
\bibitem [{\citenamefont {Fletcher}\ \emph {et~al.}(2015)\citenamefont
  {Fletcher}, \citenamefont {Lee}, \citenamefont {D{\"o}ppner}, \citenamefont
  {Galtier}, \citenamefont {Nagler}, \citenamefont {Heimann}, \citenamefont
  {Fortmann}, \citenamefont {LePape}, \citenamefont {Ma}, \citenamefont
  {Millot}, \citenamefont {Pak}, \citenamefont {Turnbull}, \citenamefont
  {Chapman}, \citenamefont {Gericke}, \citenamefont {Vorberger}, \citenamefont
  {White}, \citenamefont {Gregori}, \citenamefont {Wei}, \citenamefont
  {Barbrel}, \citenamefont {Falcone}, \citenamefont {Kao}, \citenamefont
  {Nuhn}, \citenamefont {Welch}, \citenamefont {Zastrau}, \citenamefont
  {Neumayer}, \citenamefont {Hastings},\ and\ \citenamefont
  {Glenzer}}]{Fletcher2015}%
  \BibitemOpen
  \bibfield  {author} {\bibinfo {author} {\bibfnamefont {L.~B.}\ \bibnamefont
  {Fletcher}}, \bibinfo {author} {\bibfnamefont {H.~J.}\ \bibnamefont {Lee}},
  \bibinfo {author} {\bibfnamefont {T.}~\bibnamefont {D{\"o}ppner}}, \bibinfo
  {author} {\bibfnamefont {E.}~\bibnamefont {Galtier}}, \bibinfo {author}
  {\bibfnamefont {B.}~\bibnamefont {Nagler}}, \bibinfo {author} {\bibfnamefont
  {P.}~\bibnamefont {Heimann}}, \bibinfo {author} {\bibfnamefont
  {C.}~\bibnamefont {Fortmann}}, \bibinfo {author} {\bibfnamefont
  {S.}~\bibnamefont {LePape}}, \bibinfo {author} {\bibfnamefont
  {T.}~\bibnamefont {Ma}}, \bibinfo {author} {\bibfnamefont {M.}~\bibnamefont
  {Millot}}, \bibinfo {author} {\bibfnamefont {A.}~\bibnamefont {Pak}},
  \bibinfo {author} {\bibfnamefont {D.}~\bibnamefont {Turnbull}}, \bibinfo
  {author} {\bibfnamefont {D.~A.}\ \bibnamefont {Chapman}}, \bibinfo {author}
  {\bibfnamefont {D.~O.}\ \bibnamefont {Gericke}}, \bibinfo {author}
  {\bibfnamefont {J.}~\bibnamefont {Vorberger}}, \bibinfo {author}
  {\bibfnamefont {T.}~\bibnamefont {White}}, \bibinfo {author} {\bibfnamefont
  {G.}~\bibnamefont {Gregori}}, \bibinfo {author} {\bibfnamefont
  {M.}~\bibnamefont {Wei}}, \bibinfo {author} {\bibfnamefont {B.}~\bibnamefont
  {Barbrel}}, \bibinfo {author} {\bibfnamefont {R.~W.}\ \bibnamefont
  {Falcone}}, \bibinfo {author} {\bibfnamefont {C.~C.}\ \bibnamefont {Kao}},
  \bibinfo {author} {\bibfnamefont {H.}~\bibnamefont {Nuhn}}, \bibinfo {author}
  {\bibfnamefont {J.}~\bibnamefont {Welch}}, \bibinfo {author} {\bibfnamefont
  {U.}~\bibnamefont {Zastrau}}, \bibinfo {author} {\bibfnamefont
  {P.}~\bibnamefont {Neumayer}}, \bibinfo {author} {\bibfnamefont {J.~B.}\
  \bibnamefont {Hastings}},\ and\ \bibinfo {author} {\bibfnamefont {S.~H.}\
  \bibnamefont {Glenzer}},\ }\bibfield  {title} {\bibinfo {title} {Ultrabright
  x-ray laser scattering for dynamic warm dense matter physics},\ }\href@noop
  {} {\bibfield  {journal} {\bibinfo  {journal} {Nature Photonics}\ }\textbf
  {\bibinfo {volume} {9}},\ \bibinfo {pages} {274} (\bibinfo {year}
  {2015})}\BibitemShut {NoStop}%
\bibitem [{\citenamefont {Ciricosta}\ \emph {et~al.}(2012)\citenamefont
  {Ciricosta}, \citenamefont {Vinko}, \citenamefont {Chung}, \citenamefont
  {Cho}, \citenamefont {Brown}, \citenamefont {Burian}, \citenamefont
  {Chalupsk\'y}, \citenamefont {Engelhorn}, \citenamefont {Falcone},
  \citenamefont {Graves}, \citenamefont {H\'ajkov\'a}, \citenamefont
  {Higginbotham}, \citenamefont {Juha}, \citenamefont {Krzywinski},
  \citenamefont {Lee}, \citenamefont {Messerschmidt}, \citenamefont {Murphy},
  \citenamefont {Ping}, \citenamefont {Rackstraw}, \citenamefont {Scherz},
  \citenamefont {Schlotter}, \citenamefont {Toleikis}, \citenamefont {Turner},
  \citenamefont {Vysin}, \citenamefont {Wang}, \citenamefont {Wu},
  \citenamefont {Zastrau}, \citenamefont {Zhu}, \citenamefont {Lee},
  \citenamefont {Heimann}, \citenamefont {Nagler},\ and\ \citenamefont
  {Wark}}]{PhysRevLett.109.065002}%
  \BibitemOpen
  \bibfield  {author} {\bibinfo {author} {\bibfnamefont {O.}~\bibnamefont
  {Ciricosta}}, \bibinfo {author} {\bibfnamefont {S.~M.}\ \bibnamefont
  {Vinko}}, \bibinfo {author} {\bibfnamefont {H.-K.}\ \bibnamefont {Chung}},
  \bibinfo {author} {\bibfnamefont {B.-I.}\ \bibnamefont {Cho}}, \bibinfo
  {author} {\bibfnamefont {C.~R.~D.}\ \bibnamefont {Brown}}, \bibinfo {author}
  {\bibfnamefont {T.}~\bibnamefont {Burian}}, \bibinfo {author} {\bibfnamefont
  {J.}~\bibnamefont {Chalupsk\'y}}, \bibinfo {author} {\bibfnamefont
  {K.}~\bibnamefont {Engelhorn}}, \bibinfo {author} {\bibfnamefont {R.~W.}\
  \bibnamefont {Falcone}}, \bibinfo {author} {\bibfnamefont {C.}~\bibnamefont
  {Graves}}, \bibinfo {author} {\bibfnamefont {V.}~\bibnamefont {H\'ajkov\'a}},
  \bibinfo {author} {\bibfnamefont {A.}~\bibnamefont {Higginbotham}}, \bibinfo
  {author} {\bibfnamefont {L.}~\bibnamefont {Juha}}, \bibinfo {author}
  {\bibfnamefont {J.}~\bibnamefont {Krzywinski}}, \bibinfo {author}
  {\bibfnamefont {H.~J.}\ \bibnamefont {Lee}}, \bibinfo {author} {\bibfnamefont
  {M.}~\bibnamefont {Messerschmidt}}, \bibinfo {author} {\bibfnamefont {C.~D.}\
  \bibnamefont {Murphy}}, \bibinfo {author} {\bibfnamefont {Y.}~\bibnamefont
  {Ping}}, \bibinfo {author} {\bibfnamefont {D.~S.}\ \bibnamefont {Rackstraw}},
  \bibinfo {author} {\bibfnamefont {A.}~\bibnamefont {Scherz}}, \bibinfo
  {author} {\bibfnamefont {W.}~\bibnamefont {Schlotter}}, \bibinfo {author}
  {\bibfnamefont {S.}~\bibnamefont {Toleikis}}, \bibinfo {author}
  {\bibfnamefont {J.~J.}\ \bibnamefont {Turner}}, \bibinfo {author}
  {\bibfnamefont {L.}~\bibnamefont {Vysin}}, \bibinfo {author} {\bibfnamefont
  {T.}~\bibnamefont {Wang}}, \bibinfo {author} {\bibfnamefont {B.}~\bibnamefont
  {Wu}}, \bibinfo {author} {\bibfnamefont {U.}~\bibnamefont {Zastrau}},
  \bibinfo {author} {\bibfnamefont {D.}~\bibnamefont {Zhu}}, \bibinfo {author}
  {\bibfnamefont {R.~W.}\ \bibnamefont {Lee}}, \bibinfo {author} {\bibfnamefont
  {P.}~\bibnamefont {Heimann}}, \bibinfo {author} {\bibfnamefont
  {B.}~\bibnamefont {Nagler}},\ and\ \bibinfo {author} {\bibfnamefont {J.~S.}\
  \bibnamefont {Wark}},\ }\bibfield  {title} {\bibinfo {title} {Direct
  measurements of the ionization potential depression in a dense plasma},\
  }\href@noop {} {\bibfield  {journal} {\bibinfo  {journal} {Phys. Rev. Lett.}\
  }\textbf {\bibinfo {volume} {109}},\ \bibinfo {pages} {065002} (\bibinfo
  {year} {2012})}\BibitemShut {NoStop}%
\bibitem [{\citenamefont {Gawne}\ \emph {et~al.}(2023)\citenamefont {Gawne},
  \citenamefont {Campbell}, \citenamefont {Forte}, \citenamefont {Hollebon},
  \citenamefont {Perez-Callejo}, \citenamefont {Humphries}, \citenamefont
  {Karnbach}, \citenamefont {Kasim}, \citenamefont {Preston}, \citenamefont
  {Lee}, \citenamefont {Miscampbell}, \citenamefont {van~den Berg},
  \citenamefont {Nagler}, \citenamefont {Ren}, \citenamefont {Royle},
  \citenamefont {Wark},\ and\ \citenamefont {Vinko}}]{Gawne_PRE_2023}%
  \BibitemOpen
  \bibfield  {author} {\bibinfo {author} {\bibfnamefont {T.}~\bibnamefont
  {Gawne}}, \bibinfo {author} {\bibfnamefont {T.}~\bibnamefont {Campbell}},
  \bibinfo {author} {\bibfnamefont {A.}~\bibnamefont {Forte}}, \bibinfo
  {author} {\bibfnamefont {P.}~\bibnamefont {Hollebon}}, \bibinfo {author}
  {\bibfnamefont {G.}~\bibnamefont {Perez-Callejo}}, \bibinfo {author}
  {\bibfnamefont {O.~S.}\ \bibnamefont {Humphries}}, \bibinfo {author}
  {\bibfnamefont {O.}~\bibnamefont {Karnbach}}, \bibinfo {author}
  {\bibfnamefont {M.~F.}\ \bibnamefont {Kasim}}, \bibinfo {author}
  {\bibfnamefont {T.~R.}\ \bibnamefont {Preston}}, \bibinfo {author}
  {\bibfnamefont {H.~J.}\ \bibnamefont {Lee}}, \bibinfo {author} {\bibfnamefont
  {A.}~\bibnamefont {Miscampbell}}, \bibinfo {author} {\bibfnamefont {Q.~Y.}\
  \bibnamefont {van~den Berg}}, \bibinfo {author} {\bibfnamefont
  {B.}~\bibnamefont {Nagler}}, \bibinfo {author} {\bibfnamefont
  {S.}~\bibnamefont {Ren}}, \bibinfo {author} {\bibfnamefont {R.~B.}\
  \bibnamefont {Royle}}, \bibinfo {author} {\bibfnamefont {J.~S.}\ \bibnamefont
  {Wark}},\ and\ \bibinfo {author} {\bibfnamefont {S.~M.}\ \bibnamefont
  {Vinko}},\ }\bibfield  {title} {\bibinfo {title} {Investigating mechanisms of
  state localization in highly ionized dense plasmas},\ }\href
  {https://doi.org/10.1103/PhysRevE.108.035210} {\bibfield  {journal} {\bibinfo
   {journal} {Phys. Rev. E}\ }\textbf {\bibinfo {volume} {108}},\ \bibinfo
  {pages} {035210} (\bibinfo {year} {2023})}\BibitemShut {NoStop}%
\bibitem [{\citenamefont {Hollebon}\ \emph {et~al.}(2019)\citenamefont
  {Hollebon}, \citenamefont {Ciricosta}, \citenamefont {Desjarlais},
  \citenamefont {Cacho}, \citenamefont {Spindloe}, \citenamefont {Springate},
  \citenamefont {Turcu}, \citenamefont {Wark},\ and\ \citenamefont
  {Vinko}}]{PhysRevE.100.043207}%
  \BibitemOpen
  \bibfield  {author} {\bibinfo {author} {\bibfnamefont {P.}~\bibnamefont
  {Hollebon}}, \bibinfo {author} {\bibfnamefont {O.}~\bibnamefont {Ciricosta}},
  \bibinfo {author} {\bibfnamefont {M.~P.}\ \bibnamefont {Desjarlais}},
  \bibinfo {author} {\bibfnamefont {C.}~\bibnamefont {Cacho}}, \bibinfo
  {author} {\bibfnamefont {C.}~\bibnamefont {Spindloe}}, \bibinfo {author}
  {\bibfnamefont {E.}~\bibnamefont {Springate}}, \bibinfo {author}
  {\bibfnamefont {I.~C.~E.}\ \bibnamefont {Turcu}}, \bibinfo {author}
  {\bibfnamefont {J.~S.}\ \bibnamefont {Wark}},\ and\ \bibinfo {author}
  {\bibfnamefont {S.~M.}\ \bibnamefont {Vinko}},\ }\bibfield  {title} {\bibinfo
  {title} {Ab initio simulations and measurements of the free-free opacity in
  aluminum},\ }\href@noop {} {\bibfield  {journal} {\bibinfo  {journal} {Phys.
  Rev. E}\ }\textbf {\bibinfo {volume} {100}},\ \bibinfo {pages} {043207}
  (\bibinfo {year} {2019})}\BibitemShut {NoStop}%
\bibitem [{\citenamefont {Preston}\ \emph {et~al.}(2017)\citenamefont
  {Preston}, \citenamefont {Vinko}, \citenamefont {Ciricosta}, \citenamefont
  {Hollebon}, \citenamefont {Chung}, \citenamefont {Dakovski}, \citenamefont
  {Krzywinski}, \citenamefont {Minitti}, \citenamefont {Burian}, \citenamefont
  {Chalupsk\'y}, \citenamefont {H\'ajkov\'a}, \citenamefont {Juha},
  \citenamefont {Vozda}, \citenamefont {Zastrau}, \citenamefont {Lee},\ and\
  \citenamefont {Wark}}]{PhysRevLett.119.085001}%
  \BibitemOpen
  \bibfield  {author} {\bibinfo {author} {\bibfnamefont {T.~R.}\ \bibnamefont
  {Preston}}, \bibinfo {author} {\bibfnamefont {S.~M.}\ \bibnamefont {Vinko}},
  \bibinfo {author} {\bibfnamefont {O.}~\bibnamefont {Ciricosta}}, \bibinfo
  {author} {\bibfnamefont {P.}~\bibnamefont {Hollebon}}, \bibinfo {author}
  {\bibfnamefont {H.-K.}\ \bibnamefont {Chung}}, \bibinfo {author}
  {\bibfnamefont {G.~L.}\ \bibnamefont {Dakovski}}, \bibinfo {author}
  {\bibfnamefont {J.}~\bibnamefont {Krzywinski}}, \bibinfo {author}
  {\bibfnamefont {M.}~\bibnamefont {Minitti}}, \bibinfo {author} {\bibfnamefont
  {T.}~\bibnamefont {Burian}}, \bibinfo {author} {\bibfnamefont
  {J.}~\bibnamefont {Chalupsk\'y}}, \bibinfo {author} {\bibfnamefont
  {V.}~\bibnamefont {H\'ajkov\'a}}, \bibinfo {author} {\bibfnamefont
  {L.}~\bibnamefont {Juha}}, \bibinfo {author} {\bibfnamefont {V.}~\bibnamefont
  {Vozda}}, \bibinfo {author} {\bibfnamefont {U.}~\bibnamefont {Zastrau}},
  \bibinfo {author} {\bibfnamefont {R.~W.}\ \bibnamefont {Lee}},\ and\ \bibinfo
  {author} {\bibfnamefont {J.~S.}\ \bibnamefont {Wark}},\ }\bibfield  {title}
  {\bibinfo {title} {Measurements of the $k$-shell opacity of a solid-density
  magnesium plasma heated by an x-ray free-electron laser},\ }\href@noop {}
  {\bibfield  {journal} {\bibinfo  {journal} {Phys. Rev. Lett.}\ }\textbf
  {\bibinfo {volume} {119}},\ \bibinfo {pages} {085001} (\bibinfo {year}
  {2017})}\BibitemShut {NoStop}%
\bibitem [{\citenamefont {Vinko}\ \emph {et~al.}(2020)\citenamefont {Vinko},
  \citenamefont {Vozda}, \citenamefont {Andreasson}, \citenamefont {Bajt},
  \citenamefont {Bielecki}, \citenamefont {Burian}, \citenamefont {Chalupsky},
  \citenamefont {Ciricosta}, \citenamefont {Desjarlais}, \citenamefont
  {Fleckenstein}, \citenamefont {Hajdu}, \citenamefont {Hajkova}, \citenamefont
  {Hollebon}, \citenamefont {Juha}, \citenamefont {Kasim}, \citenamefont
  {McBride}, \citenamefont {Muehlig}, \citenamefont {Preston}, \citenamefont
  {Rackstraw}, \citenamefont {Roling}, \citenamefont {Toleikis}, \citenamefont
  {Wark},\ and\ \citenamefont {Zacharias}}]{PhysRevLett.124.225002}%
  \BibitemOpen
  \bibfield  {author} {\bibinfo {author} {\bibfnamefont {S.~M.}\ \bibnamefont
  {Vinko}}, \bibinfo {author} {\bibfnamefont {V.}~\bibnamefont {Vozda}},
  \bibinfo {author} {\bibfnamefont {J.}~\bibnamefont {Andreasson}}, \bibinfo
  {author} {\bibfnamefont {S.}~\bibnamefont {Bajt}}, \bibinfo {author}
  {\bibfnamefont {J.}~\bibnamefont {Bielecki}}, \bibinfo {author}
  {\bibfnamefont {T.}~\bibnamefont {Burian}}, \bibinfo {author} {\bibfnamefont
  {J.}~\bibnamefont {Chalupsky}}, \bibinfo {author} {\bibfnamefont
  {O.}~\bibnamefont {Ciricosta}}, \bibinfo {author} {\bibfnamefont {M.~P.}\
  \bibnamefont {Desjarlais}}, \bibinfo {author} {\bibfnamefont
  {H.}~\bibnamefont {Fleckenstein}}, \bibinfo {author} {\bibfnamefont
  {J.}~\bibnamefont {Hajdu}}, \bibinfo {author} {\bibfnamefont
  {V.}~\bibnamefont {Hajkova}}, \bibinfo {author} {\bibfnamefont
  {P.}~\bibnamefont {Hollebon}}, \bibinfo {author} {\bibfnamefont
  {L.}~\bibnamefont {Juha}}, \bibinfo {author} {\bibfnamefont {M.~F.}\
  \bibnamefont {Kasim}}, \bibinfo {author} {\bibfnamefont {E.~E.}\ \bibnamefont
  {McBride}}, \bibinfo {author} {\bibfnamefont {K.}~\bibnamefont {Muehlig}},
  \bibinfo {author} {\bibfnamefont {T.~R.}\ \bibnamefont {Preston}}, \bibinfo
  {author} {\bibfnamefont {D.~S.}\ \bibnamefont {Rackstraw}}, \bibinfo {author}
  {\bibfnamefont {S.}~\bibnamefont {Roling}}, \bibinfo {author} {\bibfnamefont
  {S.}~\bibnamefont {Toleikis}}, \bibinfo {author} {\bibfnamefont {J.~S.}\
  \bibnamefont {Wark}},\ and\ \bibinfo {author} {\bibfnamefont
  {H.}~\bibnamefont {Zacharias}},\ }\bibfield  {title} {\bibinfo {title}
  {Time-resolved xuv opacity measurements of warm dense aluminum},\ }\href@noop
  {} {\bibfield  {journal} {\bibinfo  {journal} {Phys. Rev. Lett.}\ }\textbf
  {\bibinfo {volume} {124}},\ \bibinfo {pages} {225002} (\bibinfo {year}
  {2020})}\BibitemShut {NoStop}%
\bibitem [{\citenamefont {Vinko}\ \emph {et~al.}(2015)\citenamefont {Vinko},
  \citenamefont {Ciricosta}, \citenamefont {Preston}, \citenamefont
  {Rackstraw}, \citenamefont {Brown}, \citenamefont {Burian}, \citenamefont
  {Chalupsk{\'y}}, \citenamefont {Cho}, \citenamefont {Chung}, \citenamefont
  {Engelhorn}, \citenamefont {Falcone}, \citenamefont {Fiokovinini},
  \citenamefont {H{\'a}jkov{\'a}}, \citenamefont {Heimann}, \citenamefont
  {Juha}, \citenamefont {Lee}, \citenamefont {Lee}, \citenamefont
  {Messerschmidt}, \citenamefont {Nagler}, \citenamefont {Schlotter},
  \citenamefont {Turner}, \citenamefont {Vysin}, \citenamefont {Zastrau},\ and\
  \citenamefont {Wark}}]{NatCommun-colls}%
  \BibitemOpen
  \bibfield  {author} {\bibinfo {author} {\bibfnamefont {S.~M.}\ \bibnamefont
  {Vinko}}, \bibinfo {author} {\bibfnamefont {O.}~\bibnamefont {Ciricosta}},
  \bibinfo {author} {\bibfnamefont {T.~R.}\ \bibnamefont {Preston}}, \bibinfo
  {author} {\bibfnamefont {D.~S.}\ \bibnamefont {Rackstraw}}, \bibinfo {author}
  {\bibfnamefont {C.~R.~D.}\ \bibnamefont {Brown}}, \bibinfo {author}
  {\bibfnamefont {T.}~\bibnamefont {Burian}}, \bibinfo {author} {\bibfnamefont
  {J.}~\bibnamefont {Chalupsk{\'y}}}, \bibinfo {author} {\bibfnamefont {B.~I.}\
  \bibnamefont {Cho}}, \bibinfo {author} {\bibfnamefont {H.~K.}\ \bibnamefont
  {Chung}}, \bibinfo {author} {\bibfnamefont {K.}~\bibnamefont {Engelhorn}},
  \bibinfo {author} {\bibfnamefont {R.~W.}\ \bibnamefont {Falcone}}, \bibinfo
  {author} {\bibfnamefont {R.}~\bibnamefont {Fiokovinini}}, \bibinfo {author}
  {\bibfnamefont {V.}~\bibnamefont {H{\'a}jkov{\'a}}}, \bibinfo {author}
  {\bibfnamefont {P.~A.}\ \bibnamefont {Heimann}}, \bibinfo {author}
  {\bibfnamefont {L.}~\bibnamefont {Juha}}, \bibinfo {author} {\bibfnamefont
  {H.~J.}\ \bibnamefont {Lee}}, \bibinfo {author} {\bibfnamefont {R.~W.}\
  \bibnamefont {Lee}}, \bibinfo {author} {\bibfnamefont {M.}~\bibnamefont
  {Messerschmidt}}, \bibinfo {author} {\bibfnamefont {B.}~\bibnamefont
  {Nagler}}, \bibinfo {author} {\bibfnamefont {W.}~\bibnamefont {Schlotter}},
  \bibinfo {author} {\bibfnamefont {J.~J.}\ \bibnamefont {Turner}}, \bibinfo
  {author} {\bibfnamefont {L.}~\bibnamefont {Vysin}}, \bibinfo {author}
  {\bibfnamefont {U.}~\bibnamefont {Zastrau}},\ and\ \bibinfo {author}
  {\bibfnamefont {J.~S.}\ \bibnamefont {Wark}},\ }\bibfield  {title} {\bibinfo
  {title} {Investigation of femtosecond collisional ionization rates in a
  solid-density aluminium plasma},\ }\href@noop {} {\bibfield  {journal}
  {\bibinfo  {journal} {Nature Communications}\ }\textbf {\bibinfo {volume}
  {6}},\ \bibinfo {pages} {6397} (\bibinfo {year} {2015})}\BibitemShut
  {NoStop}%
\bibitem [{\citenamefont {van~den Berg}\ \emph {et~al.}(2018)\citenamefont
  {van~den Berg}, \citenamefont {Fernandez-Tello}, \citenamefont {Burian},
  \citenamefont {Chalupsk\'y}, \citenamefont {Chung}, \citenamefont
  {Ciricosta}, \citenamefont {Dakovski}, \citenamefont {H\'ajkov\'a},
  \citenamefont {Hollebon}, \citenamefont {Juha}, \citenamefont {Krzywinski},
  \citenamefont {Lee}, \citenamefont {Minitti}, \citenamefont {Preston},
  \citenamefont {de~la Varga}, \citenamefont {Vozda}, \citenamefont {Zastrau},
  \citenamefont {Wark}, \citenamefont {Velarde},\ and\ \citenamefont
  {Vinko}}]{PhysRevLett.120.055002}%
  \BibitemOpen
  \bibfield  {author} {\bibinfo {author} {\bibfnamefont {Q.~Y.}\ \bibnamefont
  {van~den Berg}}, \bibinfo {author} {\bibfnamefont {E.~V.}\ \bibnamefont
  {Fernandez-Tello}}, \bibinfo {author} {\bibfnamefont {T.}~\bibnamefont
  {Burian}}, \bibinfo {author} {\bibfnamefont {J.}~\bibnamefont {Chalupsk\'y}},
  \bibinfo {author} {\bibfnamefont {H.-K.}\ \bibnamefont {Chung}}, \bibinfo
  {author} {\bibfnamefont {O.}~\bibnamefont {Ciricosta}}, \bibinfo {author}
  {\bibfnamefont {G.~L.}\ \bibnamefont {Dakovski}}, \bibinfo {author}
  {\bibfnamefont {V.}~\bibnamefont {H\'ajkov\'a}}, \bibinfo {author}
  {\bibfnamefont {P.}~\bibnamefont {Hollebon}}, \bibinfo {author}
  {\bibfnamefont {L.}~\bibnamefont {Juha}}, \bibinfo {author} {\bibfnamefont
  {J.}~\bibnamefont {Krzywinski}}, \bibinfo {author} {\bibfnamefont {R.~W.}\
  \bibnamefont {Lee}}, \bibinfo {author} {\bibfnamefont {M.~P.}\ \bibnamefont
  {Minitti}}, \bibinfo {author} {\bibfnamefont {T.~R.}\ \bibnamefont
  {Preston}}, \bibinfo {author} {\bibfnamefont {A.~G.}\ \bibnamefont {de~la
  Varga}}, \bibinfo {author} {\bibfnamefont {V.}~\bibnamefont {Vozda}},
  \bibinfo {author} {\bibfnamefont {U.}~\bibnamefont {Zastrau}}, \bibinfo
  {author} {\bibfnamefont {J.~S.}\ \bibnamefont {Wark}}, \bibinfo {author}
  {\bibfnamefont {P.}~\bibnamefont {Velarde}},\ and\ \bibinfo {author}
  {\bibfnamefont {S.~M.}\ \bibnamefont {Vinko}},\ }\bibfield  {title} {\bibinfo
  {title} {Clocking femtosecond collisional dynamics via resonant x-ray
  spectroscopy},\ }\href@noop {} {\bibfield  {journal} {\bibinfo  {journal}
  {Phys. Rev. Lett.}\ }\textbf {\bibinfo {volume} {120}},\ \bibinfo {pages}
  {055002} (\bibinfo {year} {2018})}\BibitemShut {NoStop}%
\bibitem [{\citenamefont {Vinko}\ \emph {et~al.}(2012)\citenamefont {Vinko},
  \citenamefont {Ciricosta}, \citenamefont {Cho}, \citenamefont {Engelhorn},
  \citenamefont {Chung}, \citenamefont {Brown}, \citenamefont {Burian},
  \citenamefont {Chalupsk{\`y}}, \citenamefont {Falcone}, \citenamefont
  {Graves} \emph {et~al.}}]{vinko2012creation}%
  \BibitemOpen
  \bibfield  {author} {\bibinfo {author} {\bibfnamefont {S.}~\bibnamefont
  {Vinko}}, \bibinfo {author} {\bibfnamefont {O.}~\bibnamefont {Ciricosta}},
  \bibinfo {author} {\bibfnamefont {B.}~\bibnamefont {Cho}}, \bibinfo {author}
  {\bibfnamefont {K.}~\bibnamefont {Engelhorn}}, \bibinfo {author}
  {\bibfnamefont {H.-K.}\ \bibnamefont {Chung}}, \bibinfo {author}
  {\bibfnamefont {C.}~\bibnamefont {Brown}}, \bibinfo {author} {\bibfnamefont
  {T.}~\bibnamefont {Burian}}, \bibinfo {author} {\bibfnamefont
  {J.}~\bibnamefont {Chalupsk{\`y}}}, \bibinfo {author} {\bibfnamefont
  {R.}~\bibnamefont {Falcone}}, \bibinfo {author} {\bibfnamefont
  {C.}~\bibnamefont {Graves}}, \emph {et~al.},\ }\bibfield  {title} {\bibinfo
  {title} {Creation and diagnosis of a solid-density plasma with an x-ray
  free-electron laser},\ }\href@noop {} {\bibfield  {journal} {\bibinfo
  {journal} {Nature}\ }\textbf {\bibinfo {volume} {482}},\ \bibinfo {pages}
  {59} (\bibinfo {year} {2012})}\BibitemShut {NoStop}%
\bibitem [{ins()}]{instruments}%
  \BibitemOpen
  \href@noop {} {\bibinfo {title} {Excellence in x-ray detection: Advanced
  x-ray cameras for scientific and industrial applications.}},\ \Eprint
  {https://arxiv.org/abs/Princeton Instruments} {Princeton Instruments}
  \BibitemShut {NoStop}%
\bibitem [{\citenamefont {Lewenkopf}\ \emph {et~al.}(2008)\citenamefont
  {Lewenkopf}, \citenamefont {Mucciolo},\ and\ \citenamefont
  {Castro~Neto}}]{lewenkopf2008numerical}%
  \BibitemOpen
  \bibfield  {author} {\bibinfo {author} {\bibfnamefont {C.~H.}\ \bibnamefont
  {Lewenkopf}}, \bibinfo {author} {\bibfnamefont {E.~R.}\ \bibnamefont
  {Mucciolo}},\ and\ \bibinfo {author} {\bibfnamefont {A.}~\bibnamefont
  {Castro~Neto}},\ }\bibfield  {title} {\bibinfo {title} {Numerical studies of
  conductivity and fano factor in disordered graphene},\ }\href@noop {}
  {\bibfield  {journal} {\bibinfo  {journal} {Physical Review B—Condensed
  Matter and Materials Physics}\ }\textbf {\bibinfo {volume} {77}},\ \bibinfo
  {pages} {081410} (\bibinfo {year} {2008})}\BibitemShut {NoStop}%
\bibitem [{\citenamefont {Newberry}(1991)}]{newberry1991signal}%
  \BibitemOpen
  \bibfield  {author} {\bibinfo {author} {\bibfnamefont {M.~V.}\ \bibnamefont
  {Newberry}},\ }\bibfield  {title} {\bibinfo {title} {Signal-to-noise
  considerations for sky-subtracted ccd data},\ }\href@noop {} {\bibfield
  {journal} {\bibinfo  {journal} {Publications of the Astronomical Society of
  the Pacific}\ }\textbf {\bibinfo {volume} {103}},\ \bibinfo {pages} {122}
  (\bibinfo {year} {1991})}\BibitemShut {NoStop}%
\bibitem [{\citenamefont {Du}\ and\ \citenamefont
  {Voss}(2004)}]{du2004effects}%
  \BibitemOpen
  \bibfield  {author} {\bibinfo {author} {\bibfnamefont {H.}~\bibnamefont
  {Du}}\ and\ \bibinfo {author} {\bibfnamefont {K.~J.}\ \bibnamefont {Voss}},\
  }\bibfield  {title} {\bibinfo {title} {Effects of point-spread function on
  calibration and radiometric accuracy of ccd camera},\ }\href@noop {}
  {\bibfield  {journal} {\bibinfo  {journal} {Applied optics}\ }\textbf
  {\bibinfo {volume} {43}},\ \bibinfo {pages} {665} (\bibinfo {year}
  {2004})}\BibitemShut {NoStop}%
\bibitem [{\citenamefont {Abboud}\ \emph {et~al.}(2013)\citenamefont {Abboud},
  \citenamefont {Send}, \citenamefont {Pashniak}, \citenamefont {Leitenberger},
  \citenamefont {Ihle}, \citenamefont {Huth}, \citenamefont {Hartmann},
  \citenamefont {Str{\"u}der},\ and\ \citenamefont {Pietsch}}]{abboud2013sub}%
  \BibitemOpen
  \bibfield  {author} {\bibinfo {author} {\bibfnamefont {A.}~\bibnamefont
  {Abboud}}, \bibinfo {author} {\bibfnamefont {S.}~\bibnamefont {Send}},
  \bibinfo {author} {\bibfnamefont {N.}~\bibnamefont {Pashniak}}, \bibinfo
  {author} {\bibfnamefont {W.}~\bibnamefont {Leitenberger}}, \bibinfo {author}
  {\bibfnamefont {S.}~\bibnamefont {Ihle}}, \bibinfo {author} {\bibfnamefont
  {M.}~\bibnamefont {Huth}}, \bibinfo {author} {\bibfnamefont {R.}~\bibnamefont
  {Hartmann}}, \bibinfo {author} {\bibfnamefont {L.}~\bibnamefont
  {Str{\"u}der}},\ and\ \bibinfo {author} {\bibfnamefont {U.}~\bibnamefont
  {Pietsch}},\ }\bibfield  {title} {\bibinfo {title} {Sub-pixel resolution of a
  pnccd for x-ray white beam applications},\ }\href@noop {} {\bibfield
  {journal} {\bibinfo  {journal} {Journal of Instrumentation}\ }\textbf
  {\bibinfo {volume} {8}}\bibinfo  {number} { (05)},\ \bibinfo {pages}
  {P05005}}\BibitemShut {NoStop}%
\bibitem [{\citenamefont {Hagino}\ \emph {et~al.}(2019)\citenamefont {Hagino},
  \citenamefont {Oono}, \citenamefont {Negishi}, \citenamefont {Yarita},
  \citenamefont {Kohmura}, \citenamefont {Tsuru}, \citenamefont {Tanaka},
  \citenamefont {Uchida}, \citenamefont {Harada}, \citenamefont {Okuno} \emph
  {et~al.}}]{hagino2019measurement}%
  \BibitemOpen
\bibfield  {number} {  }\bibfield  {author} {\bibinfo {author} {\bibfnamefont
  {K.}~\bibnamefont {Hagino}}, \bibinfo {author} {\bibfnamefont
  {K.}~\bibnamefont {Oono}}, \bibinfo {author} {\bibfnamefont {K.}~\bibnamefont
  {Negishi}}, \bibinfo {author} {\bibfnamefont {K.}~\bibnamefont {Yarita}},
  \bibinfo {author} {\bibfnamefont {T.}~\bibnamefont {Kohmura}}, \bibinfo
  {author} {\bibfnamefont {T.~G.}\ \bibnamefont {Tsuru}}, \bibinfo {author}
  {\bibfnamefont {T.}~\bibnamefont {Tanaka}}, \bibinfo {author} {\bibfnamefont
  {H.}~\bibnamefont {Uchida}}, \bibinfo {author} {\bibfnamefont
  {S.}~\bibnamefont {Harada}}, \bibinfo {author} {\bibfnamefont
  {T.}~\bibnamefont {Okuno}}, \emph {et~al.},\ }\bibfield  {title} {\bibinfo
  {title} {Measurement of charge cloud size in x-ray soi pixel sensors},\
  }\href@noop {} {\bibfield  {journal} {\bibinfo  {journal} {IEEE Transactions
  on Nuclear Science}\ }\textbf {\bibinfo {volume} {66}},\ \bibinfo {pages}
  {1897} (\bibinfo {year} {2019})}\BibitemShut {NoStop}%
\bibitem [{\citenamefont {Gatti}\ \emph {et~al.}(1987)\citenamefont {Gatti},
  \citenamefont {Longoni}, \citenamefont {Rehak},\ and\ \citenamefont
  {Sampietro}}]{gatti1987dynamics}%
  \BibitemOpen
  \bibfield  {author} {\bibinfo {author} {\bibfnamefont {E.}~\bibnamefont
  {Gatti}}, \bibinfo {author} {\bibfnamefont {A.}~\bibnamefont {Longoni}},
  \bibinfo {author} {\bibfnamefont {P.}~\bibnamefont {Rehak}},\ and\ \bibinfo
  {author} {\bibfnamefont {M.}~\bibnamefont {Sampietro}},\ }\bibfield  {title}
  {\bibinfo {title} {Dynamics of electrons in drift detectors},\ }\href@noop {}
  {\bibfield  {journal} {\bibinfo  {journal} {Nuclear Instruments and Methods
  in Physics Research Section A: Accelerators, Spectrometers, Detectors and
  Associated Equipment}\ }\textbf {\bibinfo {volume} {253}},\ \bibinfo {pages}
  {393} (\bibinfo {year} {1987})}\BibitemShut {NoStop}%
\bibitem [{\citenamefont {Mullikin}\ \emph {et~al.}(1994)\citenamefont
  {Mullikin}, \citenamefont {van Vliet}, \citenamefont {Netten}, \citenamefont
  {Boddeke}, \citenamefont {Van~der Feltz},\ and\ \citenamefont
  {Young}}]{mullikin1994methods}%
  \BibitemOpen
  \bibfield  {author} {\bibinfo {author} {\bibfnamefont {J.~C.}\ \bibnamefont
  {Mullikin}}, \bibinfo {author} {\bibfnamefont {L.~J.}\ \bibnamefont {van
  Vliet}}, \bibinfo {author} {\bibfnamefont {H.}~\bibnamefont {Netten}},
  \bibinfo {author} {\bibfnamefont {F.~R.}\ \bibnamefont {Boddeke}}, \bibinfo
  {author} {\bibfnamefont {G.}~\bibnamefont {Van~der Feltz}},\ and\ \bibinfo
  {author} {\bibfnamefont {I.~T.}\ \bibnamefont {Young}},\ }\bibfield  {title}
  {\bibinfo {title} {Methods for ccd camera characterization},\ }in\ \href@noop
  {} {\emph {\bibinfo {booktitle} {Image Acquisition and Scientific Imaging
  Systems}}},\ Vol.\ \bibinfo {volume} {2173}\ (\bibinfo {organization}
  {Spie},\ \bibinfo {year} {1994})\ pp.\ \bibinfo {pages} {73--84}\BibitemShut
  {NoStop}%
\bibitem [{\citenamefont {Shor}\ \emph {et~al.}(2004)\citenamefont {Shor},
  \citenamefont {Eisen},\ and\ \citenamefont {Mardor}}]{shor2004edge}%
  \BibitemOpen
  \bibfield  {author} {\bibinfo {author} {\bibfnamefont {A.}~\bibnamefont
  {Shor}}, \bibinfo {author} {\bibfnamefont {Y.}~\bibnamefont {Eisen}},\ and\
  \bibinfo {author} {\bibfnamefont {I.}~\bibnamefont {Mardor}},\ }\bibfield
  {title} {\bibinfo {title} {Edge effects in pixelated cdznte gamma
  detectors},\ }\href@noop {} {\bibfield  {journal} {\bibinfo  {journal} {IEEE
  Transactions on Nuclear Science}\ }\textbf {\bibinfo {volume} {51}},\
  \bibinfo {pages} {2412} (\bibinfo {year} {2004})}\BibitemShut {NoStop}%
\bibitem [{\citenamefont {Morton}(1968)}]{morton1968photon}%
  \BibitemOpen
  \bibfield  {author} {\bibinfo {author} {\bibfnamefont {G.}~\bibnamefont
  {Morton}},\ }\bibfield  {title} {\bibinfo {title} {Photon counting},\
  }\href@noop {} {\bibfield  {journal} {\bibinfo  {journal} {Applied Optics}\
  }\textbf {\bibinfo {volume} {7}},\ \bibinfo {pages} {1} (\bibinfo {year}
  {1968})}\BibitemShut {NoStop}%
\bibitem [{\citenamefont {Zachariasen}(1967)}]{zachariasen1967general}%
  \BibitemOpen
  \bibfield  {author} {\bibinfo {author} {\bibfnamefont {W.}~\bibnamefont
  {Zachariasen}},\ }\bibfield  {title} {\bibinfo {title} {A general theory of
  x-ray diffraction in crystals},\ }\href@noop {} {\bibfield  {journal}
  {\bibinfo  {journal} {Acta Crystallographica}\ }\textbf {\bibinfo {volume}
  {23}},\ \bibinfo {pages} {558} (\bibinfo {year} {1967})}\BibitemShut
  {NoStop}%
\bibitem [{\citenamefont {Humphries}(2020)}]{humphries2020isochoric}%
  \BibitemOpen
  \bibfield  {author} {\bibinfo {author} {\bibfnamefont {O.}~\bibnamefont
  {Humphries}},\ }\emph {\bibinfo {title} {Isochoric generation and
  spectroscopic diagnosis of high energy-density systems}},\ \href@noop {}
  {Ph.D. thesis},\ \bibinfo  {school} {University of Oxford} (\bibinfo {year}
  {2020})\BibitemShut {NoStop}%
\bibitem [{\citenamefont {Goldstein}(1980)}]{goldstein:mechanics}%
  \BibitemOpen
  \bibfield  {author} {\bibinfo {author} {\bibfnamefont {H.}~\bibnamefont
  {Goldstein}},\ }\href@noop {} {\emph {\bibinfo {title} {Classical
  Mechanics}}}\ (\bibinfo  {publisher} {Addison-Wesley},\ \bibinfo {year}
  {1980})\BibitemShut {NoStop}%
\bibitem [{\citenamefont {Gao}\ and\ \citenamefont
  {Han}(2012)}]{gao2012implementing}%
  \BibitemOpen
  \bibfield  {author} {\bibinfo {author} {\bibfnamefont {F.}~\bibnamefont
  {Gao}}\ and\ \bibinfo {author} {\bibfnamefont {L.}~\bibnamefont {Han}},\
  }\bibfield  {title} {\bibinfo {title} {Implementing the nelder-mead simplex
  algorithm with adaptive parameters},\ }\href@noop {} {\bibfield  {journal}
  {\bibinfo  {journal} {Computational Optimization and Applications}\ }\textbf
  {\bibinfo {volume} {51}},\ \bibinfo {pages} {259} (\bibinfo {year}
  {2012})}\BibitemShut {NoStop}%
\bibitem [{\citenamefont {v.~H{\'a}mos}(1932)}]{vonHamos_1932}%
  \BibitemOpen
  \bibfield  {author} {\bibinfo {author} {\bibfnamefont {L.}~\bibnamefont
  {v.~H{\'a}mos}},\ }\bibfield  {title} {\bibinfo {title}
  {R{\"o}ntgenspektroskopie und abbildung mittels gekr{\"u}mmter
  kristallreflektoren},\ }\href {https://doi.org/10.1007/BF01494468} {\bibfield
   {journal} {\bibinfo  {journal} {Naturwissenschaften}\ }\textbf {\bibinfo
  {volume} {20}},\ \bibinfo {pages} {705} (\bibinfo {year} {1932})}\BibitemShut
  {NoStop}%
\bibitem [{\citenamefont {Forte}\ \emph {et~al.}(2024)\citenamefont {Forte},
  \citenamefont {Gawne}, \citenamefont {Alaa El-Din}, \citenamefont
  {Humphries}, \citenamefont {Preston}, \citenamefont {Cr{\'e}pisson},
  \citenamefont {Campbell}, \citenamefont {Svensson}, \citenamefont {Azadi},
  \citenamefont {Heighway} \emph {et~al.}}]{forte2024resonant}%
  \BibitemOpen
  \bibfield  {author} {\bibinfo {author} {\bibfnamefont {A.}~\bibnamefont
  {Forte}}, \bibinfo {author} {\bibfnamefont {T.}~\bibnamefont {Gawne}},
  \bibinfo {author} {\bibfnamefont {K.~K.}\ \bibnamefont {Alaa El-Din}},
  \bibinfo {author} {\bibfnamefont {O.~S.}\ \bibnamefont {Humphries}}, \bibinfo
  {author} {\bibfnamefont {T.~R.}\ \bibnamefont {Preston}}, \bibinfo {author}
  {\bibfnamefont {C.}~\bibnamefont {Cr{\'e}pisson}}, \bibinfo {author}
  {\bibfnamefont {T.}~\bibnamefont {Campbell}}, \bibinfo {author}
  {\bibfnamefont {P.}~\bibnamefont {Svensson}}, \bibinfo {author}
  {\bibfnamefont {S.}~\bibnamefont {Azadi}}, \bibinfo {author} {\bibfnamefont
  {P.}~\bibnamefont {Heighway}}, \emph {et~al.},\ }\bibfield  {title} {\bibinfo
  {title} {Resonant inelastic x-ray scattering in warm-dense fe compounds
  beyond the sase fel resolution limit},\ }\href@noop {} {\bibfield  {journal}
  {\bibinfo  {journal} {Communications Physics}\ }\textbf {\bibinfo {volume}
  {7}},\ \bibinfo {pages} {266} (\bibinfo {year} {2024})}\BibitemShut {NoStop}%
\bibitem [{\citenamefont {Taupin}(1964)}]{taupin1964theorie}%
  \BibitemOpen
  \bibfield  {author} {\bibinfo {author} {\bibfnamefont {D.}~\bibnamefont
  {Taupin}},\ }\bibfield  {title} {\bibinfo {title} {Th{\'e}orie dynamique de
  la diffraction des rayons x par les cristaux d{\'e}form{\'e}s},\ }\href@noop
  {} {\bibfield  {journal} {\bibinfo  {journal} {Bulletin de Min{\'e}ralogie}\
  }\textbf {\bibinfo {volume} {87}},\ \bibinfo {pages} {469} (\bibinfo {year}
  {1964})}\BibitemShut {NoStop}%
\bibitem [{\citenamefont {Uschmann}\ \emph {et~al.}(1993)\citenamefont
  {Uschmann}, \citenamefont {F{\"o}rster}, \citenamefont {G{\"a}bel},
  \citenamefont {H{\"o}lzer},\ and\ \citenamefont {Ensslen}}]{uschmann1993x}%
  \BibitemOpen
  \bibfield  {author} {\bibinfo {author} {\bibfnamefont {I.}~\bibnamefont
  {Uschmann}}, \bibinfo {author} {\bibfnamefont {E.}~\bibnamefont
  {F{\"o}rster}}, \bibinfo {author} {\bibfnamefont {K.}~\bibnamefont
  {G{\"a}bel}}, \bibinfo {author} {\bibfnamefont {G.}~\bibnamefont
  {H{\"o}lzer}},\ and\ \bibinfo {author} {\bibfnamefont {M.}~\bibnamefont
  {Ensslen}},\ }\bibfield  {title} {\bibinfo {title} {X-ray reflection
  properties of elastically bent perfect crystals in bragg geometry},\
  }\href@noop {} {\bibfield  {journal} {\bibinfo  {journal} {Journal of applied
  crystallography}\ }\textbf {\bibinfo {volume} {26}},\ \bibinfo {pages} {405}
  (\bibinfo {year} {1993})}\BibitemShut {NoStop}%
\bibitem [{\citenamefont {Gawne}\ \emph {et~al.}(2024)\citenamefont {Gawne},
  \citenamefont {Bellenbaum}, \citenamefont {Fletcher}, \citenamefont {Appel},
  \citenamefont {Baehtz}, \citenamefont {Bouffetier}, \citenamefont
  {Brambrink}, \citenamefont {Brown}, \citenamefont {Cangi}, \citenamefont
  {Descamps}, \citenamefont {Göde}, \citenamefont {Hartley}, \citenamefont
  {Herbert}, \citenamefont {Hesselbach}, \citenamefont {Höppner},
  \citenamefont {Humphries}, \citenamefont {Konôpková}, \citenamefont {Laso},
  \citenamefont {Lindqvist}, \citenamefont {Lütgert}, \citenamefont
  {MacDonald}, \citenamefont {Makita}, \citenamefont {Martin}, \citenamefont
  {Mishchenko}, \citenamefont {Moldabekov}, \citenamefont {Nakatsutsumi},
  \citenamefont {Naedler}, \citenamefont {Neumayer}, \citenamefont {Pelka},
  \citenamefont {Qu}, \citenamefont {Randolph}, \citenamefont {Rips},
  \citenamefont {Toncian}, \citenamefont {Vorberger}, \citenamefont
  {Wollenweber}, \citenamefont {Zastrau}, \citenamefont {Kraus}, \citenamefont
  {Preston},\ and\ \citenamefont
  {Dornheim}}]{gawne2024effectsmosaiccrystalinstrument}%
  \BibitemOpen
  \bibfield  {author} {\bibinfo {author} {\bibfnamefont {T.}~\bibnamefont
  {Gawne}}, \bibinfo {author} {\bibfnamefont {H.}~\bibnamefont {Bellenbaum}},
  \bibinfo {author} {\bibfnamefont {L.~B.}\ \bibnamefont {Fletcher}}, \bibinfo
  {author} {\bibfnamefont {K.}~\bibnamefont {Appel}}, \bibinfo {author}
  {\bibfnamefont {C.}~\bibnamefont {Baehtz}}, \bibinfo {author} {\bibfnamefont
  {V.}~\bibnamefont {Bouffetier}}, \bibinfo {author} {\bibfnamefont
  {E.}~\bibnamefont {Brambrink}}, \bibinfo {author} {\bibfnamefont
  {D.}~\bibnamefont {Brown}}, \bibinfo {author} {\bibfnamefont
  {A.}~\bibnamefont {Cangi}}, \bibinfo {author} {\bibfnamefont
  {A.}~\bibnamefont {Descamps}}, \bibinfo {author} {\bibfnamefont
  {S.}~\bibnamefont {Göde}}, \bibinfo {author} {\bibfnamefont {N.~J.}\
  \bibnamefont {Hartley}}, \bibinfo {author} {\bibfnamefont {M.-L.}\
  \bibnamefont {Herbert}}, \bibinfo {author} {\bibfnamefont {P.}~\bibnamefont
  {Hesselbach}}, \bibinfo {author} {\bibfnamefont {H.}~\bibnamefont
  {Höppner}}, \bibinfo {author} {\bibfnamefont {O.~S.}\ \bibnamefont
  {Humphries}}, \bibinfo {author} {\bibfnamefont {Z.}~\bibnamefont
  {Konôpková}}, \bibinfo {author} {\bibfnamefont {A.}~\bibnamefont {Laso}},
  \bibinfo {author} {\bibfnamefont {B.}~\bibnamefont {Lindqvist}}, \bibinfo
  {author} {\bibfnamefont {J.}~\bibnamefont {Lütgert}}, \bibinfo {author}
  {\bibfnamefont {M.~J.}\ \bibnamefont {MacDonald}}, \bibinfo {author}
  {\bibfnamefont {M.}~\bibnamefont {Makita}}, \bibinfo {author} {\bibfnamefont
  {W.}~\bibnamefont {Martin}}, \bibinfo {author} {\bibfnamefont
  {M.}~\bibnamefont {Mishchenko}}, \bibinfo {author} {\bibfnamefont {Z.~A.}\
  \bibnamefont {Moldabekov}}, \bibinfo {author} {\bibfnamefont
  {M.}~\bibnamefont {Nakatsutsumi}}, \bibinfo {author} {\bibfnamefont {J.-P.}\
  \bibnamefont {Naedler}}, \bibinfo {author} {\bibfnamefont {P.}~\bibnamefont
  {Neumayer}}, \bibinfo {author} {\bibfnamefont {A.}~\bibnamefont {Pelka}},
  \bibinfo {author} {\bibfnamefont {C.}~\bibnamefont {Qu}}, \bibinfo {author}
  {\bibfnamefont {L.}~\bibnamefont {Randolph}}, \bibinfo {author}
  {\bibfnamefont {J.}~\bibnamefont {Rips}}, \bibinfo {author} {\bibfnamefont
  {T.}~\bibnamefont {Toncian}}, \bibinfo {author} {\bibfnamefont
  {J.}~\bibnamefont {Vorberger}}, \bibinfo {author} {\bibfnamefont
  {L.}~\bibnamefont {Wollenweber}}, \bibinfo {author} {\bibfnamefont
  {U.}~\bibnamefont {Zastrau}}, \bibinfo {author} {\bibfnamefont
  {D.}~\bibnamefont {Kraus}}, \bibinfo {author} {\bibfnamefont {T.~R.}\
  \bibnamefont {Preston}},\ and\ \bibinfo {author} {\bibfnamefont
  {T.}~\bibnamefont {Dornheim}},\ }\href {https://arxiv.org/abs/2406.03301}
  {\bibinfo {title} {Effects of mosaic crystal instrument functions on x-ray
  thomson scattering diagnostics}} (\bibinfo {year} {2024}),\ \Eprint
  {https://arxiv.org/abs/2406.03301} {arXiv:2406.03301 [physics.plasm-ph]}
  \BibitemShut {NoStop}%
\bibitem [{\citenamefont {et~al.}(2020)}]{preston}%
  \BibitemOpen
  \bibfield  {author} {\bibinfo {author} {\bibfnamefont {T.~P.}\ \bibnamefont
  {et~al.}},\ }\bibfield  {title} {\bibinfo {title} {Design and performance
  characterisation of the hapg von hámos spectrometer at the high energy
  density instrument of the european xfel},\ }\bibfield  {journal} {\bibinfo
  {journal} {Journal of instrumentation}\ }\textbf {\bibinfo {volume} {13}},\
  \href {https://doi.org/0.1088/1748-0221/15/11/P11033}
  {0.1088/1748-0221/15/11/P11033} (\bibinfo {year} {2020})\BibitemShut
  {NoStop}%
\bibitem [{\citenamefont {Zastrau}\ \emph {et~al.}(2012)\citenamefont
  {Zastrau}, \citenamefont {Brown}, \citenamefont {Döppner}, \citenamefont
  {Glenzer}, \citenamefont {Gregori}, \citenamefont {Lee}, \citenamefont
  {Marschner}, \citenamefont {Toleikis}, \citenamefont {Wehrhan},\ and\
  \citenamefont {Förster}}]{2012_Zastrau_Focal}%
  \BibitemOpen
  \bibfield  {author} {\bibinfo {author} {\bibfnamefont {U.}~\bibnamefont
  {Zastrau}}, \bibinfo {author} {\bibfnamefont {C.~R.~D.}\ \bibnamefont
  {Brown}}, \bibinfo {author} {\bibfnamefont {T.}~\bibnamefont {Döppner}},
  \bibinfo {author} {\bibfnamefont {S.~H.}\ \bibnamefont {Glenzer}}, \bibinfo
  {author} {\bibfnamefont {G.}~\bibnamefont {Gregori}}, \bibinfo {author}
  {\bibfnamefont {H.~J.}\ \bibnamefont {Lee}}, \bibinfo {author} {\bibfnamefont
  {H.}~\bibnamefont {Marschner}}, \bibinfo {author} {\bibfnamefont
  {S.}~\bibnamefont {Toleikis}}, \bibinfo {author} {\bibfnamefont
  {O.}~\bibnamefont {Wehrhan}},\ and\ \bibinfo {author} {\bibfnamefont
  {E.}~\bibnamefont {Förster}},\ }\bibfield  {title} {\bibinfo {title} {{Focal
  aberrations of large-aperture HOPG von-Hàmos x-ray spectrometers}},\ }\href
  {https://doi.org/10.1088/1748-0221/7/09/P09015} {\bibfield  {journal}
  {\bibinfo  {journal} {Journal of Instrumentation}\ }\textbf {\bibinfo
  {volume} {7}}\bibinfo  {number} { (09)},\ \bibinfo {pages}
  {P09015}}\BibitemShut {NoStop}%
\bibitem [{\citenamefont {Zastrau}\ \emph {et~al.}(2013)\citenamefont
  {Zastrau}, \citenamefont {Woldegeorgis}, \citenamefont {F{\"o}rster},
  \citenamefont {Loetzsch}, \citenamefont {Marschner},\ and\ \citenamefont
  {Uschmann}}]{zastrau2013characterization}%
  \BibitemOpen
\bibfield  {number} {  }\bibfield  {author} {\bibinfo {author} {\bibfnamefont
  {U.}~\bibnamefont {Zastrau}}, \bibinfo {author} {\bibfnamefont
  {A.}~\bibnamefont {Woldegeorgis}}, \bibinfo {author} {\bibfnamefont
  {E.}~\bibnamefont {F{\"o}rster}}, \bibinfo {author} {\bibfnamefont
  {R.}~\bibnamefont {Loetzsch}}, \bibinfo {author} {\bibfnamefont
  {H.}~\bibnamefont {Marschner}},\ and\ \bibinfo {author} {\bibfnamefont
  {I.}~\bibnamefont {Uschmann}},\ }\bibfield  {title} {\bibinfo {title}
  {Characterization of strongly-bent hapg crystals for von-h{\'a}mos x-ray
  spectrographs},\ }\href@noop {} {\bibfield  {journal} {\bibinfo  {journal}
  {Journal of Instrumentation}\ }\textbf {\bibinfo {volume} {8}}\bibinfo
  {number} { (10)},\ \bibinfo {pages} {P10006}}\BibitemShut {NoStop}%
\bibitem [{\citenamefont {Glenzer}\ \emph {et~al.}(2016)\citenamefont
  {Glenzer}, \citenamefont {Fletcher}, \citenamefont {Galtier}, \citenamefont
  {Nagler}, \citenamefont {Alonso-Mori}, \citenamefont {Barbrel}, \citenamefont
  {Brown}, \citenamefont {Chapman}, \citenamefont {Chen}, \citenamefont {Curry}
  \emph {et~al.}}]{glenzer2016matter}%
  \BibitemOpen
\bibfield  {number} {  }\bibfield  {author} {\bibinfo {author} {\bibfnamefont
  {S.}~\bibnamefont {Glenzer}}, \bibinfo {author} {\bibfnamefont
  {L.}~\bibnamefont {Fletcher}}, \bibinfo {author} {\bibfnamefont
  {E.}~\bibnamefont {Galtier}}, \bibinfo {author} {\bibfnamefont
  {B.}~\bibnamefont {Nagler}}, \bibinfo {author} {\bibfnamefont
  {R.}~\bibnamefont {Alonso-Mori}}, \bibinfo {author} {\bibfnamefont
  {B.}~\bibnamefont {Barbrel}}, \bibinfo {author} {\bibfnamefont
  {S.}~\bibnamefont {Brown}}, \bibinfo {author} {\bibfnamefont
  {D.}~\bibnamefont {Chapman}}, \bibinfo {author} {\bibfnamefont
  {Z.}~\bibnamefont {Chen}}, \bibinfo {author} {\bibfnamefont {C.}~\bibnamefont
  {Curry}}, \emph {et~al.},\ }\bibfield  {title} {\bibinfo {title} {Matter
  under extreme conditions experiments at the linac coherent light source},\
  }\href@noop {} {\bibfield  {journal} {\bibinfo  {journal} {Journal of Physics
  B: Atomic, Molecular and Optical Physics}\ }\textbf {\bibinfo {volume}
  {49}},\ \bibinfo {pages} {092001} (\bibinfo {year} {2016})}\BibitemShut
  {NoStop}%
\bibitem [{\citenamefont {MacDonald}\ \emph {et~al.}(2021)\citenamefont
  {MacDonald}, \citenamefont {Saunders}, \citenamefont {Bachmann},
  \citenamefont {Bethkenhagen}, \citenamefont {Divol}, \citenamefont {Doyle},
  \citenamefont {Fletcher}, \citenamefont {Glenzer}, \citenamefont {Kraus},
  \citenamefont {Landen} \emph {et~al.}}]{MacDonald_PoP_2021}%
  \BibitemOpen
  \bibfield  {author} {\bibinfo {author} {\bibfnamefont {M.}~\bibnamefont
  {MacDonald}}, \bibinfo {author} {\bibfnamefont {A.}~\bibnamefont {Saunders}},
  \bibinfo {author} {\bibfnamefont {B.}~\bibnamefont {Bachmann}}, \bibinfo
  {author} {\bibfnamefont {M.}~\bibnamefont {Bethkenhagen}}, \bibinfo {author}
  {\bibfnamefont {L.}~\bibnamefont {Divol}}, \bibinfo {author} {\bibfnamefont
  {M.}~\bibnamefont {Doyle}}, \bibinfo {author} {\bibfnamefont
  {L.}~\bibnamefont {Fletcher}}, \bibinfo {author} {\bibfnamefont
  {S.}~\bibnamefont {Glenzer}}, \bibinfo {author} {\bibfnamefont
  {D.}~\bibnamefont {Kraus}}, \bibinfo {author} {\bibfnamefont
  {O.}~\bibnamefont {Landen}}, \emph {et~al.},\ }\bibfield  {title} {\bibinfo
  {title} {{Demonstration of a laser-driven, narrow spectral bandwidth x-ray
  source for collective x-ray scattering experiments}},\ }\href@noop {}
  {\bibfield  {journal} {\bibinfo  {journal} {Physics of Plasmas}\ }\textbf
  {\bibinfo {volume} {28}} (\bibinfo {year} {2021})}\BibitemShut {NoStop}%
\bibitem [{\citenamefont {Mozzanica}\ \emph {et~al.}(2018)\citenamefont
  {Mozzanica}, \citenamefont {Andr{\"a}}, \citenamefont {Barten}, \citenamefont
  {Bergamaschi}, \citenamefont {Chiriotti}, \citenamefont {Br{\"u}ckner},
  \citenamefont {Dinapoli}, \citenamefont {Fr{\"o}jdh}, \citenamefont
  {Greiffenberg}, \citenamefont {Leonarski} \emph
  {et~al.}}]{mozzanica2018jungfrau}%
  \BibitemOpen
  \bibfield  {author} {\bibinfo {author} {\bibfnamefont {A.}~\bibnamefont
  {Mozzanica}}, \bibinfo {author} {\bibfnamefont {M.}~\bibnamefont
  {Andr{\"a}}}, \bibinfo {author} {\bibfnamefont {R.}~\bibnamefont {Barten}},
  \bibinfo {author} {\bibfnamefont {A.}~\bibnamefont {Bergamaschi}}, \bibinfo
  {author} {\bibfnamefont {S.}~\bibnamefont {Chiriotti}}, \bibinfo {author}
  {\bibfnamefont {M.}~\bibnamefont {Br{\"u}ckner}}, \bibinfo {author}
  {\bibfnamefont {R.}~\bibnamefont {Dinapoli}}, \bibinfo {author}
  {\bibfnamefont {E.}~\bibnamefont {Fr{\"o}jdh}}, \bibinfo {author}
  {\bibfnamefont {D.}~\bibnamefont {Greiffenberg}}, \bibinfo {author}
  {\bibfnamefont {F.}~\bibnamefont {Leonarski}}, \emph {et~al.},\ }\bibfield
  {title} {\bibinfo {title} {The jungfrau detector for applications at
  synchrotron light sources and xfels},\ }\href@noop {} {\bibfield  {journal}
  {\bibinfo  {journal} {Synchrotron Radiation News}\ }\textbf {\bibinfo
  {volume} {31}},\ \bibinfo {pages} {16} (\bibinfo {year} {2018})}\BibitemShut
  {NoStop}%
\bibitem [{\citenamefont {Chalupsk\'{y}}\ \emph {et~al.}(2010)\citenamefont
  {Chalupsk\'{y}}, \citenamefont {Krzywinski}, \citenamefont {Juha},
  \citenamefont {H\'{a}jkov\'{a}}, \citenamefont {Cihelka}, \citenamefont
  {Burian}, \citenamefont {Vy\v{s}\'{i}n}, \citenamefont {Gaudin},
  \citenamefont {Gleeson}, \citenamefont {Jurek}, \citenamefont {Khorsand},
  \citenamefont {Klinger}, \citenamefont {Wabnitz}, \citenamefont
  {Sobierajski}, \citenamefont {St\"{o}rmer}, \citenamefont {Tiedtke},\ and\
  \citenamefont {Toleikis}}]{Chalupsky_Opt_2010}%
  \BibitemOpen
  \bibfield  {author} {\bibinfo {author} {\bibfnamefont {J.}~\bibnamefont
  {Chalupsk\'{y}}}, \bibinfo {author} {\bibfnamefont {J.}~\bibnamefont
  {Krzywinski}}, \bibinfo {author} {\bibfnamefont {L.}~\bibnamefont {Juha}},
  \bibinfo {author} {\bibfnamefont {V.}~\bibnamefont {H\'{a}jkov\'{a}}},
  \bibinfo {author} {\bibfnamefont {J.}~\bibnamefont {Cihelka}}, \bibinfo
  {author} {\bibfnamefont {T.}~\bibnamefont {Burian}}, \bibinfo {author}
  {\bibfnamefont {L.}~\bibnamefont {Vy\v{s}\'{i}n}}, \bibinfo {author}
  {\bibfnamefont {J.}~\bibnamefont {Gaudin}}, \bibinfo {author} {\bibfnamefont
  {A.}~\bibnamefont {Gleeson}}, \bibinfo {author} {\bibfnamefont
  {M.}~\bibnamefont {Jurek}}, \bibinfo {author} {\bibfnamefont {A.~R.}\
  \bibnamefont {Khorsand}}, \bibinfo {author} {\bibfnamefont {D.}~\bibnamefont
  {Klinger}}, \bibinfo {author} {\bibfnamefont {H.}~\bibnamefont {Wabnitz}},
  \bibinfo {author} {\bibfnamefont {R.}~\bibnamefont {Sobierajski}}, \bibinfo
  {author} {\bibfnamefont {M.}~\bibnamefont {St\"{o}rmer}}, \bibinfo {author}
  {\bibfnamefont {K.}~\bibnamefont {Tiedtke}},\ and\ \bibinfo {author}
  {\bibfnamefont {S.}~\bibnamefont {Toleikis}},\ }\bibfield  {title} {\bibinfo
  {title} {{Spot size characterization of focused non-Gaussian X-ray laser
  beams}},\ }\href {https://doi.org/10.1364/OE.18.027836} {\bibfield  {journal}
  {\bibinfo  {journal} {Opt. Express}\ }\textbf {\bibinfo {volume} {18}},\
  \bibinfo {pages} {27836} (\bibinfo {year} {2010})}\BibitemShut {NoStop}%
\bibitem [{\citenamefont {Schlesiger}\ \emph {et~al.}(2017)\citenamefont
  {Schlesiger}, \citenamefont {Anklamm}, \citenamefont {Malzer}, \citenamefont
  {Gnewkow},\ and\ \citenamefont {Kanngie{\ss}er}}]{Schlesiger_JAC_2017}%
  \BibitemOpen
  \bibfield  {author} {\bibinfo {author} {\bibfnamefont {C.}~\bibnamefont
  {Schlesiger}}, \bibinfo {author} {\bibfnamefont {L.}~\bibnamefont {Anklamm}},
  \bibinfo {author} {\bibfnamefont {W.}~\bibnamefont {Malzer}}, \bibinfo
  {author} {\bibfnamefont {R.}~\bibnamefont {Gnewkow}},\ and\ \bibinfo {author}
  {\bibfnamefont {B.}~\bibnamefont {Kanngie{\ss}er}},\ }\bibfield  {title}
  {\bibinfo {title} {{A new model for the description of X-ray diffraction from
  mosaic crystals for ray-tracing calculations}},\ }\href@noop {} {\bibfield
  {journal} {\bibinfo  {journal} {Journal of Applied Crystallography}\ }\textbf
  {\bibinfo {volume} {50}},\ \bibinfo {pages} {1490} (\bibinfo {year}
  {2017})}\BibitemShut {NoStop}%
\bibitem [{\citenamefont {Dusi}\ \emph {et~al.}(2003)\citenamefont {Dusi},
  \citenamefont {Auricchio}, \citenamefont {Brigliadori}, \citenamefont
  {Donati}, \citenamefont {Landini}, \citenamefont {Mengoni}, \citenamefont
  {Perillo},\ and\ \citenamefont {Ventura}}]{dusi2003study}%
  \BibitemOpen
  \bibfield  {author} {\bibinfo {author} {\bibfnamefont {W.}~\bibnamefont
  {Dusi}}, \bibinfo {author} {\bibfnamefont {N.}~\bibnamefont {Auricchio}},
  \bibinfo {author} {\bibfnamefont {L.}~\bibnamefont {Brigliadori}}, \bibinfo
  {author} {\bibfnamefont {A.}~\bibnamefont {Donati}}, \bibinfo {author}
  {\bibfnamefont {G.}~\bibnamefont {Landini}}, \bibinfo {author} {\bibfnamefont
  {D.}~\bibnamefont {Mengoni}}, \bibinfo {author} {\bibfnamefont
  {E.}~\bibnamefont {Perillo}},\ and\ \bibinfo {author} {\bibfnamefont
  {G.}~\bibnamefont {Ventura}},\ }\bibfield  {title} {\bibinfo {title} {A study
  of the spectroscopic response of planar cdte detectors when irradiated at
  various angles of incidence},\ }\href@noop {} {\bibfield  {journal} {\bibinfo
   {journal} {Nuclear Instruments and Methods in Physics Research Section A:
  Accelerators, Spectrometers, Detectors and Associated Equipment}\ }\textbf
  {\bibinfo {volume} {506}},\ \bibinfo {pages} {119} (\bibinfo {year}
  {2003})}\BibitemShut {NoStop}%
\bibitem [{\citenamefont {Gerlach}\ \emph {et~al.}(2015)\citenamefont
  {Gerlach}, \citenamefont {Anklamm}, \citenamefont {Antonov}, \citenamefont
  {Grigorieva}, \citenamefont {Holfelder}, \citenamefont {Kanngie{\ss}er},
  \citenamefont {Legall}, \citenamefont {Malzer}, \citenamefont {Schlesiger},\
  and\ \citenamefont {Beckhoff}}]{Gerlach_JAC_2015}%
  \BibitemOpen
  \bibfield  {author} {\bibinfo {author} {\bibfnamefont {M.}~\bibnamefont
  {Gerlach}}, \bibinfo {author} {\bibfnamefont {L.}~\bibnamefont {Anklamm}},
  \bibinfo {author} {\bibfnamefont {A.}~\bibnamefont {Antonov}}, \bibinfo
  {author} {\bibfnamefont {I.}~\bibnamefont {Grigorieva}}, \bibinfo {author}
  {\bibfnamefont {I.}~\bibnamefont {Holfelder}}, \bibinfo {author}
  {\bibfnamefont {B.}~\bibnamefont {Kanngie{\ss}er}}, \bibinfo {author}
  {\bibfnamefont {H.}~\bibnamefont {Legall}}, \bibinfo {author} {\bibfnamefont
  {W.}~\bibnamefont {Malzer}}, \bibinfo {author} {\bibfnamefont
  {C.}~\bibnamefont {Schlesiger}},\ and\ \bibinfo {author} {\bibfnamefont
  {B.}~\bibnamefont {Beckhoff}},\ }\bibfield  {title} {\bibinfo {title}
  {{Characterization of HAPG mosaic crystals using synchrotron radiation}},\
  }\href@noop {} {\bibfield  {journal} {\bibinfo  {journal} {Journal of Applied
  Crystallography}\ }\textbf {\bibinfo {volume} {48}},\ \bibinfo {pages} {1381}
  (\bibinfo {year} {2015})}\BibitemShut {NoStop}%
\bibitem [{\citenamefont {Feldhaus}\ \emph {et~al.}(1997)\citenamefont
  {Feldhaus}, \citenamefont {Saldin}, \citenamefont {Schneider}, \citenamefont
  {Schneidmiller},\ and\ \citenamefont {Yurkov}}]{feldhaus1997possible}%
  \BibitemOpen
  \bibfield  {author} {\bibinfo {author} {\bibfnamefont {J.}~\bibnamefont
  {Feldhaus}}, \bibinfo {author} {\bibfnamefont {E.}~\bibnamefont {Saldin}},
  \bibinfo {author} {\bibfnamefont {J.}~\bibnamefont {Schneider}}, \bibinfo
  {author} {\bibfnamefont {E.}~\bibnamefont {Schneidmiller}},\ and\ \bibinfo
  {author} {\bibfnamefont {M.}~\bibnamefont {Yurkov}},\ }\bibfield  {title}
  {\bibinfo {title} {Possible application of x-ray optical elements for
  reducing the spectral bandwidth of an x-ray sase fel},\ }\href@noop {}
  {\bibfield  {journal} {\bibinfo  {journal} {Optics Communications}\ }\textbf
  {\bibinfo {volume} {140}},\ \bibinfo {pages} {341} (\bibinfo {year}
  {1997})}\BibitemShut {NoStop}%
\bibitem [{\citenamefont {Wollenweber}\ \emph {et~al.}(2021)\citenamefont
  {Wollenweber}, \citenamefont {Preston}, \citenamefont {Descamps},
  \citenamefont {Cerantola}, \citenamefont {Comley}, \citenamefont {Eggert},
  \citenamefont {Fletcher}, \citenamefont {Geloni}, \citenamefont {Gericke},
  \citenamefont {Glenzer} \emph {et~al.}}]{wollenweber2021high}%
  \BibitemOpen
  \bibfield  {author} {\bibinfo {author} {\bibfnamefont {L.}~\bibnamefont
  {Wollenweber}}, \bibinfo {author} {\bibfnamefont {T.}~\bibnamefont
  {Preston}}, \bibinfo {author} {\bibfnamefont {A.}~\bibnamefont {Descamps}},
  \bibinfo {author} {\bibfnamefont {V.}~\bibnamefont {Cerantola}}, \bibinfo
  {author} {\bibfnamefont {A.}~\bibnamefont {Comley}}, \bibinfo {author}
  {\bibfnamefont {J.}~\bibnamefont {Eggert}}, \bibinfo {author} {\bibfnamefont
  {L.}~\bibnamefont {Fletcher}}, \bibinfo {author} {\bibfnamefont
  {G.}~\bibnamefont {Geloni}}, \bibinfo {author} {\bibfnamefont
  {D.}~\bibnamefont {Gericke}}, \bibinfo {author} {\bibfnamefont
  {S.}~\bibnamefont {Glenzer}}, \emph {et~al.},\ }\bibfield  {title} {\bibinfo
  {title} {High-resolution inelastic x-ray scattering at the high energy
  density scientific instrument at the european x-ray free-electron laser},\
  }\href@noop {} {\bibfield  {journal} {\bibinfo  {journal} {Review of
  scientific instruments}\ }\textbf {\bibinfo {volume} {92}} (\bibinfo {year}
  {2021})}\BibitemShut {NoStop}%
\bibitem [{\citenamefont {Voigt}\ \emph {et~al.}(2021)\citenamefont {Voigt},
  \citenamefont {Zhang}, \citenamefont {Ramakrishna}, \citenamefont
  {Amouretti}, \citenamefont {Appel}, \citenamefont {Brambrink}, \citenamefont
  {Cerantola}, \citenamefont {Chekrygina}, \citenamefont {Döppner},
  \citenamefont {Falcone}, \citenamefont {Falk}, \citenamefont {Fletcher},
  \citenamefont {Gericke}, \citenamefont {Göde}, \citenamefont {Harmand},
  \citenamefont {Hartley}, \citenamefont {Hau-Riege}, \citenamefont {Huang},
  \citenamefont {Humphries}, \citenamefont {Lokamani}, \citenamefont {Makita},
  \citenamefont {Pelka}, \citenamefont {Prescher}, \citenamefont {Schuster},
  \citenamefont {Šmíd}, \citenamefont {Toncian}, \citenamefont {Vorberger},
  \citenamefont {Zastrau}, \citenamefont {Preston},\ and\ \citenamefont
  {Kraus}}]{Voigt_PoP_20201}%
  \BibitemOpen
  \bibfield  {author} {\bibinfo {author} {\bibfnamefont {K.}~\bibnamefont
  {Voigt}}, \bibinfo {author} {\bibfnamefont {M.}~\bibnamefont {Zhang}},
  \bibinfo {author} {\bibfnamefont {K.}~\bibnamefont {Ramakrishna}}, \bibinfo
  {author} {\bibfnamefont {A.}~\bibnamefont {Amouretti}}, \bibinfo {author}
  {\bibfnamefont {K.}~\bibnamefont {Appel}}, \bibinfo {author} {\bibfnamefont
  {E.}~\bibnamefont {Brambrink}}, \bibinfo {author} {\bibfnamefont
  {V.}~\bibnamefont {Cerantola}}, \bibinfo {author} {\bibfnamefont
  {D.}~\bibnamefont {Chekrygina}}, \bibinfo {author} {\bibfnamefont
  {T.}~\bibnamefont {Döppner}}, \bibinfo {author} {\bibfnamefont {R.~W.}\
  \bibnamefont {Falcone}}, \bibinfo {author} {\bibfnamefont {K.}~\bibnamefont
  {Falk}}, \bibinfo {author} {\bibfnamefont {L.~B.}\ \bibnamefont {Fletcher}},
  \bibinfo {author} {\bibfnamefont {D.~O.}\ \bibnamefont {Gericke}}, \bibinfo
  {author} {\bibfnamefont {S.}~\bibnamefont {Göde}}, \bibinfo {author}
  {\bibfnamefont {M.}~\bibnamefont {Harmand}}, \bibinfo {author} {\bibfnamefont
  {N.~J.}\ \bibnamefont {Hartley}}, \bibinfo {author} {\bibfnamefont {S.~P.}\
  \bibnamefont {Hau-Riege}}, \bibinfo {author} {\bibfnamefont {L.~G.}\
  \bibnamefont {Huang}}, \bibinfo {author} {\bibfnamefont {O.~S.}\ \bibnamefont
  {Humphries}}, \bibinfo {author} {\bibfnamefont {M.}~\bibnamefont {Lokamani}},
  \bibinfo {author} {\bibfnamefont {M.}~\bibnamefont {Makita}}, \bibinfo
  {author} {\bibfnamefont {A.}~\bibnamefont {Pelka}}, \bibinfo {author}
  {\bibfnamefont {C.}~\bibnamefont {Prescher}}, \bibinfo {author}
  {\bibfnamefont {A.~K.}\ \bibnamefont {Schuster}}, \bibinfo {author}
  {\bibfnamefont {M.}~\bibnamefont {Šmíd}}, \bibinfo {author} {\bibfnamefont
  {T.}~\bibnamefont {Toncian}}, \bibinfo {author} {\bibfnamefont
  {J.}~\bibnamefont {Vorberger}}, \bibinfo {author} {\bibfnamefont
  {U.}~\bibnamefont {Zastrau}}, \bibinfo {author} {\bibfnamefont {T.~R.}\
  \bibnamefont {Preston}},\ and\ \bibinfo {author} {\bibfnamefont
  {D.}~\bibnamefont {Kraus}},\ }\bibfield  {title} {\bibinfo {title}
  {{Demonstration of an x-ray Raman spectroscopy setup to study warm dense
  carbon at the high energy density instrument of European XFEL}},\ }\href
  {https://doi.org/10.1063/5.0048150} {\bibfield  {journal} {\bibinfo
  {journal} {Physics of Plasmas}\ }\textbf {\bibinfo {volume} {28}},\ \bibinfo
  {pages} {082701} (\bibinfo {year} {2021})},\ \Eprint
  {https://arxiv.org/abs/https://pubs.aip.org/aip/pop/article-pdf/doi/10.1063/5.0048150/19770979/082701\_1\_online.pdf}
  {https://pubs.aip.org/aip/pop/article-pdf/doi/10.1063/5.0048150/19770979/082701\_1\_online.pdf}
  \BibitemShut {NoStop}%
\bibitem [{\citenamefont {H\"olzer}\ \emph {et~al.}(1997)\citenamefont
  {H\"olzer}, \citenamefont {Fritsch}, \citenamefont {Deutsch}, \citenamefont
  {H\"artwig},\ and\ \citenamefont {F\"orster}}]{Hoelzer_PRA_1997}%
  \BibitemOpen
  \bibfield  {author} {\bibinfo {author} {\bibfnamefont {G.}~\bibnamefont
  {H\"olzer}}, \bibinfo {author} {\bibfnamefont {M.}~\bibnamefont {Fritsch}},
  \bibinfo {author} {\bibfnamefont {M.}~\bibnamefont {Deutsch}}, \bibinfo
  {author} {\bibfnamefont {J.}~\bibnamefont {H\"artwig}},\ and\ \bibinfo
  {author} {\bibfnamefont {E.}~\bibnamefont {F\"orster}},\ }\bibfield  {title}
  {\bibinfo {title} {$k{\ensuremath{\alpha}}_{1,2}$ and
  $k{\ensuremath{\beta}}_{1,3}$ x-ray emission lines of the $3d$ transition
  metals},\ }\href {https://doi.org/10.1103/PhysRevA.56.4554} {\bibfield
  {journal} {\bibinfo  {journal} {Phys. Rev. A}\ }\textbf {\bibinfo {volume}
  {56}},\ \bibinfo {pages} {4554} (\bibinfo {year} {1997})}\BibitemShut
  {NoStop}%
\bibitem [{\citenamefont {Richardson}(1972)}]{richardson1972bayesian}%
  \BibitemOpen
  \bibfield  {author} {\bibinfo {author} {\bibfnamefont {W.~H.}\ \bibnamefont
  {Richardson}},\ }\bibfield  {title} {\bibinfo {title} {Bayesian-based
  iterative method of image restoration},\ }\href@noop {} {\bibfield  {journal}
  {\bibinfo  {journal} {JoSA}\ }\textbf {\bibinfo {volume} {62}},\ \bibinfo
  {pages} {55} (\bibinfo {year} {1972})}\BibitemShut {NoStop}%
\bibitem [{\citenamefont {Yaffe}\ and\ \citenamefont
  {Rowlands}(1997)}]{yaffe1997x}%
  \BibitemOpen
  \bibfield  {author} {\bibinfo {author} {\bibfnamefont {M.}~\bibnamefont
  {Yaffe}}\ and\ \bibinfo {author} {\bibfnamefont {J.}~\bibnamefont
  {Rowlands}},\ }\bibfield  {title} {\bibinfo {title} {X-ray detectors for
  digital radiography},\ }\href@noop {} {\bibfield  {journal} {\bibinfo
  {journal} {Physics in Medicine \& Biology}\ }\textbf {\bibinfo {volume}
  {42}},\ \bibinfo {pages} {1} (\bibinfo {year} {1997})}\BibitemShut {NoStop}%
\end{thebibliography}%

\end{document}